\documentclass[superscriptaddress,secnumarabic,amssymb,amsmath,nobibnotes,aps,prd,showkeys,nofootinbib,onecolumn,12pt]{revtex4}
\usepackage[utf8]{inputenc} 
\usepackage[T1]{fontenc}    
\usepackage{hyperref}       
\usepackage{url}            
\usepackage{booktabs}       
\usepackage{amsfonts}       
\usepackage{nicefrac}       
\usepackage{microtype}      
\usepackage{lipsum}		
\usepackage{graphicx}
\usepackage{natbib}
\usepackage{doi}
\usepackage{graphicx}
\usepackage{subfigure}
\usepackage{float}
\usepackage{epsf}
\usepackage{bm}
\usepackage{amsmath}
\usepackage{amsfonts}
\usepackage{amssymb}
\usepackage{graphicx}
\usepackage{tabularx}
\usepackage{color}%
\providecommand{\U}[1]{\protect\rule{.1in}{.1in} }

\newcommand{\be}{\begin{equation}}
\newcommand{\ee}{\end{equation}}

\newcommand{\mincir}{\raise
-3.truept\hbox{\rlap{\hbox{$\sim$}}\raise4.truept\hbox{$<$}\ }}
\newcommand{\magcir}{\raise
-3.truept\hbox{\rlap{\hbox{$\sim$}}\raise4.truept\hbox{$>$}\ }}

\usepackage{subcaption}
\usepackage{bigints}
\usepackage{braket}

\providecommand{\U}[1]{\protect\rule{.1in}{.1in}}

\usepackage{tikz,xcolor,hyperref}

\definecolor{lime}{HTML}{A6CE39}
\DeclareRobustCommand{\orcidicon}{%
	\begin{tikzpicture}
	\draw[lime, fill=lime] (0,0) 
	circle [radius=0.16] 
	node[white] {{\fontfamily{qag}\selectfont \tiny ID}};
	\draw[white, fill=white] (-0.0625,0.095) 
	circle [radius=0.007];
	\end{tikzpicture}
	\hspace{-2mm}
}

\foreach \x in {A, ..., Z}{%
	\expandafter\xdef\csname orcid\x\endcsname{\noexpand\href{https//orcid.org/\csname orcidauthor\x\endcsname}{\noexpand\orcidicon}}
}

\begin{document}

\title{Conformal and Non-Minimal Couplings in Fractional Cosmology}

\author{Kevin Marroqu\'in\orcidE{}}
\email{kevin.marroquin@alumnos.ucn.cl}
\affiliation{Departamento de Matem\'{a}ticas, Universidad Cat\'{o}lica del Norte, Avenida
Angamos 0610, Casilla 1280, Antofagasta, 1270709, Chile}

\author{Genly Leon\orcidA{}}
\email{genly.leon@ucn.cl}
\affiliation{Departamento de Matem\'{a}ticas, Universidad Cat\'{o}lica del Norte, Avenida
Angamos 0610, Casilla 1280, Antofagasta, 1270709, Chile}
\affiliation{Institute of Systems Science, Durban University of Technology, P.O. Box 1334,
\mbox{Durban 4000,   South Africa}}
\author{Alfredo D.  Millano\orcidB{}}
\email{alfredo.millano@alumnos.ucn.cl}
\affiliation{Departamento de Matem\'{a}ticas, Universidad Cat\'{o}lica del Norte, Avenida
Angamos 0610, Casilla 1280, Antofagasta, 1270709, Chile}

\author{Claudio Michea\orcidC{}}
\email{claudio.ramirez@ce.ucn.cl}
\affiliation{Departamento de Física, Universidad Católica del Norte, Avenida Angamos 0610,  
Antofagasta, 1270709, Chile}

\author{Andronikos~Paliathanasis\orcidD{}}
\email{anpaliat@phys.uoa.gr}
\affiliation{Institute of Systems Science, Durban University of Technology, P.O. Box 1334,
\mbox{Durban 4000,   South Africa}}
\affiliation{Departamento de Matem\'{a}ticas, Universidad Cat\'{o}lica del Norte, Avenida
Angamos 0610, Casilla 1280, Antofagasta, 1270709, Chile}

\begin{abstract}
Fractional differential calculus is a mathematical tool that has found applications in the study of social and physical behaviors considered ``anomalous''. It is often used when traditional integer derivatives models fail to represent cases where the power law is observed accurately. Fractional calculus must reflect non-local, frequency- and history-dependent properties of power-law phenomena. This tool has various important applications, such as fractional mass conservation, electrochemical analysis, groundwater flow problems, and fractional spatiotemporal diffusion equations. It can also be used in cosmology to explain late-time cosmic acceleration without the need for dark energy. We review some models using fractional differential equations. We look at the Einstein--Hilbert action, which is based on a fractional derivative action, and add a scalar field, $\phi$, to create a non-minimal interaction theory with the coupling, $\xi R \phi^2 $, between gravity and the scalar field, where $\xi$ is the interaction constant. By employing various mathematical approaches, we can offer precise schemes to find analytical and numerical approximations of the solutions. Moreover, we comprehensively study the modified cosmological equations and analyze the solution space using the theory of dynamical systems and asymptotic expansion methods. This enables us to provide a qualitative description of cosmologies with a scalar field based on fractional calculus formalism.
\end{abstract}


\date{\today}
\maketitle

\tableofcontents

\section{Introduction} \label{intr}
Fractional differential calculus is a mathematical tool that has found applications in the study of social and physical behaviors considered ``anomalous''. These phenomena are often described empirically using the power law, which is universal. However, some of the mathematical models that use integer derivatives, including nonlinear models, need to work better in many cases where the power law is observed. Alternative tools like fractional calculus must be introduced to accurately reflect power-law phenomena' non-local, frequency- and history-dependent properties \cite{monje2010fractional, Tarasov2013,bandyopadhyay2014stabilization,padula2014advances,herrmann2014fractional,tarasov2019applications,klafter2012fractional,malinowska2015advanced,lorenzo2016fractional,book1:2006, book2:1999, Uch:2013}.

Fractional calculus has found critical applications in fractional mass conservation. In such cases, a fractional mass conservation equation is needed to model fluid flow when the control volume is not large enough compared to the scale of heterogeneity and when the flow within the volume control is not linear \cite{Wheatcraft-Meerschaert-2008}. Other areas where fractional calculus is useful include electrochemical analysis \cite{Oldham-1972, Pospíšil2019}  and groundwater flow problems \cite{Atangana-Bildik-2013, Atangana-Vermeulen-2014}. In the latter references, Darcy's classical law is generalized by considering the water flow as a function of a non-integer-order derivative of the piezometric height. This generalized law and the law of conservation of mass are then used to derive a new equation for groundwater flow \cite{Atangana-Bildik-2013, Atangana-Vermeulen-2014}. There are various practical applications of fractional calculus, including the fractional advection-dispersion equation \cite{Benson-Wheatcraft-Meerschaert-2000a, Benson-Wheatcraft-Meerschaert-2000b, Benson-Schumer-Meerschaert-Wheatcraft-2001} and fractional spatiotemporal diffusion equations \cite{Metzler-Klafter-2000, Mainardi-Luchko-Pagnini-2007}. An extension of the fractional derivative is the fractional derivative of variable order, which is helpful in defining anomalous diffusion processes in complex media~\cite{Atangana-Kilicman-2014, Gorenflo-Mainardi-2003, Colbrook-Ma-Hopkins-Squire-2017}.  Fractional derivatives simulate viscoelastic damping in materials such as polymers~\cite{Mainardi-2022}. In quantum theory, the fractional Schr\"odinger equation \cite{Laskin-2000, Laskin-2002} and its variations, such as the variable order fractional Schr\"odinger equation \cite{Bhrawy-Zaky-2017}, are significant examples of fractional calculus. 

\textls[-5] Recent studies have shown that fractional calculus is also helpful in modeling fractional derivatives and quantum fields \cite{Lim:2006hp, LimEab+2019+237+256}, quantum gravity and cosmology \cite{El-Nabulsi:2013hsa,El-Nabulsi:2013mwa, Moniz:2020emn,Rasouli:2021lgy,VargasMoniz:2020hve, Calcagni:2020tvw}, black holes~\cite{Vacaru:2010wn,Jalalzadeh:2021gtq}, fractional dynamics and fractional action cosmology \cite{El-Nabulsi:2007wgc,el2009fractional, rami2009fractional,Vacaru:2010wj,Jamil:2011uj,Debnath2012,Debnath2013,Shchigolev:2015rei}, non-minimal coupling in fractional cosmology \cite{El-Nabulsi:2013mwa, Rami:2015kha, Giusti:2020rul}, fractal universe and quantum cosmology~\cite{Calcagni:2009kc, Calcagni:2010bj}, multiscale gravity \cite{Calcagni:2013yqa}, classical and quantum gravity with fractional operators~\cite{Calcagni:2021aap, Calcagni_2021}, multi-fractional spacetimes \cite{Calcagni:2020ads}, quantum gravity and gravitational-wave astronomy~\cite{Calcagni:2019ngc},  cosmic microwave background and inflation \cite{Calcagni:2016ofu}, fractional dynamics from Einstein gravity~\cite{Vacaru:2010fb, Vacaru:2010wn}, and fractional cosmology \cite{Roberts:2009ix, Shchigolev:2010vh,Shchigolev:2012rp,Shchigolev:2013jq,Rami:2015kha} and dark energy, among others \cite{Shchigolev:2021lbm, Jalalzadeh:2022uhl,Garcia-Aspeitia:2022uxz, Landim:2021ial,Landim:2021www, Micolta-Riascos:2023mqo,Gonzalez:2023who, LeonTorres:2023ehd}.

Some important research questions in this field include the following: Can we use a combination of local and global variables to qualitatively describe scalar field cosmologies based on fractional calculus formalism? Additionally, can we develop precise schemes for finding analytical and numerical approximations to solutions by selecting various approaches? To answer these questions, we present a new formulation of gravity with a scalar field that exhibits non-minimal or, more generally, conformal coupling to gravity. The new action includes the Ricci scalar, $R$, the scalar field, $\phi$, with its potential $V(\phi)$, the coupling constant, $\xi$, the matter Lagrangian, $L_m$, the scale factor, $a$, and the lapse function, $N$. The Gamma function, $\Gamma(\mu)$, the fractional parameter, $\mu$, and the equation of state parameter of matter, $w$, are also included. We utilize Caputo's derivative and the formalism of fractional calculus to modify the Friedmann and Klein--Gordon equations. This approach has many applications in describing natural phenomena, making it a unique and compelling framework.

Our research is comprehensive, encompassing a review of the state of the art and confirming that dynamical systems theory, methods of perturbation theory, and the main tools of fractional calculus can be combined to implement a qualitative analysis of gravity models with fractional derivatives. This comprehensive approach is particularly beneficial in cosmologies with scalar fields, conformal coupling, and non-minimal coupling to gravity, broadening previous studies' scope. Our paper thoroughly analyzes the findings from cosmologies involving scalar fields and gravity models with fractional derivatives, specifically focusing on a scalar field with conformal or, generically, non-minimal gravitational coupling. We comment on the use of dynamical systems theory and the tools of fractional calculus in the analysis of cosmological models and propose new cosmologies with scalar fields that have conformal or, in general, non-minimal coupling to gravity within the framework of fractional calculus. Using dynamical systems and asymptotic methods, our qualitative analysis of the models further enhances our understanding. It allows us to draw comprehensive and robust conclusions, instilling confidence in the validity of our findings.
Our research goes beyond theory and has significant practical implications. We start with a thorough literature review that serves as a theoretical framework. We then formulate a research hypothesis that suggests the possibility of using qualitative techniques to obtain relevant information on the flow properties associated with an autonomous system of ordinary differential equations in the cosmological context. This approach, enabled by dynamical systems theory, provides qualitative descriptions of cosmologies with a scalar field in fractional calculus, which can be a valuable tool for future cosmological studies. We also offer schemes for finding analytical approximations of solutions and exact solutions by selecting various approaches, providing a practical guide for researchers in the field. In the fractional context, we analyze cosmologies with a scalar field coupled to gravity with conformal and, generally, a non-minimal coupling. We aim to generalize previous results within the framework of the fractional formulation of gravity and describe the phase space of the models. We obtain qualitative results and explicit and approximate exact solutions with high numerical precision from a mathematical point of view. 

Our methodology first involves reviewing the literature on fractional calculus methods, solution of fractional differential equations, and their applications. The current literature on applying fractional differential calculus to theories of gravity will also be studied, obtaining modifications. The differential equations that describe the dynamics of the different models are to be derived and well-posed for mathematical analysis and physical interpretation. At this point, it is expected to select the work tools and become familiar with Caputo's fractional derivative and its application in formulating a fractional cosmological action and its respective variational calculation. Secondly, systems of ordinary differential equations will be derived and analyzed from a mathematical point of view. This analysis consists of several steps: finding equilibrium points and invariant sets; defining a linearization matrix for each system and evaluating it at each critical point to calculate the eigenvalues; classifying the stability of the points; generating phase space diagrams and numerical solutions for each system for appropriate values of the coupling and fractional parameters. It is expected that we will find each system's future and past attractors and classify each critical point according to its stability and asymptotic behavior. The dynamical behavior of the systems will also be described graphically by plotting orbits that illustrate the qualitative analysis, which allows the description of the cosmological evolution. Finally, we generalize previous results by considering a conformal and non-minimal coupling to gravity. 

The paper is organized as follows. Section \ref{cap3} covers the primary outcomes of fractional calculus, including different approaches to possible fractional derivatives such as Gr\"unwald--Letnikov, Riemann--Liouville, and Caputo derivatives. The emphasis is on defining fractional integration and the fractional derivative. We discuss fractional differential equations and use Wolfram Language 13.3 \citep{Mathematica} to solve physical problems formulated. {We also explore the quantum version of the fractional harmonic oscillator and the relation between fractional calculus and $q$-deformed Lie algebras.} 

In Section \ref{cap4}, we explore fractional cosmologies with non-minimal coupling ($\xi\neq0$), building upon previous results with minimal coupling. By studying different dynamical systems, including the invariant set $\Omega_{\Lambda}=0$, we utilize numerical integration methods to analyze the behavior of solutions to differential equations from the cosmological model. We also investigate the stability of critical points. Our conclusions are presented in Section \ref{conclusions}. 
In Appendix \ref{app1}, we define several special functions that recurrently appear in these topics.

\section{Theoretical Framework} \label{cap3}

Fractional calculus is a field of study that deals with the extension of derivatives and integrals to fractional orders. It also focuses on the methods for solving differential equations that involve these fractional derivatives and fractional integrals. It was developed by Newton and Leibniz in the 17th century and involves two fundamental operations---differentiation and integration. 
These fractional operators have memory and are more flexible when describing the dynamic behavior of phenomena and systems using fractional differential equations. On the other hand, the description with integer differential equations uses local operators and is limited in the differentiation order to a constant. Consequently, the resulting models must be sufficiently accurate in many cases~\cite{west2021fractional}. 

Although many contributions have been made to this topic over the years, it has not been applied directly for centuries. However, in recent years, its applications have grown tremendously, and it is now used universally as an empirical description of complex social and physical phenomena. Research on fractional differentiation spans multiple disciplines and has various applications, such as fractional spacetime in quantum mechanics and gravity and fractional quantum field theory~\citep{Calcagni:2010bj, Calcagni:2009kc, Lim:2006hp, LimEab+2019+237+256, VargasMoniz:2020hve, Moniz:2020emn, Rasouli:2021lgy, Jalalzadeh:2021gtq, rami2009fractional}.  
This branch of mathematics is gaining more popularity in fluid dynamics, control theory, and signal processing. Due to the importance and potential of this topic, support for fractional derivatives and fractional integrals has been added to the Wolfram Language since the release of Version 13.1 \url{https://blog.wolfram.com/2022/08/12/fractional-calculus-in-wolfram-language-13-1/} (revised on 20 February 2024). This paper will be using Wolfram Language 13.3 \citep{Mathematica}.

Niels Henrik Abel is known for contributing to fractional calculus in the early 19th century. He introduced the integration and differentiation of fractional order and their inverse relationship. Abel also unified the notation for differentiation and integration of arbitrary real order in the ``differintegral'' operation. Abel considered the generalized version of the tautochrone problem, formulated as the following integral equation: 
\begin{equation}
    f(x)= \int_0^{x} \frac{\phi(s)}{\sqrt{x-s}}\,ds,
\end{equation}
for the unknown function, $\phi(x)$. After several algebraic manipulations, this integral equation could be rewritten in the following form: 
\begin{equation}
    \phi(x)= \frac{1}{\pi} \left(\frac{f(0)}{\sqrt{x}} + \int_0^x \frac{f'(\tau)}{\sqrt{x-\tau}}\,d\tau\right),
\end{equation}
which is what we now call Caputo's fractional derivative of $f(x)$ with respect to $x$.

Scientists from various areas and backgrounds have been working on fractional calculus theory, approaching it from different perspectives. Some approaches define a fractional differentiation operation. In contrast, others consider a unified differentiation/integration operation, known as the fractional differintegral of order $\mu$, with respect to $x$ and a lower limit, $a$. Differintegrals depend on the function's value at point $a$, $f(a)$, using the function's ``history''. In practice, the lower limit is generally taken as 0. 

Summarizing, in fractional calculus, the traditional derivatives and integrals of the integer order have been generalized to derivatives and integrals of arbitrary order. This concept is presented in various sources, including~\citep{book2:1999,book1:2006,monje2010fractional,klafter2012fractional, Tarasov2013, Uch:2013,bandyopadhyay2014stabilization,padula2014advances,herrmann2014fractional,malinowska2015advanced,lorenzo2016fractional,tarasov2019applications}. The idea is that a derivative operator of order $n$, which is a natural number, has been generalized to a derivative of order $\mu$, which can be a complex number or even a complex function. The three most popular and influential definitions widely used in practice are the Gr\"unwald--Letnikov differintegral, the Riemann--Liouville fractional derivative, and Caputo's fractional derivative.

\subsection{Gr\"unwald–Letnikov Approach}

The Gr\"unwald--Letnikov differintegral provides the basic extension of classical derivatives/integrals and is based on the following limits: \vspace{-6pt}
\begin{equation}
    {}_{0} D_x^\mu f(x)= \lim _{h\rightarrow 0} h^{-\mu} \sum _{r=0}^{\infty} (-1)^{r} \begin{pmatrix}
    \mu\\
    r
    \end{pmatrix} f(x- r h).
\end{equation}

In practice, this approach is useless since it contains infinite approximations of the function at different points.

\subsection{Riemann--Liouville Approach}

The definition of the Riemann--Liouville fractional derivative is as follows:
\begin{equation}
     {}^{\text{RL}}_{0}{D}_x^\mu f(x)= \frac{1}{\Gamma(n-\mu)} \frac{d^n}{d x ^n} \left(\int_{0}^{x} (x-\tau)^{n-\mu-1} f(\tau) d \tau \right),
\end{equation}
where $n= \max \{0, \lfloor \mu\rfloor\}$, where $\lfloor x \rfloor$ denotes the floor function, $\lfloor x\rfloor= \max\{m\in\mathbb{ Z}: m \leq x\}$.

It is based on a solid and strict mathematical theory of fractional calculus. This theory is well developed, but the Riemann--Liouville approach has some limitations that make it less suitable for applications in real-world problems. 

For example, the derivative of a constant is not zero: 
\begin{equation}
  {}^{\text{RL}}_{0} D_x^\mu \text{constant} = \frac{\text{constant} }{\Gamma(1-\mu)}x^{-\mu} . \label{eq35}
\end{equation}

Table \ref{FractionalD} shows the $\mu$-th fractional derivatives and $n$-th ordinary derivatives for some standard functions (see Appendix \ref{app1} for definitions of special functions).
\begin{table}
\begin{tabularx}{\textwidth}{ccc}
\toprule
\boldmath{$f(x)$}	& \boldmath{$D^{\mu}_x(f(x))$}	& \boldmath{$f^{(n)}(x)$}\\
\midrule
$e^x$   & $- e^x \left(Q(-\mu, x) -1\right)$  & $e^x$\\
$x^2$	& $\frac{2 x^{2-\mu}}{\Gamma(3-\mu)}$ & $\frac{2 x^{2-n}}{(2-n)!}$ \\
$\log(x)$	& $\left\{\begin{array}{cc}
   -(-1)^{\mu} (\mu -1)! x^{-\mu} &  \mu \in \mathbb{Z}\land \mu>0\\
   -\frac{x^{-\mu}\left(\psi^{(0)}(1-\mu) - \log(x) + \gamma \right)}{\Gamma(1-\mu)}&  \text{True}
\end{array}\right.$			& $\left\{\begin{array}{cc}
     -(-1)^{n} (n -1)! x^{-n} &  n\geq 1\\
   \log(x)  & \text{True}
\end{array}\right.$ \\
$\sin(x)$	& $\sqrt{\pi} \, 2^{\mu-1} x^{1-\mu} \, {}_1 \tilde{F}_2 \left(1; 1 -\frac{\mu}{2}, \frac{3 -\mu}{2}; -\frac{x^2}{4}\right)$			&  $\sin\left(\frac{\pi n}{2} +x\right)$\\
$\tan^{-1}(x)$	& $\sqrt{\pi} \, 2^{\mu-1} x^{1-\mu} \, {}_3 \tilde{F}_2 \left(\frac{1}{2},1,1; 1 -\frac{\mu}{2}, \frac{3 -\mu}{2}; -x^2\right) $			& $\sqrt{\pi} \, 2^{n-1} x^{1-n} \, {}_3 \tilde{F}_2 \left(\frac{1}{2},1,1; 1 -\frac{n}{2}, \frac{3 -n}{2}; -x^2\right) $	 \\
$J_0(x)$	& $\sqrt{\pi} \, 2^{\mu} x^{-\mu} \, {}_1 \tilde{F}_2 \left(\frac{1}{2}; \frac{1- \mu}{2}, 1 - \frac{\mu}{2}; -\frac{x^2}{4}\right)$				& $2^{-n}\sum_{K_1=0}^{n} (-1)^{K_1} J_{2 K_1 -n}(x)\begin{pmatrix}
    n\\
    K_1
\end{pmatrix}$ \\
${}_1 F_1 (a; b; x)$	& $\Gamma(b) x^{-\mu} {}_2 \tilde{F}_2\left(1, a; 1-\mu, b; x\right)$			& $\left\{\begin{array}{cc}
 \Gamma(b) x^{-n} {}_2 \tilde{F}_2\left(1, a; 1-n, b; x\right)    & n\geq 1  \\
 {}_1 F_1 (a; b; x)    & \text{True}
\end{array}\right.$  \\
\bottomrule
\end{tabularx}
\end{table}

\subsubsection{Caputo's Approach}

Caputo's definition of fractional derivative is as follows:
\begin{equation}
     {}^{\text{C}}_{0}{D}_x^\mu f(x)= \frac{1}{\Gamma(n-\mu)} \int_{0}^{x} (x-\tau)^{n-\mu-1} \frac{d^n}{d \tau^n}f(\tau) d \tau,  \quad n=\left\{\begin{array}{cc}
     \lfloor \mu\rfloor +1& \mu \notin \mathbb{N}\\
     \mu & \mu \in \mathbb{N}
    \end{array} \right..\label{Caputo_Def}
\end{equation}

There is some similarity between this and the Riemann--Liouville differintegral; in fact, the Caputo differintegral can be defined via the Riemann--Liouville differintegral by the follwing:
\begin{equation}
{}^{\text{C}}_{0}{D}_x^\mu f(x)= {}^{RL}_{0} D_x^\mu - \sum _{k=0}^{ \lfloor \mu \rfloor -1} \frac{x^{k-\mu}}{\Gamma (k -\mu +1)} \frac{d^k}{d x^k} f(0).
\end{equation}

Caputo's definition of fractional and integral derivatives has many advantages compared to those of Riemann--Liouville or Gr\"unwald--Letnikov: first, it takes into account the values of the function and its derivatives at the origin (or, in general, at any lower or upper bound $a$), which automatically makes it suitable for solving fractional order initial value problems using Laplace transforms. Furthermore, the Caputo fractional derivative of a constant is $0$. Therefore, it is more consistent with the classical calculation.

The Riemann--Liouville fractional differintegral has been implemented in the Wolfram Language, using the Wolfram Language Version 13.3 \citep{Mathematica} function \textbf{FractionalD}. This function calculates the Riemann--Liouville fractional derivative of order $\mu$ of the function $f(x)$; the lower limit of the integral is considered $0$.

The Caputo fractional differential integral is \textbf{{CaputoD}} in Wolfram Language 13.1, which gives the Caputo fractional derivative of order $\mu$ of a function $f(x)$. For negative orders of $\mu$, the output of \textbf{{CaputoD}} matches \textbf{{FractionalD}}.

Table \ref{CaputoD} presents the half-order Caputo fractional derivatives of some common mathematical functions.

\begin{table}[h] 
\caption{Half-order Caputo fractional derivatives of some common mathematical functions. \label{CaputoD}}
\newcolumntype{C}{>{\centering\arraybackslash}X}
\begin{tabularx}{\textwidth}{CC}
\toprule
\boldmath{$f(x)$}	& \boldmath{$D^{\frac{1}{2}}_x(f(x))$}	\\
\midrule
$e^x$   & $e^{x} \text{erf}(\sqrt{x})$\\
$x^2$	& $\frac{8 x^{\frac{3}{2}}}{3\sqrt{\pi}}$\\
$\log(x)$	& $\frac{\log(4)}{\sqrt{\pi} \sqrt{x}}$ \\
$\sin(x)$	& $\frac{2 \sqrt{x} \; {}_1 F_2 \left(1; \frac{3}{4}, \frac{5}{4}; -\frac{x^2}{4}\right) }{\sqrt{\pi}}$\\
$\tan^{-1}(x)$	& 	$\frac{2 \sqrt{x} \; {}_2 F_3 \left(\frac{1}{2},1,1; \frac{3}{4}, \frac{5}{4}; -x^2\right) }{\sqrt{\pi}}$ \\
$J_0(x)$	& $\frac{{}_1 F_2 \left(\frac{1}{2}; \frac{1}{4}, \frac{3}{4}; -\frac{x^2}{4}\right) -1}{\sqrt{\pi} \sqrt{x}}$ \\
${}_1 F_1 (a; b; x)$	& $\frac{{}_2 F_2 \left(1,a; \frac{1}{2}, b; x\right) -1}{\sqrt{\pi} \sqrt{x}}$\\
\bottomrule
\end{tabularx}
\end{table}

\subsubsection{Rule for Successive Fractional Derivatives} 

The Caputo fractional derivative for an arbitrary $c$ is as follows: \citep{malinowska2015advanced} 
\begin{equation}
    {}^{\text{C}}_{c}D_x^\mu f(t)= \frac{1}{\Gamma(n-\mu)}\int_{c}^{x}\frac{d^n f(\tau)}{d\tau ^n} \left(x-\tau\right)^{n-\mu-1}d\tau, \quad  n=\left\{\begin{array}{cc}
     \lfloor \mu\rfloor +1& \mu \notin \mathbb{N}\\
     \mu & \mu \in \mathbb{N}
    \end{array} \right..
\end{equation}

Given the square function, i.e., $f(x)=x^2/2$, choosing $c=0$, removing the subindex $0$, and applying the derivative of order $1/2$ twice, it results in the following: $\mu=1/2 \Rightarrow n :=\lfloor \mu\rfloor+1=1$,
    \begin{equation} \nonumber
        {}^{\text{C}}D_x^{1/2}\left(x^2/2\right)=\frac{1}{\Gamma(\frac{1}{2})}\int_{0}^ {x}\frac{d}{dt}\left[\frac{t^2}{2}\right](x-t)^{-\frac{1}{2}}dt=\frac{1}{ \sqrt{\pi}}\frac{4}{3}x^{\frac{3}{2}},
    \end{equation}
    \begin{equation} \nonumber
        {}^{\text{C}}D_x^{1/2}\left(\frac{1}{\sqrt{\pi}}\frac{4}{3}x^{\frac{3}{2}}\right)=\frac{1}{\sqrt{\pi}}\frac{4}{3}\frac{1}{\Gamma(\frac{1}{2})}\int_{0}^{ x}\frac{d}{dt}\left[t^{\frac{3}{2}}\right](x-t)^{-\frac{1}{2}}dt=\frac{1} {\sqrt{\pi}}\frac{4}{3}\frac{1}{\sqrt{\pi}}\frac{3 \pi x}{4}=x.
    \end{equation}
    
The first derivative of the square function is obtained through two ``half-order fractional differentiation'' procedures. One could easily verify that two similar half-order integration procedures can obtain the anti-derivative of the square function. So, with this example, we show what fractional calculus is and how it relates to and generalizes the classical version.
However, this result is based on the choice, $c=0$. In general, the rule for successive derivatives for fractional calculus is as follows:
\begin{align}
    D_{x}^{\mu} \left[D_{x}^{\beta} f(x)\right]=D_{x}^{\mu+\beta} f(x)-\sum _{j= 1}^{n} D_{x}^{\beta-j} f(c_+) \frac{(x-c)^{-\mu-j}}{\Gamma(1-\mu-j)}.
\end{align}

\subsubsection{Fractional Derivatives as Generalizations of Integer-Order Derivatives}
From now on, we follow the exposition of \citep{herrmann2014fractional}. 
We have the formulas for the\linebreak $n$-th~derivatives,
\begin{align}
   & \frac{d^n}{dx^n}e^{k x}=k^ne^{k x}, \quad  \frac{d^n}{dx^n}\sin{(k x)}=k^n\sin{\left(k x+\frac{\pi}{2}n\right)},
 \quad  \frac{d^n}{dx^n}x^k=\frac{k!}{(k-n)!}x^{k-n},
    \label{1.2.3}
\end{align}
for $n\in \mathbb{N}$ can be extended for real and complex $\mu$ numbers,
\begin{align}
  & D_{x}^{\mu}e^{k x}=k^{\mu}e^{k x}, \hspace{20px} k\geq 0,
    \label{1.2.4}
\\
  & D_{x}^{\mu}\sin{(k x)}=k^{\mu}\sin{\left(k x+\frac{\pi}{2}\mu\right)}, \hspace{ 20px}k\geq 0,
    \label{1.2.5}
\\
  & D_{x}^{\mu}x^k=\frac{\Gamma (k+1)}{\Gamma (k+1-\mu)}x^{k-\mu}, \hspace{20px } x\geq 0, k\neq -1,-2,-3,\ldots \hspace{5px} .
    \label{1.2.6}
\end{align}

The definition of the fractional derivative of the exponential function was given by Liouville in 1832; those of trigonometric functions were proposed by Fourier in 1822, and those of powers were systematically studied by Riemann in 1847. Still, the first attempts at the latter were made by Leibniz and Euler, the latter in 1738. The Riemann--Liouville fractional derivative of a constant is given by \eqref{1.2.6}, setting $k=0$, we obtain unexpected~behavior, as follows: 
\begin{equation*}
    D_{x}^{\mu} 1 =D_{x}^{\mu}x^0=\frac{1}{\Gamma(1-\mu)}x^{-\mu}.
\end{equation*}

Therefore, as an additional postulate, Caputo in 1967 introduced the following condition: \linebreak
$D_{x}^{\mu}1=0$.

These four definitions of fractional derivatives \eqref{1.2.4}--\eqref{1.2.6} and Caputo's share several aspects. On the one hand, they satisfy the following correspondence principle:
\begin{equation}
    \lim_{\mu \to n}D_{x}^{\mu}f(x)=\frac{d^n}{dx^n}f(x), \hspace{10px}n=0, 1,2,\ldots\hspace{5px}.
    \label{1.2.9}
\end{equation}

And the following rules apply:
\begin{align}
   & D_{x}^{\mu}cf(x)=cD_{x}^{\mu}f(x),
    \quad  D_{x}^{\mu}[f(x)+g(x)]=D_{x}^{\mu}f(x)+D_{x}^{\mu}g(x).
\end{align}

Therefore, the fractional derivative of an analytic function can be calculated to its~series.

According to Liouville, we have the following:\vspace{-6pt}
\begin{equation}
    f(x)=\sum _{k=0}^{\infty}a_ke^{k x} \implies
    D_{x}^{\mu}f(x)=\sum _{k=0}^{\infty}a_kk^{\mu}e^{k x}.
    \label{1.2.13}
\end{equation}

According to Fourier, we have the following:\vspace{-6pt}
\begin{align}
   & f(x)=a_0+\sum _{k=1}^{\infty}a_k\sin{(k x)}+\sum _{k=1}^{\infty}b_k\cos{(k x)},
    \label{1.2.14}
\\
   & D_{x}^{\mu}f(x)=\sum _{k=1}^{\infty}a_kk^{\mu}\sin{\left(k x+\frac{\pi}{2}\mu\right)}+\sum_{k=1}^{\infty}b_kk^{\mu}\cos{\left(k x+\frac{\pi}{2}\mu\right)}.
    \label{1.2.15}
\end{align}

According to Riemann, we have the following:\vspace{-6pt}
\begin{align}
& f(x)=x^{\mu-1}\sum _{k=0}^{\infty}a_k x^{k\mu},
    \label{1.2.16}
\\
  & D_{x}^{\mu}f(x)=x^{\mu-1}\sum _{k=0}^{\infty}a_{k+1}\frac{\Gamma ((k+ 2)\mu)}{\Gamma ((k+1)\mu)}x^{k\mu}.
    \label{1.2.17}
\end{align}

According to Caputo, we have the following:\vspace{-6pt}
\begin{align}
   & f(x)=\sum _{k=0}^{\infty}a_k x^{k\mu},
    \label{1.2.18}
\\
   & D_{x}^{\mu}f(x)=\sum _{k=0}^{\infty}a_{k+1}\frac{\Gamma (1+(k+1)\mu)} {\Gamma (1+k\mu)}x^{k\mu}.
    \label{1.2.19}
\end{align}

Calculating the Caputo fractional derivative of order $\mu$ to the exponential function, we~have the following:
\begin{equation}
    D_{x}^{\mu}e^x=D_{x}^{\mu}\sum _{n=0}^{\infty}\frac{x^n}{n!} =\sum _{n =1}^{\infty}\frac{x^{n-\mu}}{\Gamma (1+n-\mu)} =x^{1-\mu}E(1,2-\mu, x).
\label{1.2.20}
\end{equation}

So, in contrast to Liouville's definition \eqref{1.2.13}, Caputo's definition is not obviously in terms of an exponential, but in terms of a generalized Mittag-Leffler function \eqref{1.1.2.6}. 

\subsubsection{Leibniz's Rule for the Fractional Derivative of the Product}
Let us now analyze whether all the techniques of ordinary differential calculus can be transferred to the field of fractional calculus \citep{herrmann2014fractional}. For example, if we take the ordinary Leibniz rule, as follows:
\begin{equation}
    \frac{d}{dx}(\psi \chi)=\left(\frac{d}{dx}\psi\right)\chi+\psi \left(\frac{d}{dx}\chi\right ),
    \label{1.2.1.1}
\end{equation}
it could be carelessly inferred that the relation
\begin{equation}
    D_{x}^{\mu}(\psi \chi)=\left(D_{x}^{\mu}\psi\right)\chi+\psi \left(D_{x}^{\mu}\chi\right),
    \label{1.2.1.2}
\end{equation}
remains in the fractional calculus. However, using Liouville's definition of a fractional derivative, we have the following: \vspace{-4pt}
\begin{equation}
    D_{x}^{\mu}(e^{k_1 x} e^{k_2 x})=D_{x}^{\mu}e^{(k_1+k_2) x}=(k_1+k_2)^\mu e^{(k_1+k_2) x}.
    \label{1.2.1.3}
\end{equation}

While using \eqref{1.2.1.2}, we obtain the following: 
\begin{equation}
    D_{x}^{\mu}(e^{k_1 x} e^{k_2 x})=(k_{1}^{\mu}+k_{2}^{\mu})(e^{k_1 x}e ^{k_2 x}).
    \label{1.2.1.4}
\end{equation}

Therefore, Leibniz's rule \eqref{1.2.1.2} is not valid for fractional derivatives.

We start from the rule for successive non-negative integer orders to extend Leibniz's rule to be valid for fractional calculus.
\begin{small}
\begin{equation}
\begin{split}
    &\frac{d}{dx}(\psi \chi)=\left(\frac{d}{dx}\psi\right)\chi+\psi \left(\frac{d}{dx}\chi\right),\\
    &\frac{d^2}{dx^2}(\psi \chi)=\left(\frac{d^2}{dx^2}\psi\right)\chi +2\left(\frac{ d}{dx}\psi\right)\left(\frac{d}{dx}\chi\right)+\psi \left(\frac{d^2}{dx^2}\chi\right), \\
    &\frac{d^3}{dx^3}(\psi \chi)=\left(\frac{d^3}{dx^3}\psi\right)\chi +3\left(\frac{ d^2}{dx^2}\psi\right)\left(\frac{d}{dx}\chi\right)+3\left(\frac{d}{dx}\psi\right)\left (\frac{d^2}{dx^2}\chi\right)+\psi \left(\frac{d^3}{dx^3}\chi\right),
\end{split}
\label{1.2.1.5}
\end{equation}
\end{small}
et cetera. We have the following formula for an arbitrary integer, as follows: 
\begin{equation}
   \frac{d^n}{d x^n} (\psi \chi)=\sum _{j=0}^{n}
    \begin{pmatrix}
    n\\
    j
    \end{pmatrix}
    \left(\frac{d^{n-j}}{d x^{n-j}} \psi \right) \left(\frac{d^{j}}{d x^{j}} \chi\right), \hspace{20px} n\in \mathbb{N}.
    \label{1.2.1.7}
\end{equation}\vspace{-8pt}

According to  \eqref{1.1.1.6}, this expression can be extended to an arbitrary $\mu$:
\begin{equation}
    D_{x}^{\mu}(\psi \chi)=\sum _{j=0}^{\infty}
   \frac{\Gamma(\mu +1)}{\Gamma(j +1)\Gamma(\mu - j +1)}
    (D_{x}^{\mu -j}\psi)(D_{x}^{j}\chi),
    \label{1.2.1.8}
\end{equation}\vspace{-8pt}

So, this form of the Leibniz product rule is also valid for fractional derivatives.

\subsection{Fractional Integral}

The Cauchy formula for successive integrations, which can be extended to fractional integrals, is used as the first step. A satisfactory definition of a fractional derivative can be obtained after appropriately deriving a fractional integral, $I^\mu$. If we assume the following: 
\begin{equation}
    D_{x}^{\mu}=I^{-\mu},
    \label{1.3.1}
\end{equation}
this implies that a fractional derivative is the inverse of a fractional integral, and conversely, a fractional integral is the inverse of a fractional derivative.

If ${} ^{}_{a}{I}$ is used for the integration operator, it is defined as follows: \vspace{-4pt}
\begin{equation}
    {} ^{}_{a}{I}f(x)=\int _{a}^{x}f(\xi)\,d\xi,
    \label{1.3.8}
\end{equation}

It is observed that the multiple integral is as follows: 
\begin{equation}
    {}^{}_{a}{I^n}f(x)=\int_{a}^{x}\int_{a}^{x_{n-1}}\cdots\int_{a}^ {x_1}f(x_0)\,dx_0\dotsb dx_{n-1},
    \label{1.3.10}
\end{equation}
which, using Cauchy's repeated integration formula, can be reduced to a single integral, as follows:
\begin{equation}
    {}^{}_{a}{I^n}f(x)=\frac{1}{(n-1)!}\int_{a}^{x}(x-\xi)^{n -1}f(\xi)\,d\xi.
    \label{1.3.11}
\end{equation}

This formula can be easily extended to the fractional case, as follows:\vspace{-4pt}
\begin{equation}
    {}^{}_{a}{I_+^\mu}f(x)=\frac{1}{\Gamma (\mu)}\int_{a}^{x}(x-\xi)^ {\mu -1}f(\xi)\,d\xi, \hspace{20px} x>a,
    \label{1.3.12}
\end{equation}
\begin{equation}
    {}^{}_{b}{I_-^\mu}f(x)=\frac{1}{\Gamma (\mu)}\int_{x}^{b}(\xi-x)^ {\mu -1}f(\xi)\,d\xi, \hspace{20px} x<b.
    \label{1.3.13}
\end{equation}

Constants $a$ and $b$ determine the lower and upper limits of the integral, respectively, and can initially be chosen arbitrarily. The two cases are denoted as the left and right, respectively. This can be understood because the left case deals with values $\xi<x$, values to the left of $x$. The right case deals with values $\xi>x$, values to the right of $x$. If $x$ is a time-type coordinate, then the integral of the left case is causal, and that of the right case is anti-causal.

The value of the integral depends on the specific choice of the constants $(a,b)$. This will be the notable difference between the two most widely used definitions of a fractional integral, the so-called Liouville fractional integral and the Riemann fractional integral, which are discussed below.

\subsubsection{Liouville Fractional Integral}
The Liouville fractional integral is defined by making $a=-\infty$ and $b=+\infty$ in \eqref{1.3.12} and \eqref{1.3.13}:\vspace{-8pt}
\begin{align}
  &{}^{}_{\text{L}}{I_+^\mu}f(x)=\frac{1}{\Gamma (\mu)}\int_{-\infty}^{x} (x-\xi)^{\mu -1}f(\xi)\,d\xi,
    \label{1.3.1.1}
\\
  &{}^{}_{\text{L}}{I_-^\mu}f(x)=\frac{1}{\Gamma (\mu)}\int_{x}^{+\infty} (\xi-x)^{\mu -1}f(\xi)\,d\xi.
    \label{1.3.1.2}
\end{align}
\subsubsection{Riemann Fractional Integral}
The fractional Riemann integral is defined by making $a=0$ and $b=0$ in \eqref{1.3.12} and \eqref{1.3.13}, as follows:
\begin{align}
   &{}^{}_{\text{R}}{I_+^\mu}f(x)=\frac{1}{\Gamma (\mu)}\int_{0}^{x}(x -\xi)^{\mu -1}f(\xi)\,d\xi,
    \label{1.3.2.1}
\\
   &{}^{}_{\text{R}}{I_-^\mu}f(x)=\frac{1}{\Gamma (\mu)}\int_{x}^{0}(\xi-x)^{\mu -1}f(\xi)\,d\xi.
    \label{1.3.2.2}
\end{align}

To see the difference between Liouville's definitions, \eqref{1.3.1.1} and \eqref{1.3.1.2}, and Riemann's definitions, \eqref{1.3.2.1} and \eqref{1.3.2.2}, we apply both definitions to $f (x)=e^{k x}$, as follows:
\begin{align*}
   &{}^{}_{\text{L}}{I_{+}^{\mu}}e^{k x}=k^{-\mu}e^{k x}, x>0, \quad 
    {}^{}_{\text{L}}{I_{-}^{\mu}}e^{k x}=(-k)^{-\mu}e^{k x}, k<0,
\\
 &{}^{}_{\text{R}}{I_{+}^{\mu}}e^{k x}=k^{-\mu}e^{k x}\left(1-\frac {\Gamma (\mu,k x)}{\Gamma (\mu)}\right), x>0,
   \quad {}^{}_{\text{R}}{I_{-}^{\mu}}e^{k x}=(-k)^{-\mu}e^{k x}\left(1 -\frac{\Gamma (\mu,k x)}{\Gamma (\mu)}\right), x<0,
\end{align*}
where $\Gamma (\mu,x)$ is the incomplete Gamma function. Using Liouville's definition, the function $e^{k x}$ vanishes for $k>0$ in the lower limit of the domain of the integral, so there is no additional contribution. On the other hand, when using the Riemann definition, an additional contribution is obtained because, at $a=0$, the exponential function does not vanish \citep{herrmann2014fractional}.

\subsection{Link between Integration and Fractional Differentiation}
We use the following abbreviation: 
\begin{equation}
    D_{x}^{\mu}=D^\mu,
    \label{1.4.1}
\end{equation}

It is shown that the concepts of fractional integration and fractional differentiation are closely related. Separating the fractional derivative operator into factors, we have the following: 
\begin{equation}
    D^\mu =D^{m}D^{\mu -m}=\frac{d^m}{dx^m} \hspace{0px}{}^{}_{a}{I^{m -\mu}}, \hspace{20px} m\in \mathbb{N}.
    \label{1.4.2}
\end{equation}

A fractional derivative can be interpreted as a fractional integral followed by an ordinary derivative. Once the fractional integral is specified, the fractional derivative is automatically determined. We can also reverse the sequence of the operators, resulting in an alternative decomposition, as follows:\vspace{-6pt}
\begin{equation}
    D^\mu=D^{\mu -m}D^m =\hspace{0px} {}^{}_{a}{I^{m-\mu}} \frac{d^m}{dx ^m}, \hspace{20px} m\in \mathbb{N}.
    \label{1.4.3}
\end{equation}

Decompositions \eqref{1.4.2} and \eqref{1.4.3} lead to different results.
Thus, we can understand the mechanism by which non-locality enters fractional calculus. The standard derivative, of course, is a local operator, but the fractional integral certainly is not. The fractional derivative should be understood as the inverse of fractional integration, a non-local operation. Consequently, fractional differentiation and integration produce the same difficulty \citep{herrmann2014fractional}.

In the previous section, two different definitions of the fractional integral have been given, the Liouville and the Riemann. For each of these definitions, according to the two different decompositions given above, four different realizations follow for the fractional derivative definition, which are presented below.

\subsubsection{Liouville Fractional Derivative}
For the simple case of $0<\mu <1$, using the Liouville definitions \eqref{1.3.1.1} and \eqref{1.3.1.2}, and the sequence of operators, \eqref{1.4.2}, we have the following: 
\begin{align}
   &{}^{}_{\text{L}}{D_{+}^{\mu}}f(x)=\frac{d}{dx}\hspace{0px}{}^{}_{ \text{L}}{I_{+}^{1-\mu}}f(x) =\frac{1}{\Gamma (1-\mu)} \frac{d}{dx}\int _ {-\infty}^{x}(x-\xi)^{-\mu}f(\xi)\,d\xi,
\label{1.4.1.1}
\\
   &{}^{}_{\text{L}}{D_{-}^{\mu}}f(x) =\frac{d}{dx}\hspace{0px}{}^{}_{ \text{L}}{I_{-}^{1-\mu}}f(x)=\frac{1}{\Gamma (1-\mu)} \frac{d}{dx}\int _ {x}^{+\infty}(\xi-x)^{-\mu}f(\xi)\,d\xi.
\label{1.4.1.2}
\end{align}

\subsubsection{Fractional Riemann Derivative}
Using \eqref{1.3.2.1} and \eqref{1.3.2.2}, and the sequence of operators, \eqref{1.4.2}, we have the following Riemann fractional~derivative:\vspace{-6pt}
\begin{align}
   &{}^{}_{\text{R}}{D_{+}^{\mu}}f(x)=\frac{d}{dx}\hspace{0px}{}^{}_{ \text{R}}{I_{+}^{1-\mu}}f(x)=\frac{1}{\Gamma (1-\mu)}\frac{d}{dx}\int _ {0}^{x}(x-\xi)^{-\mu}f(\xi)\,d\xi,
\label{1.4.2.1}
\\
  &{}^{}_{\text{R}}{D_{-}^{\mu}}f(x)=\frac{d}{dx}\hspace{0px}{}^{}_{ \text{R}}{I_{-}^{1-\mu}}f(x)
    =\frac{1}{\Gamma (1-\mu)}\frac{d}{dx}\int _{x}^{0}(\xi-x)^{-\mu}f(\xi )\,d\xi.
\label{1.4.2.2}
\end{align}

\subsubsection{Liouville--Caputo Fractional Derivative}
In the case that the sequence of operators is inverted according to \eqref{1.4.3}, it is based on the definition of the Liouville fractional integrals \eqref{1.3.1.1} and \eqref{1.3.1.2}, as follows:

\begin{align}
    {}^{}_{\text{LC}}{D_{+}^{\mu}}f(x)=\hspace{0px}{}^{}_{\text{L}}{I_{ +}^{1-\mu}}\frac{d}{dx}f(x) =\frac{1}{\Gamma (1-\mu)}\int _{-\infty}^{x} (x-\xi)^{-\mu}\frac{df(\xi)}{d\xi}\,d\xi,
\label{1.4.3.1}
\\
    {}^{}_{\text{LC}}{D_{-}^{\mu}}f(x) =\hspace{0px}{}^{}_{\text{L}}{I_{ -}^{1-\mu}}\frac{d}{dx}f(x) =\frac{1}{\Gamma (1-\mu)}\int _{x}^{+\infty} (\xi-x)^{-\mu}\frac{df(\xi)}{d\xi}\,d\xi.
\label{1.4.3.2}
\end{align}

It has been observed that, for a function $f(x)=$ constant, the definition of the Liouville fractional derivative \eqref{1.4.1.1} leads to a divergent result, while with the Liouville--Caputo definition \eqref{1.4.3.1}, the result is zero.  The conditions for the Liouville fractional derivative to converge are more restrictive than for the Liouville--Caputo counterpart \citep{herrmann2014fractional}.

\subsubsection{Caputo Fractional Derivative}
The inverted sequence of operators \eqref{1.4.3}, the definition of the Riemann fractional integrals \eqref{1.3.2.1} and \eqref{1.3.2.2}, leads to the following:\vspace{-4pt}
\begin{align}
   &{}^{\text{C}}{D_{+}^{\mu}}f(x)=\hspace{0px}{}^{}_{\text{R}}{I_ {+}^{1-\mu}}\frac{d}{dx}f(x)=\frac{1}{\Gamma (1-\mu)}\int _{0}^{x}( x-\xi)^{-\mu}\frac{df(\xi)}{d\xi}\,d\xi,
\label{1.4.4.1}
\\
   &{}^{\text{C}}{D_{-}^{\mu}}f(x) =\hspace{0px}{}^{}_{\text{R}}{I_ {-}^{1-\mu}}\frac{d}{dx}f(x) =\frac{1}{\Gamma (1-\mu)}\int _{x}^{0}( \xi-x)^{-\mu}\frac{df(\xi)}{d\xi}\,d\xi.
\label{1.4.4.2}
\end{align}

Then, for $f(x)=$ constant, the Riemann fractional derivative is \linebreak ${}^{}_{\text{R}}{D_{+}^{\mu}}\text{constant }=\frac{\text{constant}}{\Gamma (1-\mu)}x^{-\mu}$, while when applying the Caputo derivative, we have \linebreak ${}^{\text{C}}{D_{+}^{\mu}}\text{constant}=0$. Therefore, in this section, we present the standard definitions of a fractional derivative, which satisfy the following condition:
\begin{equation}
    \lim_{\mu \to 1}D^{\mu}f(x)=\frac{df(x)}{dx}.
    \label{1.4.4.5}
\end{equation}

\textls[-15]{Multiple applications of the first-order derivative can generate higher-order derivatives, as follows:}
\begin{equation}
    \frac{d^n}{dx^n}=\frac{d}{dx}\frac{d}{dx}\dotsb \frac{d}{dx} \hspace{20px} \text{$n$ times}.
    \label{1.5.1}
\end{equation}

After applying different definitions of the fractional derivative to the same function, it has been found that they do not give the same result. This can be non-comforting, but all these definitions are equally well-founded. Therefore, physicists can test different definitions of the fractional derivative by comparing theoretical results with experimental data. Depending on the specific problem, one of the definitions will show the most significant agreement with the experiment.

\subsection{Fractional Differential Equations}

Fractional differential equations (FDEs), which involve fractional derivatives and generalize ordinary differential equations (ODEs), have received much attention. FDEs have been widely used in engineering, physics, chemistry, biology, and other fields involving relaxation and oscillation models.

Let us consider, for example, the FDE, as follows:
\begin{align}
 &  {}^{\text{C}}_{0}D_{t}^{\frac{21}{10}}y(t) + 10 y(t)  =0, \quad  y(0)=1, \quad y'(0)= 0, \quad y''(0)=0. \label{EqFDE1}
\end{align}

 Solving \eqref{EqFDE1} using the Wolfram Language 13.3 \citep{Mathematica} function \textbf{DSolve}, 
 we obtain the following: 
\begin{equation}
 y(t)= E \left(\frac{21}{10}, -10 t^{21/10}\right). \label{solutionFEDE1}
\end{equation}

\begin{figure}[h]
 \centering
 \includegraphics[width=0.6\textwidth]{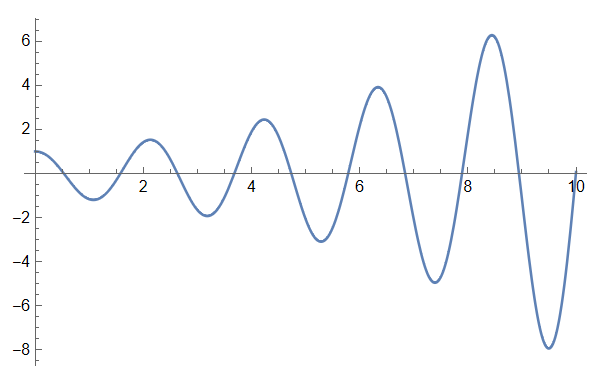}
     \caption{Solution \eqref{solutionFEDE1} of the fractional harmonic oscillator of order $2.1$ given by \eqref{EqFDE1}. For $\mu>2$, the fractional harmonic oscillator mimics an excited oscillator. }
     \label{FDE1}
\end{figure}

The solution presented in Figure \ref{FDE1} is given in terms of the  \textbf{MittagLefflerE} function, the primary function for fractional calculus applications. Its role in FDE solutions is similar in importance to the \textbf{Exp} function for ODE solutions: any FDE with constant coefficients can be solved in terms of Mittag-Leffler functions. 

An interesting example is the solution of a fractional harmonic oscillator of order 1.9, as follows:  
\begin{align}
 &  {}^{\text{C}}_{0}D_{t}^{\frac{19}{10}}x(t) + x(t)  =0,  \quad x(0)=1, \quad x'(0)= 0. \label{EqFDE2}
\end{align}

Solving \eqref{EqFDE2} using the Wolfram Language 13.3 \citep{Mathematica} function \textbf{DSolve}, we obtain the following solution: 
\begin{equation}
 \textbf{{fracOsc}}: \; x(t)= E \left(\frac{19}{10}, - t^{19/10}\right). \label{fracOsc}
\end{equation}

This type of oscillator behaves similarly to the ordinary damped harmonic oscillator, as follows:
\begin{align}
 &  x''(t) +\frac{3}{20} x'(t) + x(t)  =0,  \quad x(0)=1, \quad x'(0)= 0, \label{DAMPED2}
\end{align}
with the following solution:
\begin{equation}
\textbf{dampedOsc}: \;  x(t)=   \frac{e^{-3 t/40} \left(3 \sqrt{1591} \sin
   \left(\frac{\sqrt{1591} t}{40}\right)+1591 \cos
   \left(\frac{\sqrt{1591} t}{40}\right)\right)}{1591}. \label{dampedOsc}
\end{equation}

This example highlights how the order of an FDE can act as a control parameter to model complex systems.

\begin{figure}[h]
 \centering
    \includegraphics[width=0.7\textwidth]{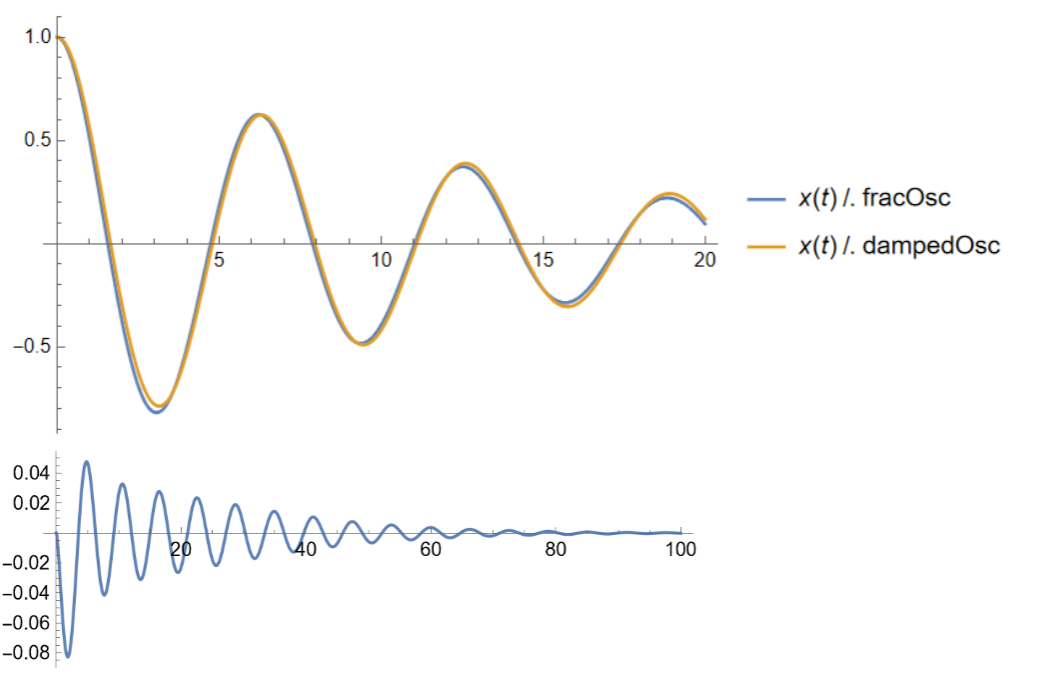}
    \caption{Comparison  between the fractional oscillator \eqref{fracOsc}, which satisfies the equation of a fractional harmonic oscillator of order $1.9$ \eqref{EqFDE2}  with the damping oscillator \eqref{dampedOsc}. The error function is drawn in the bottom plot. This shows that the error between both solutions goes to zero as $t$ becomes large. }
    \label{FDE2}
\end{figure}

Figure \ref{FDE2} shows a comparison between the solution of the fractional oscillator \eqref{fracOsc} and the damping oscillator \eqref{dampedOsc}. The error function is drawn in the bottom plot. 

As a final example of FDE, we consider the fractional friction model, as follows:
\begin{equation}
  \ddot{x}(t) +\frac{1}{4} \; {}^{\text{C}}_{0}D_{t}^{\frac{1}{2}}x(t) =0, \quad x(0)=x_0, \quad x'(0)= v_0 \label{Newton_Modification-Caputo}
\end{equation}
involving the Caputo fractional differintegral of order $1/2$.
 Solving \eqref{Newton_Modification-Caputo} using the Wolfram Language 13.3 \citep{Mathematica} function \textbf{DSolve}, we obtain the following solution: 
\begin{equation}
    x(t)= x_0 + t v_0 E \left(\frac{3}{2}, 2, -\frac{1}{4} t^{3/2}\right). 
\end{equation}

\begin{figure}[h]
 \centering
    \includegraphics[width=0.6\textwidth]{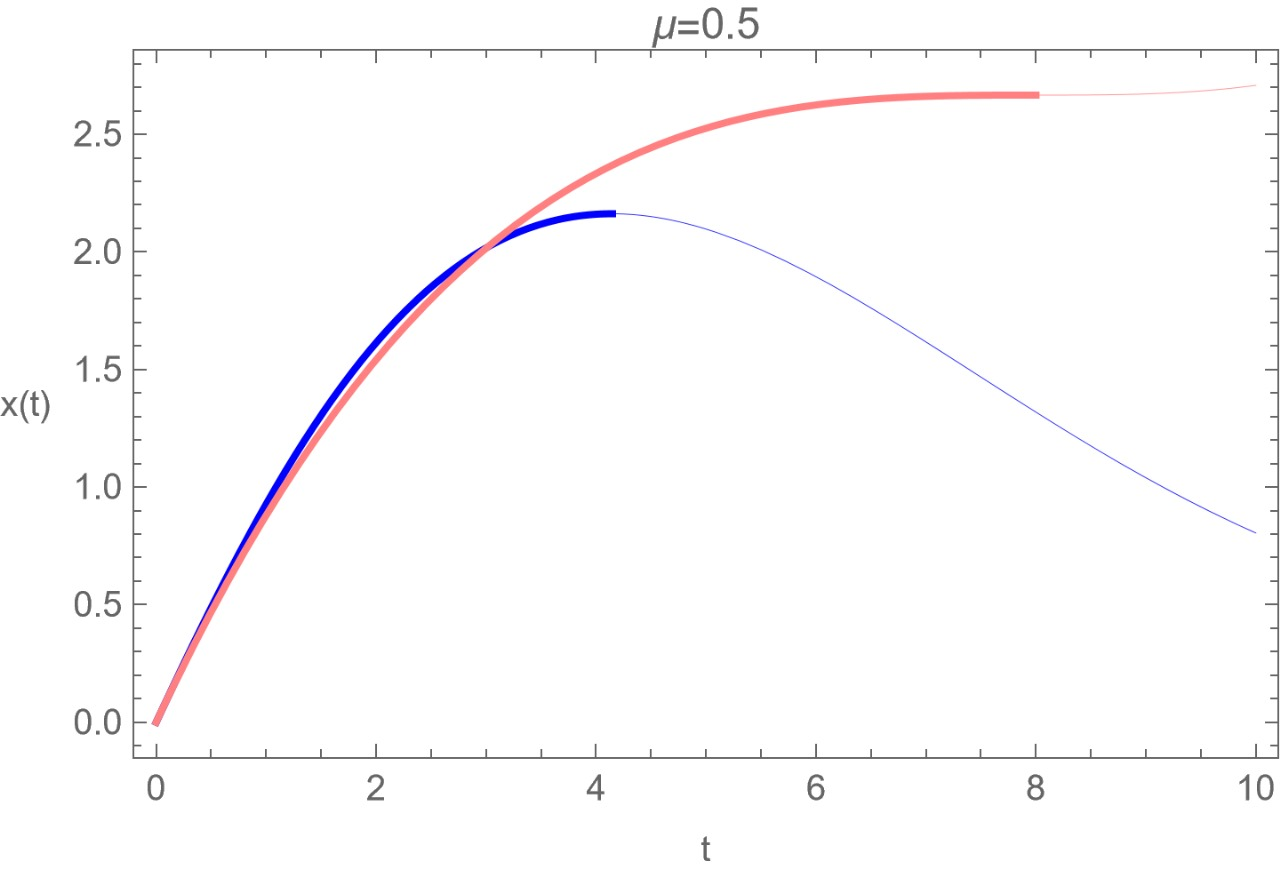}
    \caption{Solution of the FDE \eqref{Newton_Modification-Caputo} for the Caputo derivative for $\mu=1/2$,  $x_0=0 $, and $v_0=1$ (blue line), and a comparison with the solution \eqref{classical-friction}, which gives the position of a point particle under the influence of a velocity-dependent force (pink line).}
    \label{FDE3}
\end{figure}

This solution is given using the \textbf{MittagLefflerE} functions and is represented in Figure~\ref{FDE3} by a blue line. The pink line corresponds to the following solution:\begin{equation}
    x(t)=\frac{1}{192} (t-24) t^2+t\label{classical-friction0}
\end{equation}
of the classical Newtonian equation with a friction term, as follows: \begin{equation}
    \ddot{x}= -\frac{1}{4} \dot{x}^{\frac{1}{2}}, \dot{x}>0, \label{newton0}
\end{equation}
which determines the motion of a point particle influenced by a velocity-dependent force.

The thick lines describe the valid motion until the body stops ($v=\dot x=0$). The interval is extended to values beyond the physical region (thin lines) for both solutions.

The fractional equivalent of the Newtonian equation of motion, combined with a classical friction term, allows for a broader range of applications than the classical approach. Damped oscillations can be accurately described with a suitable set of fractional initial conditions. From the perspective of fractional calculus, the harmonic oscillator is an example of a fractional friction phenomenon. This implies that friction and damping, described using separate approaches in a classical context, can be combined in the fractional approach. This feature is widespread in the fractional approach and can be applied to various phenomena \citep{herrmann2014fractional}.

\subsection{Fractional Harmonic Oscillator}
This section shows the behavior of fractional derivatives with the property.
\begin{equation}
    \lim_{\mu \to 2}D^\mu=\frac{d^2}{dx^2}.
    \label{1.5.3}
\end{equation}

In classical mechanics, the equation that describes free oscillations is as follows: 
\begin{equation}
    \left(\frac{d^2}{dt^2}+\frac{k}{m}\right)x(t)=0,
    \label{2.1C}
\end{equation}
whose solution is well-known, as follows:
\begin{equation}
    x(t)=c_1\cos{(wt)}+c_2\sin{(wt)}, \quad w=\sqrt{k/m}.
    \label{2.2C}
\end{equation}

The fractional formulation of the harmonic oscillator is given by the following: 
\begin{equation}
    \left(\frac{d^{2\mu}}{dt^{2\mu}}+\frac{k}{m}\right)x(t)=0.
    \label{2.6C}
\end{equation}
the second-order derivative was directly replaced by one order, $2\mu$, where $\mu$ is a fractional number and the standard harmonic oscillator for $\mu=1$ is recovered. 

The solutions of this fractional differential equation for different types of fractional derivatives are presented below, and the specific differences of the solutions obtained are shown.

\subsubsection{Harmonic Oscillator According to Fourier}

When the typical \emph{ansatz} is proposed for this type of equation \citep{herrmann2014fractional}, as follows:
\begin{equation}
    x(t)=e^{wt},
    \label{2.1.1}
\end{equation}
the solution to the fractional harmonic oscillator reduces to examining the zeros of\linebreak the polynomial, as follows:
\begin{equation}
    w^{2\mu}+\frac{k}{m}=0.
    \label{2.1.2}
\end{equation}

In the region, $1/2<\mu \leq 3/2$, there are exactly two complex conjugate solutions, as follows:
\begin{equation}
    w_{1,2}=\left|\frac{k}{m}\right|^{\frac{1}{2\mu}}\left[\cos{\left(\frac{\pi}{ 2\mu}\right)}\pm i\sin{\left(\frac{\pi}{2\mu}\right)}\right]=\left|\frac{k}{m}\right| ^{\frac{1}{2\mu}}e^{\pm i(\frac{\pi}{2\mu})}.
    \label{2.1.3}
\end{equation}

For $\mu=1$, two pure imaginary solutions are obtained, which correspond to a free and undamped oscillation. If $\mu$ begins to decrease, an increasing negative real part occurs, which, from a classical point of view, can be interpreted as increasing damping. For $\mu>1$, we have an increasing positive real part corresponding to increasing excitation \citep{herrmann2014fractional}.

On the other hand, the ordinary differential equation for a damped harmonic oscillator~is as follows:
\begin{equation}
    m\Ddot{x}(t)=-k x(t)-\gamma \Dot{x}(t),
    \label{2.1.4}
\end{equation}
where $m$ is the mass, $k$ is the spring constant, and $\gamma$ is the damping coefficient.

With \eqref{2.1.1}, we have a quadratic equation for the frequency, $w$, as follows:
\begin{equation}
    w^2=-\frac{\gamma}{m}w-\frac{k}{m},
    \label{2.1.5}
\end{equation}

Hence, we have two solutions, as follows:\vspace{-8pt}
\begin{equation}
    w_{1,2}=-\frac{\gamma}{2m}\pm \sqrt{\frac{\gamma ^2}{4m^2}-\frac{k}{m}}.
    \label{2.1.6}
\end{equation}

Compared with \eqref{2.1.3}, we have the following: \vspace{-4pt}
\begin{equation}
    -\frac{\gamma}{2m}=\left|\frac{k}{m}\right|^{\frac{1}{2\mu}}\cos{\left(\frac{\pi} {2\mu}\right)},
    \label{2.1.7}
\end{equation}
and with a Taylor series expansion for $\cos{(\pi/2\mu)}$ around $\mu =1$, we have the following: 
\begin{equation}
    -\frac{\gamma}{2m}\approx \left|\frac{k}{m}\right|^{\frac{1}{2\mu}}\frac{1}{2}\pi ( \mu -1).
    \label{2.1.8}
\end{equation}

Near $\mu=1/2$, there is no longer any oscillating contribution. The time evolution of this system is dominated by exponential decay. On the other hand, $\mu>1$ corresponds to a negative classical friction coefficient. Consequently, we have the surprising result that the behavior of the solutions of the fractional harmonic oscillator under the variation of the fractional derivative parameter, $\mu$, can be interpreted from a classical point of view as damping and excitation phenomena, respectively \citep{herrmann2014fractional}.

The following initial conditions are introduced, as follows:
\begin{align}
   & x(t=0)=x_0,
    \label{2.1.9}
    \\
   & D^{\mu}x(t)|_{t=0}=\Tilde{v}(t=0)=\Tilde{v}_0
    \label{2.1.10}
\end{align}
to compare solutions for different types of fractional derivatives. 

Only for $\mu=1$ do these initial conditions correspond to the classical initial conditions. For all other cases, the physical interpretation of a fractional velocity remains open.

The solutions of the fractional harmonic oscillator can be reformulated: A first set with the fractional initial conditions, $x_0=1$, $\Tilde{v}_0=0$, can be considered as a fractional extension of the standard cosine function, while the fractional initial conditions, $x_0=0$, $\Tilde{v}_0=1$, characterize the fractional extension of the sine function, as follows:
\begin{equation}
    {}^{}_{\text{F}}{\sin{(\mu,t)}}=\frac{1}{2i}\left[e^{\frac{1}{2}(e ^{\frac{i\pi}{2\mu}})\pi t}-e^{\frac{1}{2}(e^{-\frac{i\pi}{2\mu}} )\pi t}\right],
    \label{2.1.11}
\end{equation}
\begin{equation}
    {}^{}_{\text{F}}{\cos{(\mu,t)}}=\frac{1}{2}\left[e^{\frac{1}{2}(e ^{\frac{i\pi}{2\mu}})\pi t}+e^{\frac{1}{2}(e^{-\frac{i\pi}{2\mu}} )\pi t}\right].
    \label{2.1.12}
\end{equation}

\begin{figure}[h]
  \subfigure{
  \includegraphics[width=0.7\textwidth]{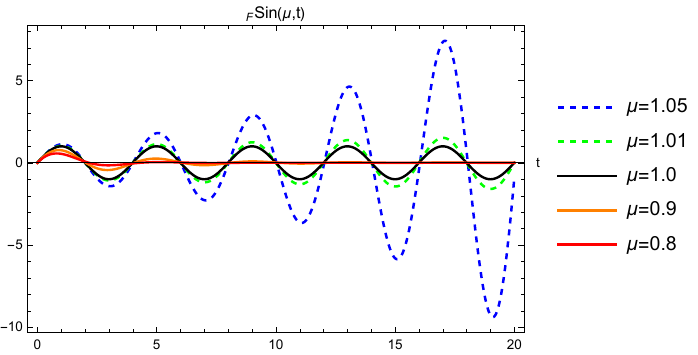}}
  \subfigure{
    \includegraphics[width=0.7\textwidth]{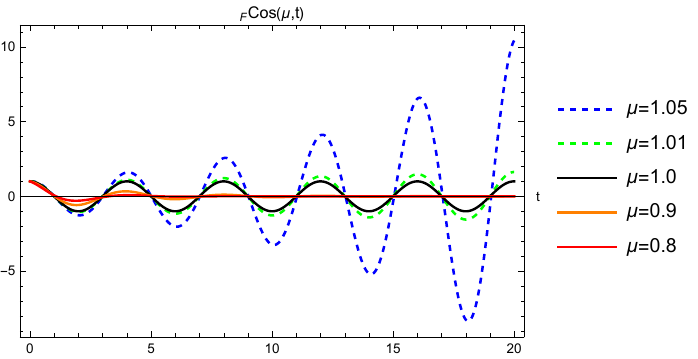}}
 \caption{Solutions ${}^{}_{\text{F}}{\sin{(\mu,t)}}$ and ${}^{}_{\text{F}}{\cos{(\mu,t)}}$ of the fractional harmonic oscillator using the Fourier derivative with different values of $\mu$.}
 \label{fig 3}
\end{figure}

An appropriately \textls[-10]{chosen linear combination of these two extended trigonometric functions will again satisfy the classical initial conditions, $x(t=0)=x_0$ and $v(t=0)=v_0$. A graphical representation of these solutions, \eqref{2.1.11} and \eqref{2.1.12}, for different $\mu$ values, is shown in Figure~\ref{fig 3}.}

\vspace{-3pt}

\subsubsection{Harmonic Oscillator According to Riemann}

To solve the differential equation of the harmonic oscillator, based on the Riemann fractional derivative, a series expansion is used according to \eqref{1.2.16}, as follows: 
\begin{equation}
    {}^{}_{\text{R}}{D^\mu}t^{n\mu}=\frac{\Gamma (1+n\mu)}{\Gamma (1+(n-1 )\mu)}t^{(n-1)\mu},
    \label{2.2.1}
\end{equation}
there are two linearly independent solutions, which are called ${}^{}_{\text{R}}{\sin{(\mu,t)}}$ and ${}^{}_{\text{ R}}{\cos{(\mu,t)}}$, in analogy to trigonometric functions:\vspace{-8pt}
\begin{align}
   &{}^{}_{\text{R}}{\sin{(\mu,t)}}=t^{\mu -1}\sum_{n=0}^{\infty}(-1 )^n\frac{t^{(2n+1)\mu}}{\Gamma ((2n+2)\mu)},
    \label{2.2.2} \\
   &{}^{}_{\text{R}}{\cos{(\mu,t)}}=t^{\mu -1}\sum_{n=0}^{\infty}(-1 )^n\frac{t^{2n\mu}}{\Gamma ((2n+1)\mu)},
    \label{2.2.3}
\end{align}
with the property\vspace{-8pt}
\begin{align}
   &{}^{}_{\text{R}} D^{\mu} {}^{}_{\text{R}}{\sin(\mu,wt)}=w^\mu {}^{}_{\text{R}}{\cos{(\mu,wt)}},
    \label{2.2.4}\\
   &{}^{}_{\text{R}} D^{\mu} {}^{}_{\text{R}}{\cos(\mu,wt)}=-w^\mu {}^{}_{\text{R}}{\sin{(\mu,wt)}}.
    \label{2.2.5}
\end{align}

These functions are related to the Mittag-Leffler function \eqref{1.1.2.6}, as follows:
\begin{align}
   & {}^{}_{\text{R}}{\sin{(\mu,t)}}=t^{2\mu -1}E(2\mu,2\mu,-t^{ 2\mu}),
    \label{2.2.6} \\
   &{}^{}_{\text{R}}{\cos{(\mu,t)}}=t^{\mu -1}E(2\mu,\mu,-t^{2\mu}).
    \label{2.2.7}
\end{align}

For $\mu <1$, the following holds: 
\begin{equation}
    \lim_{t \to 0}{}^{}_{\text{R}}{\cos{(\mu,wt)}}=\infty,
    \label{2.2.8}
\end{equation}
which means that a non-singular solution that satisfies the general initial conditions \eqref{2.1.9} cannot be given. At first glance, this seems to be a serious drawback for practically applying the fractional Riemann derivative. Nevertheless, it must be considered that the solutions presented can be useful for problems not determined by the initial conditions but are formulated in terms of boundary conditions, as is the case with the solutions of a wave equation \citep{herrmann2014fractional}.

\begin{figure}[h]
\centering
  \subfigure{
    \includegraphics[width=0.7\textwidth]{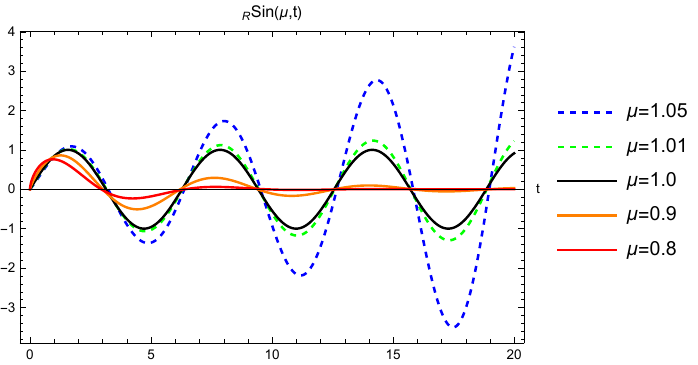}}\\
  \subfigure{
    \includegraphics[width=0.7\textwidth]{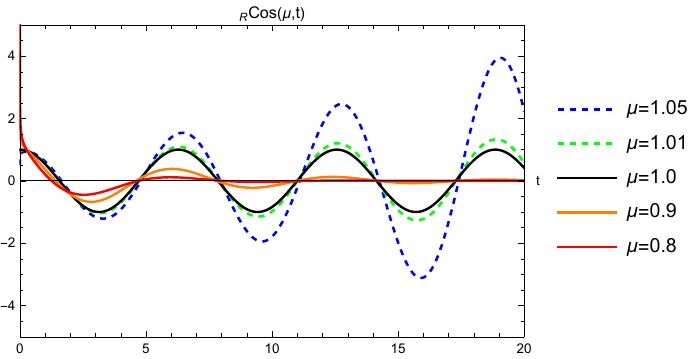}}
 \caption{Solutions  ${}^{}_{\text{R}}{\sin{(\mu,t)}}$ and ${}^{}_{\text{R}}{\cos{(\mu,t)}}$ of the fractional harmonic oscillator using the Riemann derivative for different values of $\mu$.}
 \label{fig 4}
\end{figure}

Solutions ${}^{}_{\text{R}}{\sin{(\mu,t)}}$ and ${}^{}_{\text{R}}{\cos{(\mu,t)}}$ 
 to the fractional harmonic oscillator using the Riemann derivative are presented in Figure \ref{fig 4} for different values of $\mu$.

\subsubsection{Harmonic Oscillator According to Caputo}

This solution can be written as a series according to \eqref{1.2.18}. The Caputo derivative for a power is given by the following: 
\begin{equation}
    {}^{\text{C}}{D^\mu}t^{n\mu}= \left\{ \begin{array}{lcc}
             \frac{\Gamma (1+n\mu)}{\Gamma (1+(n-1)\mu)}t^{(n-1)\mu}, & n>0,\\
             \\ 0, &n=0.
             \end{array}
   \right.
   \label{2.3.1}
\end{equation}

Therefore, we have two linearly independent solutions, which are called ${}_{\text{C}}{\sin{(\mu,t)}}$ and ${}^{}_{ \text{C}}{\cos{(\mu,t)}}$, in analogy to trigonometric functions, as follows:
\begin{align}
    & {}_{\text{C}}{\sin{(\mu,t)}}=\sum _{n=0}^{\infty}(-1)^n\frac{t^ {(2n+1)\mu}}{\Gamma (1+(2n+1)\mu)},
    \label{2.3.2}
\\
    & {}_{\text{C}}{\cos{(\mu,t)}}=\sum _{n=0}^{\infty}(-1)^n\frac{t^ {2n\mu}}{\Gamma (1+2n\mu)}.
    \label{2.3.3}
\end{align}

The most important property of this series is as follows:
\begin{align}
    &{}^{\text{C}}{D^\mu}{}_{\text{C}}{\sin{(\mu,wt)}}=w^\mu {}_{\text{C}}{\cos{(\mu,wt)}},
    \label{2.3.4}
\\
    &{}^{\text{C}}{D^\mu}{}_{\text{C}}{\cos{(\mu,wt)}}=-w^\mu {}_{\text{C}}{\sin{(\mu,wt)}}.
    \label{2.3.5}
\end{align}

\begin{figure}[h]
 \centering
  \subfigure{
    \includegraphics[width=0.7\textwidth]{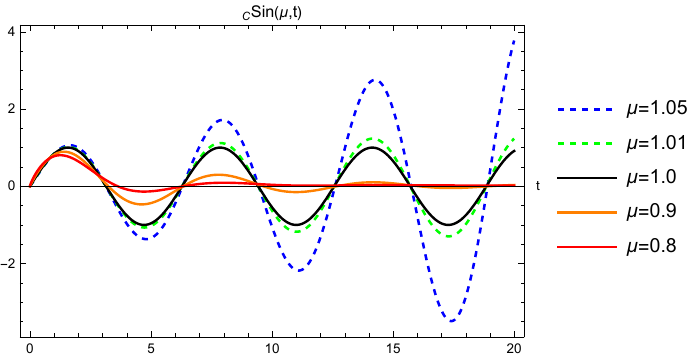}}\\
  \subfigure{
    \includegraphics[width=0.7\textwidth]{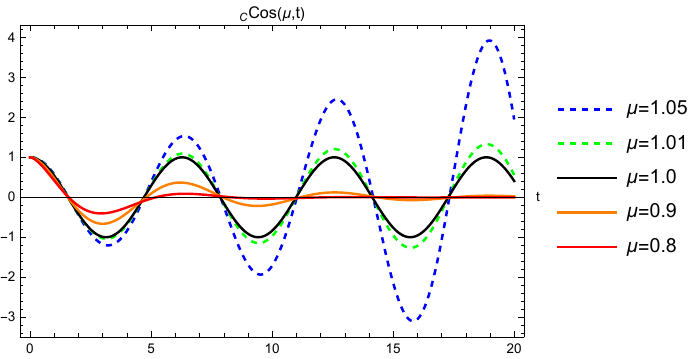}}
 \caption{Solutions ${}_{\text{C}}{\sin{(\mu,t)}}$ and ${}_{\text{C}}{\cos{(\mu,t)}}$ 
     of the fractional harmonic oscillator using the Caputo derivative for different values of $\mu$.}
 \label{Caputo}
\end{figure}

Solutions ${}_{\text{C}}{\sin{(\mu,t)}}$ and ${}_{\text{C}}{\cos{(\mu,t)}}$ of the fractional harmonic oscillator using the Caputo derivative are presented in Figure \ref{Caputo} for different values of $\mu$.

Upon comparing the solutions obtained from different definitions of the fractional derivative for the differential equation of the harmonic oscillator, we notice a remarkable similarity among the presented solutions. They all display oscillatory behavior with exponential decay when $\mu<1$, which can be interpreted as a classical damping phenomenon. Conversely, for $\mu>1$, the fractional harmonic oscillator acts as an excited oscillator. The damping and excitation properties are inherent to the fractional oscillator and do not stem from external factors. In a study by Stanislavsky (2004) \citep{Stanislavsky-Fractional-Oscillator-2004}, a multi-particle statistical interpretation was proposed to physically explain the fractional harmonic oscillator.

Furthermore, the frequency of these damped oscillations is quite similar for the various definitions of fractional derivatives presented, especially for a value of $\mu$ approximating 1. Therefore, it can be regarded as an independent property of the definition \citep{herrmann2014fractional}.

\subsection{$q$-Deformed Lie Algebras and Fractional Calculus}

The combination and integration of ideas and techniques developed across various domains of physics and mathematics have always led to new perspectives and improvements. Lie algebras are mathematical structures used to study symmetries in various branches of mathematics and physics, particularly in the realm of symmetry groups and transformation groups. $q$-deformed Lie algebras, also known as quantum deformations of Lie algebras, are generalizations of classical Lie algebras that arise in the context of quantum groups and non-commutative geometry. See, for example, \cite{Arraut:2016dlm,Hou-Xu1995,Bonatsos:1999xj}. The $q$-deformed Lie algebras theory allows for generalizations of classical symmetry groups and a broader description of natural phenomena. In what follows, we will briefly illustrate the strong relationship between fractional calculus and $q$-deformed Lie algebras. This relationship arises from the fact that both theories are based on a generalized derivative definition. It has been shown that fractional $q$-deformed Lie algebras can describe small deviations from classical Lie symmetries and bridge different Lie symmetries. See, for example, \citep{herrmann2014fractional}.

\subsubsection{$q$-Deformed Lie Algebras}

Following reference \citep{herrmann2014fractional}, section 18.1, a $q$-deformed Lie algebra is described by a parameter, $q$ (do not confuse the \emph{q}-deformation parameter with the deceleration parameter in Cosmology) and define the mapping from ordinary real numbers, $x$, to $q$-numbers, by the following: 
\begin{equation}\label{q-deform}
    \left[x\right]_q = \frac{q^x-q^{-x}}{q-q^{-1}},
\end{equation}
which has a limit
\begin{equation}
    \lim_{q\to 1} \left[x\right]_q =x.
\end{equation}

By definition, we have $\left[0\right]_q=0$. Now, a $q$-derivative may be defined as follows: 
\begin{equation}\label{q-deriv}
D_{x}^{q}f(x) = \frac{f(qx)-f(q^{-1}x)}{\left(q-q^{-1}\right)x}.
\end{equation}

This particular definition is chosen such that the $q$-derivative of the function $f(x)=x^{n}$ has a similar form to its counterpart in ordinary calculus; that is,
\begin{equation}\label{q-poli}
    D_{x}^{q}x^{n} = \left[n\right]_q x^{n-1}.
\end{equation}

This is the first common aspect of $q$-deformation and fractional calculus: they introduce a generalized derivative operator.
As an example of $q$-deformed Lie algebras, we shall review the $q$-deformed harmonic oscillator. The creation and annihilation operators, $a^{\dagger}$ and $a$, and the number operator, $N$, generate the algebra, as follows:
\begin{align}\label{q-operator1}
\left[N,a^{\dagger}\right]&=a^{\dagger}, \\ \label{q-operator2}
    \left[N,a\right]&=-a, \\ \label{q-operator3}
    aa^{\dagger}-q^{\pm1 }a^{\dagger}a&=q^{\mp N}.
\end{align}

According to definition \eqref{q-deform}, we also have the following: 
\begin{equation}
    a^{\dagger}a = \left[N\right]_q\qquad\text{and}\qquad aa^{\dagger} = \left[N+1\right]_q.
\end{equation}

Defining a vacuum state with $a\ket{0}=0$, the action of the operators $a$, $a^{\dagger}$, and $N$, on the basis $\ket{n}$ of a Fock space, defined by the repeated action of the creation operator on the vacuum state, is given by the following: \vspace{-9pt}
\begin{align}
    N\ket{n}&=n\ket{n},  \\
    a^{\dagger}\ket{n-1}&=\sqrt{\left[n\right]_q}\ket{n}, \\
    a\ket{n}&=\sqrt{\left[n\right]_q}\ket{n-1}.
\end{align}

Since the $q$-deformed algebras \eqref{q-operator1}--\eqref{q-operator3} have the same form as the non-deformed counterpart, the Hamiltonian for the $q$-deformed harmonic oscillator remains unchanged; that is,
\begin{equation}
    H=\frac{\hbar\omega}{2}\left(aa^{\dagger}+a^{\dagger}a\right),
\end{equation}
however, its eigenvalues in the $\ket{n}$ basis are the $q$-numbers: 
\begin{equation}
    E^{q}(n)=\frac{\hbar\omega}{2}\left(\left[n\right]_q+\left[n+1\right]_q\right). \label{eq:1.18.15}
\end{equation}

There is a common aspect in the concepts of fractional calculus and $q$-deformed Lie algebras; they both extend the definition of the standard derivative operator.

In the following, we establish a connection between the $q$-derivative \eqref{q-deriv} and the fractional derivative definition. Consider the Caputo definition of the fractional derivative given in the following form: \vspace{-6pt}
\begin{equation}
    D_x^{\alpha}f(x)=\left\{ \begin{array}{cc} \frac{1}{\Gamma(1-\alpha)}\int_{0}^{x}\left(x-\xi\right)^{-\alpha}\frac{\partial}{\partial \xi}f(\xi)d\xi, &  0\leq \alpha<1 \\ 
    \frac{1}{\Gamma(2-\alpha)}\int_{0}^{x}\left(x-\xi\right)^{1-\alpha}\frac{\partial^2}{\partial \xi^2}f(\xi)d\xi,  & 1\leq \alpha<2 \end{array} \right..
\end{equation}

For a function set, $f(x)=x^{n\alpha}$, we obtain the following:
\begin{equation}
    D_x^{\alpha}f(x)=\left\{ \begin{array}{cc} \frac{\Gamma(1+n\alpha)}{\Gamma(1+(n-1)\alpha)}x^{(n-1)\alpha}, & n>0 \\ 
    0, & n=0 \end{array} \right..
\end{equation}

Now, according to Equation \eqref{q-poli} and the fact that $\left[0\right]_q=0$, we can interpret the fractional derivative parameter, $\alpha$, as a deformation parameter in the sense of $q$-deformed Lie algebras. Setting $\ket{n}=x^{n\alpha}$, we define the following: 
\begin{equation}\label{q-alfa}
   \left[n\right]_\alpha \ket{n}=\left\{ \begin{array}{cc} \frac{\Gamma(1+n\alpha)}{\Gamma(1+(n-1)\alpha)}\ket{n}, & n>0 \\ 
    0, & n=0 \end{array} \right..
\end{equation}

Indeed, we have the following:  
\begin{equation}
    \lim_{\alpha\to 1}\left[n\right]_\alpha=n.
\end{equation}

The more-or-less abstract $q$-number is now interpreted within the mathematical context of fractional calculus as the fractional derivative parameter $\alpha$ with a well-understood meaning. 
Definition \eqref{q-alfa} looks like one more definition for a $q$-deformation. However, a significant difference makes the definition based on fractional calculus unique. Standard $q$ numbers are usually defined in a heuristic manner. No physical or mathematical framework exists that determines their explicit structure. Consequently, many different definitions have been proposed in the literature \cite{Bonatsos:1999xj}. In contrast to this diversity, the $q$-deformations based on the definition of the fractional derivative are uniquely determined once a set of basis functions is given.
As an example, we will briefly illustrate how to derive the $q$-numbers for the quantum version of the fractional harmonic oscillator in the next subsection. For further details, the reader should check the reference \citep{herrmann2014fractional}.

\subsubsection{The Fractional $q$-Deformed Quantum Harmonic Oscillator}

It is well-known that in quantum mechanics, the classical conjugated observables $x$ and $p$ are replaced by quantum mechanical observables $\hat{x}$ and $\hat{p}$, which are introduced as derivative operators on a Hilbert space of square-integrable wave functions. The space coordinate representations of these operators are as follows: 
\begin{align}
    \hat{x} f(x) & = x f(x), \label{classical-limits1} \\
    \hat{p} f(x) & = - i \hbar \partial_x f(x), \label{classical-limits2} 
\end{align} 
which satisfies the commutation relation, as follows:  
\begin{equation}
    [\hat{x}, \hat{p}]= i \hbar. 
\end{equation}

Now, we consider a derivative of an arbitrary (fractional) order $\alpha$.  
Then, we are introduced to two canonically conjugated operators in space representation, as follows: 
\begin{align}
    \hat{X} f(\hat{x}^{\alpha}) & = \left(\frac{\hbar}{m c}\right)^{(1-\alpha)} \hat{x}^{\alpha} f(\hat{x}^{\alpha})  \label{Fract-limits1}\\
    \hat{P} f(\hat{x}^{\alpha}) & = - i \left(\frac{\hbar}{m c}\right)^{\alpha} m c \hat{D}^{\alpha} f(\hat{x}^{\alpha}) \label{Fract-limits2}. 
\end{align}
which recover \eqref{classical-limits1} and \eqref{classical-limits2} as $\alpha \rightarrow 1$.

The classical nonrelativistic, $N$, particle Hamilton function is as follows: 
\begin{align}
    H_c = \sum_{j= 1}^{3 N} \frac{p_j^2}{2 m}  + V(x^1, \ldots, x^j, \ldots, x^{3N})
\end{align}
it is, therefore, replaced by the fractional version, as follows: 
\begin{align}
    H^{\alpha} &= \sum_{j= 1}^{3 N} \frac{\hat{P}_j^2}{2 m}  + V(\hat{X}^1, \ldots, \hat{X}^j, \ldots, \hat{X}^{3N})\\
               &= \underbrace{-\frac{1}{2} m c^2 \left(\frac{\hbar}{m c}\right)^{2\alpha} \hat{D}^{\alpha j} \hat{D}^{\alpha}_j  + V(\hat{X}^1, \ldots, \hat{X}^j, \ldots, \hat{X}^{3N})}_{\text{Using Einstein's summation convention $\sum_{j=1}^{N} x_j^2 = x^j x_j$.}}, 
\end{align}
by following the canonical quantization method.

The time-dependent Schr\"odinger type equation for fractional derivatives operators is~as follows: \vspace{-9pt}
\begin{align}
    H^{\alpha}\Psi &= i \hbar \frac{\partial}{\partial t} \Psi\\
               &= \underbrace{\left(-\frac{1}{2} m c^2 \left(\frac{\hbar}{m c}\right)^{2\alpha} \hat{D}^{\alpha j} \hat{D}^{\alpha}_j  + V(\hat{X}^1, \ldots, \hat{X}^j, \ldots, \hat{X}^{3N})\right) \Psi}_{\text{Using Einstein's summation convention $\sum_{j=1}^{N} x_j^2 = x^j x_j$.}}.
\end{align}

In the limit, $\alpha \rightarrow 1$, the classical Schr\"odinger equation is recovered.

Now, we apply the above arguments to the classical Hamilton function of the one-dimensional quantum harmonic oscillator (see section 19.1 of \citep{herrmann2014fractional}), as follows: 
\begin{equation}
    H_{\text{class}}= \frac{p^2}{2m}+\frac{1}{2}m\omega^2 x^2.
\end{equation}

Following the canonical quantization procedure, we replace the classical observables $\{x,p\}$ with the fractional derivative operators $\{\hat{X},\hat{P}\}$ according to \eqref{Fract-limits1} and \eqref{Fract-limits2}. Hence, the quantized Hamilton operator is given by the following: 
\begin{equation}
    H^{\alpha}= \frac{\hat{P}^2}{2 m} + \frac{1}{2}m\omega^2  \hat{X}^2.
\end{equation}

The stationary Schr\"odinger equation is given by the following: 
\begin{align}
    H^{\alpha} \Psi = \left(-\frac{1}{2 m}\left(\frac{\hbar}{m c}\right)^{2 \alpha} m^2 c^2 D_{x}^{\alpha}D_{x}^{\alpha} + \frac{1}{2} m \omega^2 \left(\frac{\hbar}{m c}\right)^{2 (1-\alpha)} x^{2 \alpha}\right) \Psi = E \Psi.  
\end{align}

To simplify the notation, we introduce the variable $\xi$ and the scaled energy, $E$, as follows:
\begin{equation}
    \xi^{\alpha} = \sqrt{\frac{m\omega}{\hbar}}\left(\frac{\hbar}{mc}\right)^{1-\alpha}x^\alpha \qquad\text{and}\qquad E=\hbar\omega E'
\end{equation}
are introduced.  Finally, the stationary Schr\"odinger equation for the fractional harmonic oscillator in the canonical form then reads as follows:
\begin{equation}
    H^{\alpha}\Psi_n(\xi) = \frac{1}{2}\left(-D_{\xi}^{2\alpha}+\left|\xi\right|^{2\alpha}\right)\Psi_n(\xi)=E'(n,\alpha)\Psi_n(\xi).
\end{equation}

In contrast to the classical harmonic oscillator, this fractional Schr\"odinger equation has not been solved analytically until now. Nevertheless, an approximation procedure based on the Sommerfeld quantization rule \cite{Laskin-2002} can be performed to obtain an analytical expression for the energy levels. The result is as follows: 
\begin{equation}
    E'(n,\alpha)=\left(\frac{1}{2}+n\right)^{\alpha}\pi^{\alpha/2}\left(\frac{\alpha\Gamma\left(\frac{1+\alpha}{2\alpha}\right)}{\Gamma\left(\frac{1}{2\alpha}\right)}\right)^{\alpha},\qquad n=0,1,2,\ldots.
\end{equation}

This is an analytically derived approximation of the energy spectrum for the fractional harmonic oscillator. Additionally, this equation is not dependent on a particular definition of the fractional derivative. Considering the $q$-deformed Lie algebras, we can use this analytical result to calculate the corresponding $q$-number. Using \eqref{eq:1.18.15}, the $q$-number is defined by the recursion relation, with the appropriate initial condition:  
\begin{align}
    [n+1]_{\alpha} & = 2 E'(n,\alpha) - [n]_{\alpha}.
\end{align}

This recurrence relation is fulfilled in the case of  $\alpha \in \mathbb{N}$ by the Euler polynomials \eqref{A.268}. Consequently, we have to extend their definition reasonably to the fractional case.
The initial condition $[0]_{\alpha} =0$ leads to oscillatory behavior for $\alpha<1$. 
Another choice, say,
\begin{align*}
    [0]_\alpha = 2^{1+\alpha} \pi^{\alpha/2}\left(\frac{\alpha \Gamma \left(\frac{1+\alpha}{2 \alpha}\right)}{\Gamma\left(\frac{1}{2\alpha}\right)}\right)^{\alpha}\left(\zeta\left(-\alpha, \frac{1}{4}\right) - \zeta\left(-\alpha, \frac{3}{4} \right)\right)
\end{align*}
leads to \vspace{-6pt}
\begin{align*}
[n]_\alpha  &= 2^{1+\alpha} \pi^{\alpha/2}\left(\frac{\alpha \Gamma \left(\frac{1+\alpha}{2 \alpha}\right)}{\Gamma\left(\frac{1}{2\alpha}\right)}\right)^{\alpha}\left(\zeta\left(-\alpha, \frac{1}{4} + \frac{n}{2}\right) - \zeta\left(-\alpha, \frac{3}{4} + \frac{n}{2}\right)\right)\\
&= \pi^{\alpha/2}\left(\frac{\alpha \Gamma \left(\frac{1+\alpha}{2 \alpha}\right)}{\Gamma\left(\frac{1}{2\alpha}\right)}\right)^{\alpha}E_{\alpha}\left(n+\frac{1}{2}\right),
\end{align*}
where the Hurwitz $\zeta$ function is given by \eqref{A18} and the ad hoc extension of the Euler polynomials to the fractional case is \eqref{A19}.

\subsection{Fractional Friction}
The evolution of the position $x(t)$ and the velocity $v(t)=\dot{x}(t)= \frac{d}{d t} x(t)$ of a point particle over time, influenced by an external force, $F$, is determined by Newton's second law, as follows: 
\begin{equation}
    F= m a= m \ddot{x}(t) = m \frac{d^2 }{d t^2} x(t).
\end{equation}

In the absence of external forces, the solution continues to integrate twice from $\ddot{x}=0$, as follows:
\begin{equation}
    x(t)= x_0 + v_0 t,
\end{equation}
where the constants of integration are the position $x(0)=x_0$ and initial velocities, $v(0)=v_0$ and $\dot{x}(t)=v_0$, for all $t$ values. This means the point mass maintains its initial velocity forever without external forces. This is what is known as Newton's first law. However, the daily experience of movement contradicts this result. Sooner or later, without external intervention, all types of movement end at rest. Within the framework of Newtonian theory, all these observed phenomena are due to friction forces, $F_R$.

All friction forces point in the opposite direction of the particle's speed. Empirically, a power-law function can be proposed as follows: 
\begin{equation}
    F_R= -\alpha \text{sgn}(v) |v|^{\mu}, \label{friction_G}
\end{equation}
where $\mu$ is an arbitrary real exponent. Some special cases are $\mu\approx 0$, which is observed for static and kinetic pressure in solids, while $\mu=1$ corresponds to the Stokes friction in a liquid with high viscosity, and $\mu=2$ is indicative of a general trend toward high speed \citep{herrmann2014fractional}.

Let us consider the following differential equation: 
\begin{equation}
    m \ddot{x}= -\alpha \text{sgn}(v) |v|^{\mu},
\end{equation}
where the mass, $m$, is measured in kilogram [kg], 
 and the friction coefficient, $\alpha$, is measured in {$\text{kg/s}^{2-\mu}$}, which describes the dynamics of a point particle under the effect of a friction force of the type \eqref{friction_G}.

Assuming that the velocity is positive, we have the following: 
\begin{equation}
    m \ddot{x}= -\alpha \dot{x}^{\mu}, \dot{x}>0. \label{newton}
\end{equation}

Imposing the following initial conditions: \vspace{-6pt}
\begin{align}
    & x(t=0)= x_0, \\
    & v(t=0)=v_0,
\end{align}
we obtain the general solution of \eqref{newton} given by the following: \vspace{-6pt}
\begin{equation}
    x(t)=x_0+\frac{m v_0^{2-\mu}}{\alpha(2-\mu)}-\frac{m\left[(\mu-1)(-\frac{v_0^ {1-\mu}}{1-\mu}+\frac{\alpha}{m}t)\right]^\frac{2-\mu}{1-\mu}}{\alpha(2- \mu)}. \label{classical-friction}
\end{equation}

Equation \eqref{classical-friction} determines the motion of a point particle under the influence of a velocity-dependent force for an arbitrary $\mu$.

Special cases $\mu=0$, $\mu=1$, and $\mu=2$ are included as special limits \citep{herrmann2014fractional}, as follows:
\begin{align}
   & \lim _{\mu\rightarrow 0} x(t) = x_0 + v_0 t -\frac{1}{2} \frac{\alpha}{m} t^2, \\
   & \lim _{\mu\rightarrow 1} x(t) = x_0 + \frac{m}{\alpha} v_0 \left(1 - e^{-\frac{\alpha}{m} t}\right) , \\
   & \lim _{\mu\rightarrow 2} x(t) = x_0 + \frac{m}{\alpha} \ln \left(1+ \frac{\alpha}{m} v_0 t\right).
\end{align}

Finally, the modification of Newton's Equation \eqref{newton} is studied
according to \citep{herrmann2014fractional} where the friction force
$F_R(\dot{x})= -\alpha \dot{x} (t)^{\mu}$ has been replaced by the fractional friction force, as follows:
\begin{equation}
    F_R=-\alpha D_{t}^{\mu}x(t), \hspace{20px} \Dot{x}(t)>0,
    \label{fff}
\end{equation}
where $\mu$ is the order of the fractional derivative. For $\mu\neq 1$ and $t>0$, we have the following: 
\begin{equation}
    \Dot{x}^\mu=\left(\frac{d}{dt}x(t)\right)^\mu\neq D_{t}^{\mu}x(t),
    \label{x dot alpha}
\end{equation}
so it is expected that the solutions of the fractional equation
\begin{equation}
    m \ddot{x}(t) +\alpha D_{t}^{\mu}x(t) =0 \label{Newton_Modification}
\end{equation}
have behaviors different from the standard. Note that, to compare with the classical damped harmonic oscillator given by \eqref{2.1.4}, we renamed the friction constant, $\alpha$.

Replacing the \emph{{ansatz} 
}$x(t)=e^{wt}$ into \eqref{Newton_Modification}, and applying Liouville's definition, we~have the following:
\begin{equation}
    D_{t}^{\mu}e^{w t}=w^\mu e^{w t}.
    \label{fff ansatz}
\end{equation}

Therefore, the characteristic polynomial of \eqref{Newton_Modification} is as follows: 
\begin{equation}
    w^2=-\frac{\alpha}{m}w^\mu.
    \label{w2}
\end{equation}

In addition to the trivial solution, we have the following solution: 
\begin{equation}
    w=(-1)^{\frac{1}{2-\mu}}\left|\frac{\alpha}{m}\right|^{\frac{1}{2-\mu}}.
    \label{w sol}
\end{equation}

For $0<\mu<1$, we use the following abbreviation: 
\begin{equation}
    \kappa=\left|\frac{\alpha}{m}\right|^{\frac{1}{2-\mu}},
    \label{kappa}
\end{equation}

There are two complex conjugate solutions, as follows:
\begin{equation}
    w_{1,2}=\kappa\left[\cos{\left(\frac{\pi}{2-\mu}\right)\pm i\sin{\left(\frac{\pi}{2 -\mu}\right)}}\right].
    \label{w_1solyw_2sol}
\end{equation}

So, the general solution of \eqref{Newton_Modification} is as follows: 
\begin{equation}
    x(t)=c_1+c_2e^{w_1t}+c_3e^{w_2t}, \hspace{20px} 0<\mu<1.
    \label{sol ff}
\end{equation}

It is observed that there are three different constants, $c_i$, two of which can be determined using the initial conditions, $x(t=0)=x_0$ and $v(t=0):= \dot{x}(0)= v_0$. Consequently, the freedom to specify an additional reasonable initial condition indicates that the fractional differential Equation \eqref{Newton_Modification} describes a broader range of phenomena than the classical equation \citep{herrmann2014fractional}.

Starting from \eqref{Newton_Modification}, the condition can be considered as follows:
\begin{equation}
    \Ddot{x}(t=0)=-\frac{\alpha}{m}v_0^\mu.
    \label{condition 1}
\end{equation}

Therefore, the set of equations is obtained as follows: 
\begin{equation}
    \begin{pmatrix}
         1&1&1 \\
         0&w_1&w_2\\
         0&w_1^2&w_2^2
    \end{pmatrix}\begin{pmatrix}
         c_1\\
         c_2\\
         c_3
    \end{pmatrix}=\begin{pmatrix}
         x_0\\
         v_0\\
         -\frac{\alpha}{\mu}v_0^\mu
    \end{pmatrix}.
    \label{sis eq fff}
\end{equation}

Then, the solution of the fractional differential Equation \eqref{Newton_Modification} is as follows: 
\begin{align}
    & x(t)\approx x_0+v_0t-\frac{1}{2}\frac{\alpha}{\mu}v_0^{\mu} t^2-\frac{t^3 \left(2 \alpha \cos \left(\frac{\pi }{\mu -2}\right) v_0^{\mu} \left| \frac{\alpha}{m}\right|
   ^{\frac{1}{2-\mu}}+m v_0 \left| \frac{m}{\alpha}\right| ^{\frac{2}{\mu -2}}\right)}{6 m}+o(t^4).
    \label{sol x(t) fff}
\end{align}

If the condition is chosen, i.e.,
\begin{equation}
    \Ddot{x}(t=0)=0,
    \label{cond 2}
\end{equation}
then the set of equations is obtained, as follows: 
\begin{equation}
    \begin{pmatrix}
         1&1&1 \\
         0&w_1&w_2\\
         0&w_1^2&w_2^2
    \end{pmatrix}\begin{pmatrix}
         c_1\\
         c_2\\
         c_3
    \end{pmatrix}=\begin{pmatrix}
         x_0\\
         v_0\\
         0
    \end{pmatrix},
    \label{sis eq fff 2}
\end{equation}
and in this case, the complete solution is as follows: 
\begin{equation}
    x(t)=x_0+\frac{v_0}{\kappa \sin{\left(\frac{\pi}{2-\mu}\right)}}\left[e^{\kappa \cos{\left (\frac{\pi}{2-\mu}\right)}t}\sin{\left(\frac{2\pi}{2-\mu}-\kappa \sin{\left(\frac{ \pi}{2-\mu}\right)}t\right)}-\sin{\left(\frac{2\pi}{2-\mu}\right)}\right],
    \label{sol x(t) fff 2}
\end{equation}
which is valid from $t=0$ until a final time, $t_f$, in which the mass comes to rest, which is determined by the following equation: \vspace{-9pt}
\begin{equation}
    t_f=\frac{\pi}{(2-\mu)\kappa \sin{\left(\frac{\pi}{2-\mu}\right)}}.
    \label{t_f fff}
\end{equation}

Let us compare the classical damped harmonic oscillator given by \eqref{2.1.4}. The two solutions of the classical damped harmonic oscillator are as follows:
\begin{equation}
    w_{1,2}=-\frac{\gamma}{2m}\pm \sqrt{\frac{\gamma^2}{4m^2}-\frac{k}{m}},
    \label{w_1,2OAA}
\end{equation}
where $\gamma$ is the damping coefficient in \eqref{2.1.4}. By comparing the $w_i$ of \eqref{w_1solyw_2sol} and \eqref{w_1,2OAA}, the relations are obtained, as follows:
\begin{equation}
    \frac{k}{m}=\left(\frac{\alpha}{\mu}\right)^{\frac{2}{2-\mu}}
    \label{omega1relations}
\end{equation}
and
\begin{equation}
    \frac{\gamma}{m}=-2\left(\frac{\alpha}{\mu}\right)^{\frac{1}{2-\mu}}\cos{\left(\frac {\pi}{2-\mu}\right)}.
    \label{omega2relations}
\end{equation}

These relations describe the behavior of the solutions to the fractional friction differential equation, as investigated in \citep{herrmann2014fractional}.

\begin{figure}[h]
\centering
    \includegraphics[scale=0.3]{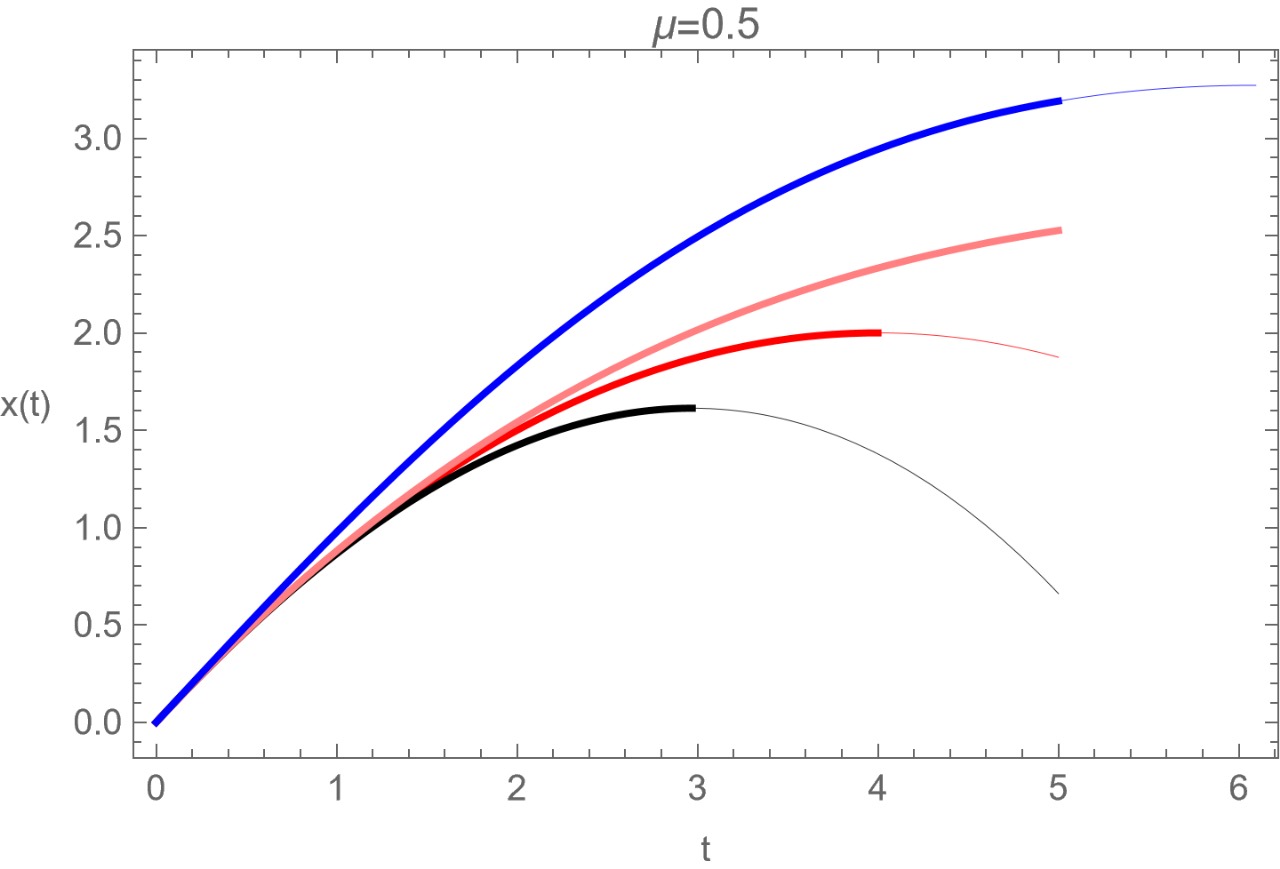}
    \caption{Different solutions of the equation of motion \eqref{Newton_Modification} with $\mu=0.5$, $x_0=0$, $v_0=1$, $m=1$, $\alpha=0.25$ \citep{herrmann2014fractional}.}
    \label{fractional friction}
\end{figure}

In Figure \ref{fractional friction}, different solutions to the fractional friction Equation \eqref{Newton_Modification} and the solution in the classical equation (pink) $m\ddot x=-\alpha \dot x^{\mu}$ given by \eqref{classical-friction} are presented. In red (up to order~2) and black (up to order 3), the solutions to the fractional friction Equation~\eqref{Newton_Modification}, given by Equation \eqref{sol x(t) fff}, are presented. The blue line represents the solution \eqref{sol x(t) fff 2} with the initial condition, $\ddot x(0)=0$ \citep{herrmann2014fractional}.

It is observed that the classical solution and the corresponding fractional solution with the initial condition \eqref{condition 1} coincide up to the second order in $t$, and there is a difference only at the end of the movement. The thick lines describe the valid motion until the body stops ($v=\dot x=0$). Furthermore, for both fractional solutions, the plot interval is extended to values $t>t_f$ (thin lines) beyond the physical region.

When the fractional parameter, $\mu$, approaches zero, the behavior approaches that of damping. 
Near $\mu=2$, the fractional friction acts as an acceleration term, and for the ideal case, $\mu=2$, the differential equation for a free particle is recovered. 

Using fractional calculus, we can expand beyond the classical approach by combining the fractional equivalent of the Newtonian equation of motion with a classical friction term. This allows us to describe a broader range of phenomena, including damped oscillations with fractional initial conditions. The harmonic oscillator is an example of a fractional friction phenomenon, where friction and damping can be unified.

It is important to note that the fractional differential equation represented by \eqref{Newton_Modification} combines two different types of classical differential equations. Nonlinear differential equations generally have complex solutions, whereas a simple second-order differential equation with constant coefficients can be solved using the same fractional equation. This observation suggests that a wide range of complex problems can be solved analytically using fractional calculus with minimal effort. On the other hand, a solution obtained from a classical theory may be more complicated.

In this section, we build upon previous research in the field \citep{herrmann2014fractional} and emphasize the importance of fractional calculus in explaining physical phenomena. We then explore the applications of this concept in theories of gravity to provide a better understanding of the complex mathematical structure of the universe. The next section aims to use advanced mathematical techniques to create new formulations and tools, which will help us better comprehend the cosmos. This is a crucial aspect of our research.

\section{Gravity Models with Fractional Derivatives}
	 \label{cap4}

In the previous section, we discussed the importance of fractional calculus in understanding physical phenomena. Now, we will focus on its applications to gravity theories to better understand the universe's mathematical structure.

Fractional derivative cosmology has been established using two methods in the classical regime. The first method is called the last-step modification method, where the fractional field equations replace the given cosmological field equations for a specific model. An example of this method is explained in \cite{Barrientos:2020kfp}. On the other hand, the second method is called the first-step modification method, which is a more fundamental approach. In this method, a fractional derivative geometry is first established, and then the variational principle for fractional action is applied to derive a modified cosmological model. This method involves defining the fractional derivative and establishing the variational principle of fractional action \citep{El-Nabulsi:2007wgc, 10.5555/1466940.1466942, Roberts:2009ix} to obtain a modified cosmological model.  For example, a fractional theory of gravitation has been developed for fractional spacetime, leading to new classes of cosmological models. These models have been studied in great detail~\citep{rami2009fractional, Vacaru:2010fb, Vacaru:2010wn, Vacaru:2010wj, Jamil:2011uj, El-Nabulsi:2007wgc, el2009fractional, El-Nabulsi:2013hsa, El-Nabulsi:2013mwa,  Debnath2012, Debnath2013, Roberts:2009ix, Shchigolev:2010vh, Shchigolev:2012rp, Shchigolev:2013jq, Shchigolev:2015rei, Shchigolev:2021lbm, Rami:2015kha, Calcagni:2009kc, Calcagni:2010bj, Calcagni:2013yqa, Calcagni:2021aap, Calcagni_2021, Calcagni:2016ofu, Calcagni:2019ngc, Calcagni:2020tvw, Calcagni:2020ads, Giusti:2020rul, Jalalzadeh:2022uhl, Jalalzadeh:2021gtq, Landim:2021ial, Landim:2021www, Garcia-Aspeitia:2022uxz, Micolta-Riascos:2023mqo}.

It is possible to create fractional versions of Newtonian mechanics and Friedman--Robertson--Walker cosmology by replacing partial and fractional derivatives in familiar equations. In cosmology, the fractional order of differentiation (represented by the symbol $\mu$) itself could be considered a variable that changes with time, although this concept has yet to be explored. The results obtained using fractional derivatives are what one might have expected, but for consistency, it is necessary to start with fractional derivatives at the first step. This leads to the subject of fractional derivative geometry. What fractional derivative geometry should look like needs to be clarified. In \cite{Roberts:2009ix}, a first attempt was made to guess what curvature and line elements might look like in two dimensions. It was found that this leads to a line element that depends on the Gamma function.  In such a framework, fractional calculus defines Riemann curvature and the Einstein tensor, dependent on the fractional parameter, $\mu$. The fractional Einstein tensor equation,  $G_{\alpha \beta}(\mu)=8 \pi G T_{\alpha \beta}(\mu)$, where $G_{\alpha \beta}(\mu)$ is the fractional Einstein tensor, was modified to study various cosmological events. 

In this section, we introduce the concept of fractional differential calculus for computing specific physical quantities. Of particular importance are the applications of these techniques to cosmology. Instead of using covariant fractional derivatives to replace the usual covariant derivatives, we will use the point-like Lagrangian formulation of cosmology in the flat FLRW metric and then generalize such action to the fractional framework. Obtaining specific approximate values for cosmological quantities is not surprising, assuming this can be defined. This is because the fractional modification of the concept of derivative can be manipulated to yield results that can be compared to cosmological measurements. The Lagrangian density is a mathematical tool used to model the dynamic properties of fields \citep{Baleanu-Muslih-2005, Agrawal-2007, El-Nabulsi-Torres-2008, Baleanu-Trujillo-2010, Odzijewicz-Malinowska-Torres-2013a, Odzijewicz-Malinowska-Torres-2013b, Odzijewicz-Malinowska-Torres-2013c}. Fractional Lagrangian densities have gained popularity in addressing cosmological problems.

In reference~\citep{Garcia-Aspeitia:2022uxz}, a~joint analysis was performed using cosmic chronometers and type Ia supernovae data. This comparison with observational tests was used to find the best-fit values for the fractional order of the derivative. These methods are a robust scheme to investigate the physical behavior of cosmological models~\citep{Hernandez-Almada:2020uyr, Leon:2021wyx, Hernandez-Almada:2021rjs, Hernandez-Almada:2021aiw, Garcia-Aspeitia:2022uxz}. They can be used in new contexts, such as~\citep{Micolta-Riascos:2023mqo}, where dynamical systems were used to analyze a fractional cosmology for different matter contents, obtaining a cosmology with late acceleration without including dark energy. References \citep{Garcia-Aspeitia:2022uxz, Gonzalez:2023who, LeonTorres:2023ehd} have explored the potential of fractional cosmology to address the $H_0$ tension \citep{DiValentino:2021izs, Efstathiou:2021ocp} from the Supernova data and the Planck value for $z<1.5$ and have reported a trend of $H_0$ that aligns with these values. However, a discrepancy exists between the $1.5<z<2.5$ range values, indicating that the tension $H_0$ has not been entirely resolved.

Based on the work presented in \citep{Garcia-Aspeitia:2022uxz}, two research paths were identified for the cosmological model without a scalar field. The first path involves comparison with the standard model, assuming that the Universe's components are cold dark matter and radiation. The second path entails deducing the equation of state for one of the matter sources based on compatibility conditions \citep{Micolta-Riascos:2023mqo}, which had not been previously analyzed in~\citep{Garcia-Aspeitia:2022uxz}.

\subsection{Cosmological Model in the Fractional Formulation of Gravity}

Given the fractional integral action,
\begin{align}
 S (\tau)
   & = \frac{1}{\Gamma(\mu)}\int_0^\tau \mathcal{L}\left(\theta, q_i(\theta), \dot{q}_i(\theta), \ddot {q}_i(\theta)\right)(\tau-\theta)^{\mu-1} d\theta, \label{GenCFL}
\end{align}
\noindent
where $\Gamma(\mu)$ is the Gamma function, $\mathcal{L}$ is the Lagrangian, $\mu$ is the constant fractional parameter, and $\tau$ and $ \theta$ are the physical and intrinsic times, respectively. Varying the action \eqref{GenCFL} with respect to $q_i$, we obtain the Euler--Poisson equations \citep{10.5555/1466940.1466942}: \vspace{-6pt}

  \begin{align}
&   \frac{\partial \mathcal{L}\left(\theta, q_i(\theta), \dot{q}_i(\theta), \ddot{q}_i(\theta)\right)}{\partial q_i}   -\frac{d}{d \theta} \frac{\partial \mathcal{L}\left(\theta, q_i(\theta), \dot{q}_i(\theta), \ddot{q}_i(\theta)\right) }{\partial \dot{q}_i} + \frac{d^2}{d \theta^2} \frac{\partial \mathcal{L}\left(\theta, q_i(\theta), \dot{q}_i(\theta), \ddot{q}_i(\theta)\right) }{\partial \ddot{q}_i}  \nonumber \\
& =  \frac{1-\mu}{\tau-\theta} \left(\frac{\partial \mathcal{L}\left(\theta, q_i(\theta), \dot{q}_i(\theta), \ddot{q}_i(\theta)\right)}{\partial \dot{q}_i}   -2 \frac{d}{d \theta} \frac{\partial  \mathcal{L}\left(\theta, q_i(\theta), \dot{q}_i(\theta), \ddot{q}_i(\theta)\right)}{\partial \ddot{q}_i} \right) \nonumber \\
& \;\;\; - \frac{\left(1-\mu\right)\left(2-\mu\right)}{\left(\tau-\theta\right)^2} \frac{\partial  \mathcal{L}\left(\theta, q_i(\theta), \dot{q}_i(\theta), \ddot{q}_i(\theta)\right)}{\partial \ddot{q}_i}. \label{EP}
 \end{align}

In cosmology, it is assumed that the flat Friedmann--Lema\^{i}tre--Robertson--Walker (FLRW)~metric,
\begin{equation}
  ds^2=-N^2 (t) dt^2+ a^2(t)(dx^2+ dy^2+dz^2), \label{FLRWm}
\end{equation}
is the spacetime metric, where $a(t)$ denotes the scale factor and $N(t)$ is the lapse function.
This result is supported observationally with data from the Planck's~\citep{Planck:2018vyg} probe. For the metric \eqref{FLRWm}, the ~Ricci scalar depends on (up to) second derivatives of $a$ and first derivatives of $N$, written as follows: \vspace{-6pt}
\begin{equation}
R(t)= 6\Bigg(\frac{\ddot{a}(t)}{a(t) N^2(t)} +\frac{{\dot{a}}^2(t)} {a^2(t) N^2(t)}-\frac{\dot{a}(t) \dot{N}(t)}{a(t) N^3(t)}\Bigg). \label{Rscalar}
\end{equation}

Thus, we can consider the integral action in units where $ 8\pi G=c=1$:
\begin{align}
 S (\tau) & = \int_0^\tau \left[\frac{R(\theta)}{2} + \frac{ \Dot{\phi}^2(\theta)}{2N^2(\theta)}-V(\phi(\theta))+ \xi R(\theta) \phi^2(\theta)+ {L}_{\text{matter}}(\theta)\right] a^ 3(\theta) N(\theta) d\theta, \label{CFL0}
\end{align}
where $R(\theta)$ is the Ricci scalar \eqref{Rscalar}.
In cosmology, the Einstein--Hilbert Lagrangian density is related to the Ricci scalar. Generically, integration by parts is taken so that a total derivative in the action is eliminated, as well as the derivatives $\ddot{a}(t)$ and $\dot{N}(t)$. However, since we will use a fractional version of the Lagrangian \eqref{CFL0}, we will not follow the standard procedure and will keep the higher-order derivatives. Using the formulation \eqref{GenCFL}, the Euler--Poisson Equation~\eqref{EP} is obtained, which involves fractional variational calculus with classical and Caputo derivatives. If ${L}_{\text{matter}}(\theta)=-\rho_{0} a(\theta)^{-3(1+w)}$ for~$w\neq -1$, we recover the usual Lagrangian density of matter of a perfect fluid as in~\citep{Wald, Carroll, carroll2004spacetime} containing derivatives of the integer order in the Lagrangian. To extend the theory given by the effective fractional action used in~\citep{Garcia-Aspeitia:2022uxz}, we can assume the Einstein--Hilbert action and add a scalar field $\phi$ to create a scalar field theory with coupling $ \xi R \phi^2$ between gravity and the scalar field, with $\xi$ being the coupling constant.
The simplest and most natural case is the minimal coupling, where $\xi=0$. Another viable option is $\xi=1/6$, known as conformal coupling because the action does not change under conformal transformations of the metric. Any value of the coupling parameter, $\xi \neq 0$, is a non-minimal coupling. 
 
The resulting action is  
\begin{small}
\begin{align}
 S_{{\text{eff}}} (\tau)
   & = \frac{1}{\Gamma(\mu)}\int_0^\tau \mathcal{L}\left(\theta, N(\theta), \dot{N}(\theta), a(\theta), \dot{a}(\theta), \ddot{a}(\theta), \phi(\theta), \dot{\phi}(\theta)\right) (\tau-\theta)^{\mu-1}  d\theta, \label{CFL}
\end{align}
\end{small}
\noindent
where $\Gamma(\mu)$ is the Gamma function, and the Lagrangian density is as follows: \vspace{-5pt}
 
\begin{small}
\begin{align}
 & \mathcal{L}\left(\theta, N(\theta), \dot{N}(\theta), a(\theta), \dot{a}(\theta), \ddot{a}(\theta)\right):=  
  \frac{3 a(\theta ) \left(N(\theta ) \left(a(\theta ) \ddot{a}(\theta )+\dot{a}^2(\theta
   )\right)-a(\theta ) \dot{a}(\theta ) \dot{N}(\theta )\right)}{ N^2(\theta )}  \nonumber \\
 & \;\;\; +a^3(\theta) N(\theta) \left(\frac{ \Dot{\phi}^2(\theta)}{2N^2(\theta)}-V(\phi(\theta))\right)  - \rho_{0} N(\theta ) a(\theta )^{-3 w} (\tau -\theta )^{-(\mu -1) (w+1)}\nonumber \\
 & \;\;\; - \frac{3 \xi \phi^2(\theta) a(\theta ) \left(N(\theta ) \left(a(\theta ) \ddot{a}(\theta )+\dot{a}^2(\theta
   )\right)-a(\theta) \dot{a}(\theta ) \dot{N}(\theta )\right)}{ N^2(\theta )},
\end{align}
\end{small}
 
\noindent
$w=p/\rho$ is a constant state equation parameter for matter, $\phi$ is the scalar field, and $V$ is its interaction potential, depending on the scalar field.

The Euler--Poisson Equation \eqref{EP} is derived after varying the action \eqref{CFL} with respect to $q_i\in \{N, a, \phi\}$, obtaining the equations, as follows: \vspace{-9pt}

\begin{align}
& \left(1-\xi  \phi^2 (\theta )\right)\left[\left(\frac{\dot{a}(\theta )}{a(\theta )}\right)^2 +\left(\frac{\dot{a}(\theta )}{a(\theta )}\right)\left(\frac{1-\mu}{\tau-\theta}\right)\right] \nonumber \\
& =\frac{1}{3}  \rho_{m0} a(\theta )^{-3 w-3} (\tau -\theta )^{-(\mu -1) (w+1)}+\frac{1}{6} \left(\dot{\phi}^2(\theta )+2 V(\phi (\theta ))\right)  + 2  \xi \frac{\dot{a}(\theta ) \phi (\theta ) \dot{\phi}(\theta )}{a(\theta )}, \label{FEQ1}
\end{align}
\begin{align}
&\left(1- \xi  \phi^2 (\theta )\right)\left[\frac{\ddot{a}(\theta )}{a(\theta )}+\frac{1}{2}\left(\frac{\dot{a}(\theta )}{a(\theta )}\right)^2+ \frac{(1-\mu)}{(t-\theta )} \frac{\dot{a}(\theta )}{a(\theta )}+\frac{(\mu -2) (\mu -1)}{2 (\tau-\theta )^2}\right] \nonumber \\
&=-\frac{1}{2} \rho_{0} w a(\theta )^{-3(1+w)}
   (\tau -\theta )^{-(\mu -1) (w+1)} -\frac{1}{4} \dot{\phi}^2(\theta )+\frac{1}{2} V(\phi (\theta )) \nonumber \\
   &  + \xi  \left[\frac{2 \phi (\theta ) \dot{a}(\theta ) \dot{\phi}(\theta )}{a(\theta )}+\frac{2 (1-\mu) \phi
   (\theta ) \dot{\phi}(\theta )}{\tau-\theta}+\phi (\theta ) \ddot{\phi}(\theta )+\dot{\phi}^2(\theta)\right], \label{REQ1}
\\
& \ddot{\phi}(\theta ) + \frac{3 \dot{a}(\theta ) \dot{\phi}(\theta )}{a(\theta )} + \frac{(1-\mu) \dot{\phi}(\theta )}{\tau-\theta}+ V'(\phi (\theta )) =  - 6 \xi  \phi (\theta ) \left[\frac{\ddot{a}(\theta )}{a(\theta )}+\left(\frac{\dot{a}(\theta )}{a(\theta )}\right)^2\right], \label{REQ2}
\end{align}
 
and the conservation of matter equation as follows: 
\begin{small}
\begin{align}
  & \dot{\rho}(\theta) = -3 \left(\frac{ \dot{a}(\theta)}{a(\theta )}+\frac{1-\mu}{3(\tau-\theta)}\right) (\rho(\theta)+p(\theta)),\label{cons}
\end{align}
\end{small}
where we replaced $N=1$ after the variation in the action \eqref{CFL}.

For fixed $\tau$, the~expressions 
\begin{equation}
    \rho(\theta)=\rho_{0} a(\theta )^{-3(1+w)} (\tau -\theta )^{-(\mu -1) (w+1)}, \label{rho(theta-tau)}
\end{equation}
and
\begin{equation}
p(\theta)=w \rho_{0} a(\theta )^{-3(1+w)} (\tau -\theta )^{-(\mu -1) (w+1)}, \label{p(theta-tau)}
\end{equation}
 define the energy density and isotropic pressure of matter fields.
\noindent

To designate the temporal independent variables, the rule $(\tau, \theta) \mapsto (2t,t)$ is applied, where the new cosmological time, $t$, \citep{Shchigolev:2010vh} is used. Hereafter, dots mean derivatives with respect to $t$. Furthermore, the parameter~Hubble is $H\equiv\dot{a}/a$. Therefore, Equations~\eqref{FEQ1} and \eqref{REQ1}, the modified Klein--Gordon Equation \eqref{REQ2}, and the conservation Equation \eqref{cons} can be written as follows: \vspace{-9pt}
 
\begin{small}
\begin{align}
& \left(1-\xi  \phi^2 (t)\right) \left(\dot{H}(t)+\frac{(1-\mu ) H(t)}{t}+\frac{3 H^2(t)}{2}+\frac{(\mu -2)
   (\mu -1)}{2 t^2}\right) \nonumber \\
   & =-\frac{1}{2}  p(t) -\frac{1}{4} \dot{\phi}(t)^2 +\frac{1}{2}
   V(\phi (t)) \nonumber \\
   & + \xi  \phi (t)  \left(2 H(t)\dot{\phi}(t)-\frac{2 (\mu -1)  \dot{\phi}(t)}{t}+  \ddot{\phi}(t)+\dot{\phi}(t)\right) \label{Raych1},\\
& \left(1-\xi  \phi^2 (t) \right) \left( H^2(t)+\frac{(1-\mu ) H(t)}{t}\right) \nonumber \\
&  = \frac{1}{3} \rho (t) +\frac{1}{6} \dot{\phi}^2(t) +\frac{1}{3} V(\phi (t))+ 2 \xi  H(t) \phi (t) \dot{\phi}(t), \label{Fried1}
\end{align}
\begin{align}
& \ddot{\phi}(t)+3 H(t) \dot{\phi}(t)-\frac{(\mu -1) \dot{\phi }(t)}{t}+V'(\phi (t)) = -\xi  \phi (t) \left(6 \dot{H}(t)+12 H(t)^2\right), \label{KG}\\
\dot{\rho}(t) &= -3 \left(H(t) +\frac{(1-\mu)}{3 t}\right)(p(t)+ \rho(t)). \label{Cons} 
\end{align}
\end{small}

{By comparing Equations \eqref{KG} and \eqref{Cons} with the conservation equations of general relativity for a non-minimally coupled scalar field to gravity, $\xi\neq 0$, we observe the following: 
\begin{align}
    \ddot{\phi}(t) + 3 H(t) \dot{\phi} + V'(\phi (t))  & =  -\xi  \phi (t) \left(6 \dot{H}(t)+12 H(t)^2\right), \label{NewREQ2}\\
    \dot{\rho}(t) & = -3 H(\rho(t)+p(t)),\label{Newcons}
\end{align}

The extra terms $- \frac{(1-\mu) \dot{\phi}(t)}{t}$ and $-\left(\frac{1-\mu}{t}\right)(\rho(t)+p(t))$, which are nontrivial for $\mu\neq 1$, can be interpreted as non-conservation equations. However, for $\mu=1$, the standard conservation equations are restored. In the general case, $\mu\neq 1$, the non-conservation equations imply the following: 
\begin{equation}
    \frac{d}{d t} \left[\left(1-\xi \phi^2 \right) \left( H^2+\frac{(1-\mu ) H}{t}\right) - \frac{1}{3} \rho -\frac{1}{6} \dot{\phi}^2 -\frac{1}{3} V(\phi)- 2 \xi H \phi \dot {\phi}\right]\neq 0.
\end{equation}

This feature of fractional cosmology leads to some constraints on the matter fields in the universe that were explored in~\citep{Micolta-Riascos:2023mqo, Gonzalez:2023who} when $\xi=0$.}

We observe that Equation \eqref{Cons} is satisfied for any value of the equation of state of the matter parameter. Furthermore, by eliminating the higher-order derivatives, we obtain Equation \eqref{Cons}, and have the following: \vspace{-9pt}
 
 \begin{small}
\begin{align}
 \dot{H} &= H \left(\frac{(1-\mu ) \left(\xi  \phi^2-1\right)}{\xi  (6 \xi -1) t \phi^2+t}-\frac{\xi  \phi  \dot{\phi}}{\xi  (6 \xi -1) \phi^2+1}\right) \nonumber \\
 & \;\;\; +\frac{H^2 \left(3 (1-8 \xi ) \xi  \phi^2-3\right)}{2 \xi  (6 \xi -1) \phi^2+2}-\frac{p}{2 \xi  (6 \xi -1) \phi^2+2} \nonumber \\ 
   &  +\frac{(\mu -2) (\mu -1) \left(\xi  \phi^2-1\right)}{2 t^2 \left(\xi  (6 \xi -1) \phi^2+1\right)} +\frac{\dot{\phi} \left((4 \xi -1) t
   \dot{\phi}-4 (\mu -1) \xi  \phi\right)}{4 \left(\xi  (6 \xi -1) t \phi^2+t\right)} \nonumber \\
 & \;\;\;  -\frac{\xi  \phi V'(\phi)}{\xi  (6 \xi -1) \phi^2+1}+\frac{V(\phi)}{2 \xi  (6 \xi -1) \phi^2(t)+2}, \label{NewRaych}
\\
 \ddot{\phi} & = H \left(\frac{6 (\mu -1) \xi  \phi \left(\xi  \phi^2-1\right)}{\xi 
   (6 \xi -1) t \phi^2+t}-\frac{3 \left(\xi  (4 \xi -1) \phi^2+1\right) \dot{\phi}}{\xi  (6 \xi -1) \phi^2+1}\right)\nonumber \\
   & +\frac{3 \xi  H^2 \phi \left(\xi  \phi^2-1\right)}{\xi  (6 \xi -1) \phi^2+1}+\frac{3 \xi  p \phi}{\xi  (6 \xi -1) \phi^2+1}-\frac{3 (\mu -2)
   (\mu -1) \xi  \phi \left(\xi  \phi^2-1\right)}{t^2 \left(\xi  (6 \xi -1) \phi^2+1\right)} \nonumber \\
   & +\frac{\dot{\phi} \left(2 (\mu -1)+\xi  \phi
   \left(2 (\mu -1) (12 \xi -1) \phi+3 (1-4 \xi ) t \dot{\phi}\right)\right)}{2 \left(\xi  (6 \xi -1) t \phi^2+t\right)} \nonumber \\
 & \;\;\;  +\frac{\left(\xi  \phi^2-1\right) V'(\phi)}{\xi  (6 \xi -1) \phi^2+1}-\frac{3 \xi  \phi V(\phi)}{\xi  (6 \xi -1) \phi^2+1}. \label{NewKG}
\end{align}
\end{small}

Solving \eqref{Fried1} for $\rho$, we obtain the following:
\begin{equation}
  \rho=  \frac{3 H (-t H+\mu -1) \left(\xi  \phi^2-1\right)}{t}-6 \xi  H\phi \dot{\phi}-V(\phi)-\frac{1}{2} \dot{\phi}^2. \label{NewFried}
\end{equation}

In \citep{Micolta-Riascos:2023mqo}, Equations \eqref{Raych1}--\eqref{Cons} for $\xi=0$ were studied. Using a procedure similar to that of RG, \eqref{Fried1} is required to be conserved over time, i.e.,
\begin{equation}
    \frac{d}{d t} \left[\left(1-\xi \phi^2 \right) \left( H^2+\frac{(1-\mu ) H}{t}\right) - \frac{1}{3} \rho -\frac{1}{6} \dot{\phi}^2 -\frac{1}{3} V(\phi)- 2 \xi H \phi \dot {\phi}\right]=0.
\end{equation}

Calculating the corresponding derivatives and substituting \eqref{Cons} and \eqref{NewRaych}--\eqref{NewFried}, we~obtain the following: \vspace{-9pt}
 
\begin{align}
   & \frac{(\mu -1)}{12 t^3 \left(\xi (6 \xi -1) \phi^2+1\right)} \Bigg[-6 t^2 H^2 \left(\xi \phi^2-1\right)^2 \nonumber \\
 & \;\;\; +\left(\xi \phi^2-1\right) \Big(-12 t H\left(-\mu +\xi \phi \left((\mu +12 \xi -3) \phi+ t \dot{\phi}\right)+3\right) \nonumber \\
   & -12 \xi t^2 \phi V'(\phi)+t \dot{\phi} \left(-12 (\mu -1) \xi \phi-t \dot{\phi}\right) \Big) -2 t^2 p \left(\xi (12 \xi +1) \phi^2-1\right)\nonumber \\
 & \;\;\; -12 \xi t^2 \dot{\phi}^2 +2 t^2 \left(\xi (12 \xi +1) \phi^2-1\right) V(\phi) +6 (\mu -2) (\mu -1) \left(\xi \phi^2-1\right)^2\Bigg]=0.
\end{align}

This equation is an identity for $\mu=1$, as expected, in standard cosmology. However, for ~$\mu\neq 1$, we acquire the new relation for the fluid pressure, as follows: \vspace{-10pt}
 
\begin{align}
   &p = -\frac{1}{2 t^2 \left(\xi (12 \xi +1)
   \phi^2-1\right)} \Bigg[6 t^2 H^2 \left(\xi \phi^2-1\right)^2 \nonumber \\
 & \;\;\; -t \left(\xi \phi^2-1\right) \Big(-12 H \left(-\mu +\xi \phi\left((\mu
   +12 \xi -3) \phi+t \dot{\phi}\right)+3\right) \nonumber \\
   & -12 \xi \phi\left((\mu -1) \dot{\phi}+t V'(\phi)\right)-t \dot{\phi}^2\Big)  +12 \xi t^2 \dot{\phi}^2
   \nonumber \\
 & \;\;\; -2 t^2 \left(\xi (12 \xi +1) \phi^2-1\right) V(\phi)   -6 (\mu -2) (\mu -1) \left(\xi \phi^2-1\right)^2\Bigg]. \label{NewP}
\end{align}

Then, using a procedure similar to GR, a new Equation~\eqref{NewP} is obtained, 
demonstrating that two of the three equations are independent. This characteristic of fractional cosmology imposes certain constraints on the matter fields in the Universe, as explored in~\citep{Micolta-Riascos:2023mqo}. Therefore, the equation of state is modified\vspace{-6pt}
\begin{align}
\label{matterEoS}
    w(t)= \frac{p(t)}{\rho(t)}.
\end{align}

Replacing the expression for $p$ given by \eqref{NewP} into \eqref{NewRaych}, \eqref{NewKG}, and \eqref{Cons}, we obtain the following: \vspace{-9pt}

\begin{align}
\dot{H} &= \left(\xi  \phi^2-1\right) \left(\frac{(\mu -2) (\mu -1)}{t^2 \left(\xi  (12 \xi +1) \phi^2-1\right)}-\frac{2 (\mu -4) H}{t
   \left(\xi  (12 \xi +1) \phi^2-1\right)}\right) \nonumber \\
   & -\frac{2 \xi  H \phi \dot{\phi}}{\xi  (12 \xi +1) \phi^2-1}  +\frac{H^2 \left(3-3 \xi  (8 \xi +1) \phi^2\right)}{\xi  (12 \xi +1) \phi^2-1}
   +\frac{2 \xi  t \dot{\phi}^2-2 \xi  \phi \left((\mu -1) \dot{\phi}+t V'(\phi)\right)}{t
   \left(\xi  (12 \xi +1) \phi^2-1\right)}, \label{NRay} \\
\ddot{\phi} & = \left(\xi  \phi^2-1\right) \Bigg[\frac{12 (\mu -4) \xi  H \phi}{t \left(\xi  (12
   \xi +1) \phi^2-1\right)}+\frac{6 \xi  H^2 \phi}{\xi  (12 \xi +1) \phi^2-1} -\frac{t^2 V'(\phi)+6 (\mu -2) (\mu -1) \xi  \phi}{t^2
   \left(\xi  (12 \xi +1) \phi^2-1\right)}\Bigg]  \nonumber \\
   & -\frac{3 H \left(\xi  (8 \xi +1) \phi^2-1\right) \dot{\phi}}{\xi  (12 \xi +1) \phi ^2-1} 
   +\frac{\dot{\phi} \left(-\mu +\xi  \phi \left((\mu -1) (24 \xi +1) \phi-12 \xi  t \dot{\phi}\right)+1\right)}{t \left(\xi  (12 \xi +1) \phi^2-1\right)}, \label{NKG}
 \end{align}

This leads to the auxiliary equation, as follows:  

\begin{align}
    & \dot{\rho}= \left(\xi  \phi^2-1\right) \Bigg[ \frac{3 H}{t^2 \left(\xi  (12 \xi +1) \phi^2-1\right)}  \nonumber \\
 & \;\;\;  \times \Big((\mu -5) (\mu -1)+\xi  \phi \Big((\mu -1) (12 (\mu -3) \xi -\mu +5)
    +6 t \left((\mu -1) \dot{\phi}+t V'(\phi)\right)\Big)\Big)\nonumber \\
 & \;\;\; +\frac{H^2 \left(-3 \mu +3 \xi  (-48 \mu  \xi +\mu +120 \xi -13) \phi^2+39\right)}{t \left(\xi  (12 \xi +1) \phi^2-1\right)} +\frac{18 H^3 \left(\xi  (6 \xi +1) \phi
  ^2-1\right)}{\xi  (12 \xi +1) \phi^2-1} \nonumber \\
 & \;\;\; -\frac{6 (\mu -1) \xi  \phi \left((\mu -1) \dot{\phi}+t V'(\phi)\right)}{t^2 \left(\xi  (12 \xi +1)
   \phi^2-1\right)}\Bigg] \nonumber \\
 & \;\;\; +\left(\xi  \phi^2-1\right)^2 \left(\frac{3 (\mu -2) (\mu -1)^2}{t^3 \left(\xi  (12 \xi +1) \phi^2-1\right)}-\frac{9
   (\mu -2) (\mu -1) H}{t^2 \left(\xi  (12 \xi +1) \phi^2-1\right)}\right)  \nonumber \\
 & \;\;\; +\frac{3 H \dot{\phi} \left(t \left(6 \xi +\xi  (6 \xi +1) \phi
  ^2-1\right) \dot{\phi}-4 (\mu -1) \xi  \phi \left(\xi  (6 \xi +1) \phi^2-1\right)\right)}{t \left(\xi  (12 \xi +1) \phi
  ^2-1\right)} \nonumber \\
 & \;\;\; +\frac{36 \xi  H^2 \phi \left(\xi  (6 \xi +1) \phi^2-1\right) \dot{\phi}}{\xi  (12 \xi +1) \phi^2-1}-\frac{(\mu -1) \left(6
   \xi +\xi  (6 \xi +1) \phi^2-1\right) \dot{\phi}^2}{t \left(\xi  (12 \xi +1) \phi^2-1\right)}. \label{NFried} 
 \end{align}

 These equations allow one to deduce an effective equation of state for matter without imposing equations of state on each matter field.
In the case of a minimal coupling, $\xi=0$, this allows the system to be investigated as follows: 
\begin{align}
   & \dot{H}= \frac{2(4- \mu ) H}{t}-3 H ^2+\frac{(\mu -2) (\mu -1)}{t^2},\label{NewIntegrableH}\\
   & \ddot{\phi}= -3 H  \dot{\phi}+\frac{(\mu -1) \dot{\phi}}{t}-  V'(\phi). \label{NNKG}
\end{align}

The Equation \eqref{NewIntegrableH} is the Riccati ordinary differential equation for $\mu \neq 1$ and $\xi=0$. The analytical solution of \eqref{NewIntegrableH} is  (for an analogous case, see Equation (36) in \citep{Shchigolev:2012rp}): 
\begin{equation}
 H(t)=\frac{1}{3 t }\left[\frac{9-2 \mu +r}{2}-\frac{c r \alpha_0^r}{c \alpha_0^r+ (H_0 t )^r}\right], \label{solution}
\end{equation}
where
\begin{align}
c & = \frac{-2 \mu +r-6 \alpha_0+9}{2 \mu +r+6 \alpha_0-9},
\;r = \sqrt{8 \mu (2 \mu -9)+105},
\end{align}%
and $\alpha_0=H_{0}t_0$, where $t_0$ is the value of $t$ today. $H_0$ is the current value of the Hubble factor, and $\alpha_0$ is the current age parameter, for which we obtain the best-fit values.
For large $t$, the asymptotic scaling factor can be expressed as follows:
\begin{equation}
    a(t)\simeq t^{\frac{1}{6} \left(+9- 2\mu +r\right)},
\end{equation}

Thus, the acceleration of the late universe can be obtained with a power-law scale factor without a cosmological constant. These results are independent of the matter source and~anisotropy.

\subsection{Minimal Coupling}
\label{RRResults}

In this section, we discuss the results of \citep{Gonzalez:2023who} for minimal coupling ($\xi= 0$) and a constant potential $V(\phi)= \Lambda$.

Introducing the logarithmic independent variable, $\tau= -\ln(1+z)= \ln a$, with~$\tau\rightarrow -\infty $ when $z\rightarrow \infty$, $\tau\rightarrow 0 $ when $z\rightarrow 0$, and~ $\tau\rightarrow \infty $ when $z\rightarrow -1$, and~defining the age parameter of the universe as $\alpha = t H$, we obtain the initial value problem, i.e.,  
\begin{align}
   & \alpha '(\tau)= 9-2 \mu -3 \alpha (\tau)+\frac{(\mu -2) (\mu -1)}{\alpha (\tau)}, \label {eq(134)}\\
   &t'(\tau)={t(\tau)}/{\alpha(\tau)}, \label{eq(135)}\\
   & \alpha(0)=t_0 H_0, t(0)=t_0, \label{eq(136)}
\end{align}
plus the auxiliary equation, as follows: 
\begin{small}
\begin{equation}
    \phi ''(\tau )=-\frac{(\mu -1) \phi '(\tau ) (\mu -t(\tau) \phi (\tau )-2)}{\alpha (\tau )^2}-\frac{\phi '(\tau ) (-2 \mu +3 t(\tau) \phi (\tau )+8)}{\alpha (\tau
   )}+3 \phi '(\tau ).
\end{equation}
\end{small}

Equation \eqref{eq(134)} gives a one-dimensional dynamical system analyzed in \citep{Micolta-Riascos:2023mqo}. There is an asymptotic behavior for large $\tau$, which is consistent with~\citep{Micolta-Riascos:2023mqo}, in~which the attractor solution has an asymptotic age parameter
$\lim_{t\rightarrow \infty} t H= \frac{1}{6} (9-2 \mu +r)$. 

The exact solution of this system is as follows: 
\begin{small}
\begin{align}
    t(\tau )=t_0 \exp \left(\frac{2 \left(\tan ^{-1}\left(\frac{6 H_0 t_0+2 \mu -9}{\sqrt{8 (9-2 \mu ) \mu -105}}\right)-\tan
   ^{-1}\left(\frac{6 \alpha (\tau )+2 \mu -9}{\sqrt{8 (9-2 \mu ) \mu -105}}\right)\right)}{\sqrt{8 (9-2 \mu ) \mu -105}}\right),
\end{align}
\end{small}
where $\alpha$ is obtained implicitly from the following: \vspace{-13pt}
 \begin{small}
\begin{align}
   &  \frac{1}{6} \ln \left(-2 \mu  \alpha (\tau )-3 \alpha (\tau )^2+9 \alpha (\tau )+\mu ^2-3 \mu +2\right)-\frac{(2 \mu -9) \tan ^{-1}\left(\frac{6 \alpha
   (\tau )+2 \mu -9}{\sqrt{-16 \mu ^2+72 \mu -105}}\right)}{3 \sqrt{-16 \mu ^2+72 \mu -105}}\nonumber \\
   & =\frac{1}{6} \left(\ln \left(-2 H_0 \mu  t_0-3
   H_0 t_0 (H_0 t_0-3)+\mu ^2-3 \mu +2\right)+\frac{2 (9-2 \mu ) \tan ^{-1}\left(\frac{6 H_0 t_0+2 \mu -9}{\sqrt{8
   (9-2 \mu ) \mu -105}}\right)}{\sqrt{8 (9-2 \mu ) \mu -105}}-6 \tau \right),
\end{align}
 \end{small}
where the parameter $\epsilon_0$ is introduced, such that we have the following:
\begin{small}
\begin{equation}
 \epsilon_0= \frac{1}{2}\lim_{t\rightarrow \infty} \left(\frac{t_0 H_0 - t H}{t H}\right), \quad \alpha_0=  \frac{1}{6} \left(9 -2 \mu +\sqrt{8 \mu  (2 \mu -9)+105}\right)(1+ 2 \epsilon_0).\label{alpha_0}
\end{equation} 
\end{small}
$\epsilon_0$ is a measure of the limiting value of the relative error in the age parameter $t H$ when approximated by $t_0 H_0$.

In a recent study \cite{Gonzalez:2023who}, we investigated the potential of using fractional cosmology to account for the universe's accelerated expansion without the requirement of exotic fluids in the model. We found self-accelerated solutions where the matter--energy density is zero. In this article, we will provide an overview of how a model with cold dark matter incorporates these possibilities following the previous reference.

\subsubsection{Model with cold dark~matter} 

Assume
\begin{equation}
    H\left( t\right) =\frac{\alpha}{t},\label{12}
\end{equation}
where
 $\alpha$ is a constant and $\mu>1$. If $\alpha=\mu-1$, we have a power-law solution, as follows: 
\begin{equation}
H\left( t\right) =\frac{\mu -1}{t}\implies a\left( t\right) =\left( 
\frac{t}{t_{0}}\right) ^{\mu -1}.\label{3}
\end{equation}

The conservation equation for matter, as described in \eqref{Cons}, applies to cold dark matter ($p_{\text{CDM}}=w_{\text{CDM}}\rho _{\text{CDM}}$, and $w_{\text{DM}}=0$); when $H$ is defined by \eqref{12}, this equation simplifies to the following: 
\begin{equation}
\dot{\rho}_{\text{CDM}} +  \frac{(3 \alpha -\mu +1) {\rho}_{\text{CDM}}}{t}=0.
\end{equation}
Hence, for the matter--energy density, we have the following: 
\begin{align}
\rho _{\text{CDM}} & = \rho_{CDM\left( 0\right) }(t/t_0)^{^{-3 \alpha +\mu -1}}.
\end{align}%
Choosing
\begin{equation}
 \alpha = \frac{\mu +1}{3}, \label{zeta}
\end{equation}
we obtain the following: 
\begin{equation}
\rho _{\text{CDM}}  = \rho_{CDM\left( 0\right) }(t/t_0)^{-2}.
\end{equation}
Then, from~Equation \eqref{NewFried}, we obtain the following: 
\begin{align}
\rho_{CDM\left( 0\right) }= \frac{2 (2-\mu) (\mu +1)}{3 t_0^2}. \label{13}
\end{align}%
Using the redshift parameter, $1+z=1/a$, we obtain the following:
\begin{align}
H\left( t\right) & = \frac{\alpha}{t}\implies  a\left(
t\right) =\left( \frac{t}{t_{0}}\right) ^{\alpha},  \label{17} \\
\; \text{and}\;1+z & = \frac{1}{a}\Longrightarrow \left( \frac{t}{t_{0}}%
\right)=\left( 1+z\right) ^{-\frac{1}{\alpha}},  \label{18}
\end{align}%
then, we have the following: 
\begin{equation}
\rho _{\text{CDM}}\left( z\right) =\rho _{\text{CDM}}\left( 0\right) \left(
1+z\right) ^{\frac{2}{\alpha}}\; \text{and}\; H\left( z\right)
=H_0 \left( 1+z\right) ^{\frac{1}{\alpha}}, 
\label{19}
\end{equation}%
where $\alpha$ is defined by \eqref{zeta}. 
 EoS $w_{\text{eff}}$ is defined through
\begin{equation}
\rho _{\text{CDM}}\left( z\right) =\rho _{\text{CDM}}\left( 0\right)  (1+z)^{3(1+w_{\text{eff}})},\label{20}
\end{equation}
we have the following: 
\begin{equation}
w_{\text{eff}} =  -1 + \frac{2}{3 \alpha}= -1+\frac{2}{\mu +1} \; \text{and}\; 
q  = -1+\frac{1}{\alpha}= -1+\frac{3}{\mu +1}. \label{18c}
\end{equation}

Similar to GR, we have the usual relation,  $q= \frac{1}{2} \left(1 + 3 w_{\text{eff}}\right)$. 
Therefore, in~fractional cosmology, we have an acceleration ($\ddot{a}>0, q<0$) as is present in GR when the effective fluid has $w_{\text{eff}}<-1/3$. 
Hence, 
\begin{equation}
\ddot{a}\left( t\right) <0, q>0, w_{\text{eff}}>-1/3 \Longleftrightarrow 1< \mu<2,  \label{19b}
\end{equation}
and
\begin{equation}
\ddot{a}\left( t\right) > 0, q<0, w_{\text{eff}}<-1/3 \Longleftrightarrow \mu >2. \label{19c}
\end{equation}
Finally, we have an accelerated expansion if $\mu >2$, caused by the fractional derivative correction, and not by the matter content. That is the powerful advantage of fractional cosmology over GR. This is consistent as $\rho _{\text{CDM}}\rightarrow 0$ with the asymptotic solution \mbox{$H(t)=\frac{\mu -1}{t}$}, where~ $q=-\frac{\mu -2}{\mu -1}$, which is a power-law solution  $a(t)= \left(t/t_0\right)^{\mu-1}$. It is accelerated if $\mu>2$ and decelerated if $1<\mu <2$, as proven in~\cite{Garcia-Aspeitia:2022uxz}.

\subsubsection{Interpretation of the Fractional Term as a Dark Energy~Source}

We write  \eqref{Fried1} as follows: 
\begin{align}
& 3H^{2} = \rho _{\text{CDM}}+\rho _{\text{frac}},  \label{23} \\
& \; \text{where}\; \rho _{\text{CDM}}\left( z\right) =\rho _{\text{CDM}}\left( 0\right) \left(
1+z\right) ^{\frac{2}{\alpha}}= \frac{2 (2-\mu) (\mu +1)}{3 t_0^2}  \left(
1+z\right) ^{\frac{6}{1+\mu}}, \\
& \; \text{where}\;\rho _{\text{frac}}\left( t\right) = \frac{3\left( \mu -1\right) }{t
}H \underbrace{\implies}_{\text{using}\; \eqref{12}\; \text{and} \; \eqref{18}} \rho _{\text{frac}}\left( z\right) =\rho _{\text{frac}}\left( 0\right)
\left( 1+z\right) ^{\frac{2}{\alpha }},  \label{24} \\
& \; \text{and}\;\rho _{\text{frac}}\left( 0\right) = \frac{3\left( \mu -1\right) }{
t_{0}}H_{0}=\frac{\left( \mu -1\right) }{t_{0}H_{0}}3H_{0}^{2}.   \label{25}
\end{align}
Using  $H=d\ln a/dt$ and $1+z=1/a$, we obtain the following: 
\begin{equation}
t_{0}H_{0}=\int_{0}^{\infty }\frac{dz}{\left( 1+z\right) E\left( z\right) }.
\label{41}
\end{equation}
Substituting (see Equation \eqref{12}), we have the following: 
\begin{equation}
E\left( z\right) =\left( 1+z\right) ^{\frac{1}{\alpha }}= \left( 1+z\right) ^{\frac{3}{1+\mu}}, 
\label{42}
\end{equation}
with $\alpha$ defined by \eqref{zeta}, 
we obtain the following:
\begin{equation}
H_0t_0=(1+\mu)/3. \label{H0t0mu}
\end{equation}
On the other hand, using \eqref{23}, \eqref{24}, and \eqref{25}, we obtain the following: 
\begin{align}
& E^{2}\left( z\right)  = \frac{3H^{2}\left( z\right) }{3H_{0}^{2}}=\frac{\rho
_{\text{CDM}}\left( z\right) }{3H_{0}^{2}}+\frac{\rho _{\text{frac}}\left( z\right) }{
3H_{0}^{2}}\Longleftrightarrow E^{2}\left( z\right) =\Omega _{\text{DM}}\left(
z\right) +\Omega _{\text{frac}}\left( z\right),   \label{26} \\
& \implies \Omega _{\text{frac}}\left( 0\right) =1-\Omega _{\text{DM}}\left( 0\right) =
\frac{\mu -1}{t_{0}H_{0}}= \frac{3(\mu -1)}{\mu+1}\sim 0.744\%, \;  \mu \sim 1.65957,  \label{28}
\end{align}
where $\Omega_{\text{DM}}\left( 0\right) \sim 0.256\%$. We compare the value $\mu\sim 1.66$, with~the observational tests performed in~\cite{Garcia-Aspeitia:2022uxz} for a flat prior $1< \mu<3$, where the best-fit value for $\mu$ is $\mu^*= 1.71$. 
 
We will inspect the nature of $\rho _{\text{frac}}$ as an effective fluid in GR, i.e.,
\begin{align}
\rho _{\text{frac}} & = \frac{3\left( \mu -1\right) }{t}H,\; q= -1-\frac{\dot{H}}{H^2}
\implies  \dot{\rho}_{\text{frac}} =-H\left( 1+q+\frac{1}{Ht}\right) \rho _{\text{frac}},  \label{32} \\
&\implies \dot{\rho}_{\text{frac}}+3H\left( 1+w _{\text{frac}}\right) \rho
_{\text{frac}}=0,  \label{33}
\end{align}
where $w_{\text{frac}} = \frac{1}{3}\left(q-2+\frac{1}{\alpha}\right)$. According to \eqref{18c}, we again deduce \eqref{20},
\begin{align}
w_{\text{frac}} & = -1+\frac{2}{3\alpha },  \label{36}
\end{align}
corresponding to the quintessence ($-1<w_{\text{frac}}<-1/3$) if $\mu >2$. 

\subsubsection{Tests against cosmological observations}

The observational tests performed in~\cite{Garcia-Aspeitia:2022uxz} were improved in \cite{Gonzalez:2023who}. In this reference, the (theoretical) Hubble parameter as a function of redshift is tested against cosmological observations. Therefore, for the adjustment, systems \eqref{eq(134)} and \eqref{eq(135)} are numerically integrated, representing a system for the variables $(\alpha,t)$ as a function of $\tau=-\ln{\left(1+z\right)}$, and~for which we consider the initial conditions $\alpha(\tau=0)\equiv\alpha_{0}=t_{0} H_{0 }$ and $t(\tau=0)\equiv t_{0}=\alpha_{0}/H_{0}$. Then, the parameter ~Hubble is obtained numerically by $H_{th}(z)=\alpha(z)/t(z)$.

Furthermore, for a better comparison, the free parameters of the $\Lambda$CDM model are also fitted, whose Hubble parameter as a function of redshift is given by \citep{Gonzalez:2023who}, as follows: 

\begin{equation}\label{HLCDM}
    H(z)=H_{0}\sqrt{\Omega_{m,0}(1+z)^{3}+1-\Omega_{m,0}}.
\end{equation}

The free parameters of the fractional cosmological model are $\theta=\{h,\mu,\epsilon_{0}\}$ and~the free parameters of the $\Lambda$CDM model are $\theta=\{h ,\Omega_{m,0}\}$. For~the free parameters, $\mu$, $\epsilon_{0}$ and~$\Omega_{m,0}$, we consider the following flat priors: $\mu\in F(1,4) $, $ \epsilon _{0}\in F(-0.1,0.1)$ and~$\Omega _{m,0}\in F(0,1)$. It is important to mention that, due to a degeneracy between $H_{0}$ and $\mathcal{M}$, the ~SNe Ia data cannot constrain the free parameter, $h$ (as a reminder, $H_{ 0}=100 \frac{\text{km/s}}{\text{Mpc}}h$), contrary to the case of OHD and, consequently, in ~the joint analysis.
Therefore, the posterior distribution of $h$ for the SNe Ia data is expected to cover all prior distributions. On the other hand, the prior is chosen as $\epsilon_{0}=0$, which is a measure of the limiting value of the relative error in the age parameter, $t H$, when approximated by $t_0 H_0 $, according to Equation \eqref{alpha_0}. For~the mean value of $\epsilon_0=0$, $\alpha_0=\frac{1}{6} (-2 \mu +r+9)$ was acquired, which implies $c=0$.

The analysis of the SNe Ia, OHD data, and the joint analysis with SNe Ia plus OHD data lead, respectively, to $h=0.696_{-0.295}^{+0.302}$, $\mu=1.340_{-0.339} ^ {+2.651}$ and $\epsilon_0=\left(1.976_{-2.067}^{+1.709}\right)\times 10^{-2}$, $h=0.675_{-0.021}^{+ 0.041 }$, $\mu=2.239_{-1.190}^{+1.386}$, and $\epsilon_0=\left(0.865_{-0.773}^{+0.793}\right)\times 10^{-2} $, and $h=0.84_{-0.027}^{+0.031}$, $\mu=1.840_{-0.773}^{+1.446}$, and $\epsilon_0=\left(1.213_{-1.057}^ { +0.482}\right)\times 10^{-2}$, where best-fit values are calculated at $3\sigma$ CL.

In this model, the deceleration parameter is calculated using specific values for $\alpha(\tau)$ and $\mu$, in closed form, as follows: \vspace{-9pt}
\begin{equation}
q(\alpha(\tau))= 2 + \frac{2 (\mu -4) }{\alpha(\tau)}-\frac{(\mu -2) (\mu -1)}{\alpha^ 2(\tau)}.\label{DecelerationFinal}
\end{equation}

Using best-fit estimates from Table \ref{tab:bestfits}, we show a transition at $z_{t}\gtrapprox 1$, with a larger transition redshift than the $\Lambda$CDM model. The current deceleration parameter for the fractional cosmological model is $q_{0}=-0.37_{-0.11}^{+0.08}$ at $3\sigma$ CL.

\begin{table}[ht!]
  \caption{Best fit values and criteria $\chi^{2}_{min}$ for the $\Lambda$CDM model with free parameters $h$ and $\Omega_{m,0}$ for the fractional cosmological model (dust + radiation) \citep{Garcia-Aspeitia:2022uxz} and the fractional cosmological model with free parameters $h$, $\mu$ and $\epsilon_{0}$ \citep{Gonzalez:2023who}. Using the MCMC analysis \cite{Foreman:2013}, the best-fit values and their joint analysis were obtained from the SNe Ia, OHD (or CC) data. The $\Lambda$CDM model was used as a reference model. (Taken from \citep{LeonTorres:2023ehd}).}
      \setlength{\tabcolsep}{4mm}
      \resizebox{\textwidth}{!}{
    \begin{tabular}{cccccc}
        \hline
         & \multicolumn{4}{c}{\textbf{Best fit values}} & \\ \\
         \cline{2-5}
         \\
        \textbf{Data} & \boldmath{$h$} & \boldmath{$\Omega_{m,0}$} & \boldmath{$\mu$} & \boldmath{$\epsilon_{0}\times 10^{2}$} & \boldmath{$\chi_{min}^{2}$} \\
        \hline
        \multicolumn{6}{c}{\boldmath{$\Lambda$}\textbf{CDM} Model} \\ \\
        \textbf{SNe Ia} & $0.692_{-0.120\;-0.278\;-0.292}^{+0.209\;+0.296\;+0.307}$ & $0.299_{-0.021\;-0.042\;-0.059}^{+0.022\;+0.046\;+0.068}$ & $\cdots$ & $\cdots$ & $1026.9$ \\ \\
        \textbf{OHD} & $0.706_{-0.012\;-0.024\;-0.036}^{+0.012\;+0.024\;+0.035}$ & $0.259_{-0.017\;-0.033\;-0.047}^{+0.018\;+0.038\;+0.059}$ & $\cdots$ & $\cdots$ & $27.5$ \\ \\
        \textbf{SNe Ia+OHD} & $0.696_{-0.010\;-0.020\;-0.029}^{+0.010\;+0.020\;+0.029}$ & $0.276_{-0.014\;-0.027\;-0.040}^{+0.014\;+0.030\;+0.043}$ & $\cdots$ & $\cdots$ & $1056.3$ \\ \\
        \hline
        \multicolumn{6}{c}{\textbf{Fractional cosmological model (dust + radiation) \citep{Garcia-Aspeitia:2022uxz} (The uncertainties presented correspond to $1\sigma(68.3\%)$ CL)}} \\ \\
        \textbf{SNe Ia} & $0.599^{+0.275}_{-0.269}$ & $0.160^{+0.050}_{-0.072}$ & $2.771^{+0.161}_{-0.214}$ & $\cdots$ & $54.83$ \\ \\
        \textbf{CC} & $0.629^{+0.027}_{-0.027}$ & $0.399^{+0.093}_{-0.122}$ & $2.281^{+0.492}_{-0.433}$ & $\cdots$ & $16.14$ \\ \\
        \textbf{SNe Ia+CC} & $0.692^{+0.019}_{-0.018}$ & $0.228^{+0.035}_{-0.040}$ & $2.839^{+0.117}_{-0.193}$ & $\cdots$ & $78.69$ \\ \\
        \hline
        \multicolumn{6}{c}{\textbf{Fractional cosmological model \citep{Gonzalez:2023who} (The uncertainties presented correspond to $1\sigma(68.3\%)$, $2\sigma(95.5\%)$, and $3\sigma(99.7\%)$ CL)}} \\ \\
        \textbf{SNe Ia} & $0.696_{-0.204\;-0.284\;-0.295}^{+0.215\;+0.293\;+0.302}$ & $\cdots$ & $1.340_{-0.245\;-0.328\;-0.339}^{+0.492\;+2.447\;+2.651}$ & $1.976_{-0.905\;-1.848\;-2.067}^{+0.599\;+1.133\;+1.709}$ & $1028.1$ \\ \\
        \textbf{OHD} & $0.675_{-0.008\;-0.015\;-0.021}^{+0.013\;+0.029\;+0.041}$ & $\cdots$ & $2.239_{-0.457\;-0.960\;-1.190}^{+0.449\;+0.908\;+1.386}$ & $0.865_{-0.407\;-0.657\;-0.773}^{+0.395\;+0.650\;+0.793}$ & $29.7$ \\ \\
        \textbf{SNe Ia+OHD} & $0.684_{-0.010\;-0.020\;-0.027}^{+0.011\;+0.021\;+0.031}$ & $\cdots$ & $1.840_{-0.298\;-0.586\;-0.773}^{+0.343\;+1.030\;+1.446}$ & $1.213_{-0.310\;-0.880\;-1.057}^{+0.216\;+0.383\;+0.482}$ & $1061.1$ \\ \\
        \hline
    \end{tabular}}
    \label{tab:bestfits}
\end{table}
We use the jerk to determine the type of dark energy in the fractional cosmological model. Its formula is based on the value of $q$ in \eqref{DecelerationFinal} using equation $j(\tau)=q(\tau)(2q(\tau)+1)-\frac{dq(\tau)}{d\tau} $. Entering $\alpha(\tau)$, we obtain the following:
\begin{align}
    &j(\alpha(\tau))= \frac{12 (\mu -4)}{\alpha (\tau)}+\frac{(\mu -21) \mu +50}{\alpha
   (\tau)^2}-\frac{2 (\mu -3) (\mu -2) (\mu -1)}{\alpha (\tau)^3}+10. \label{Jerk}
\end{align}

If $j$ deviates from unity, this may suggest a different cosmology with a dynamical equation of state during the late times.

The matter density parameter in the fractional cosmological model has significant uncertainties due to the absence of the matter equation of the state parameter in the Hubble parameter used for the reconstruction. The current value of this parameter is $\Omega_{m,0}=0.531_{-0.260}^{+0.195}$ at $1\sigma$ CL, which aligns with the asymptotic value $\Omega_{m,t\to\infty}=0.519_{-0.262}^{+0.199}$ determined at the same confidence level through joint analysis.
 
The fractional cosmological model suggests that a larger $\Omega_{m,0}$ could explain the smaller deceleration parameter $q_{0}$ and the excess matter at $\rho_{\text{frac}}=3(\mu-1)t^{-1}H$ with $\Omega_{\text{frac}}(\alpha(\tau))=(\mu-1)/\alpha(\tau)$. The current value of $\Omega_{\text{frac},0}$ is $0.469$, which satisfies $\Omega_{m,0}+\Omega_{\text{frac},0}=1$ and potentially alleviates the problem of coincidence.
On the other hand, these best-fit values lead to an age of the universe with a value of $t_0=\alpha_0/H_0=25.62_{-4.46}^{+6.89}\;\text{Gyrs}$ \citep{Gonzalez:2023who}.

In this case, the expressions \eqref{NewFried} and \eqref{NewP} are used to calculate $\rho(t)$ and $p(t)$. Substituting all expressions in system \eqref{NewRaych}--\eqref{NewFried} leads to identities.
There is an arbitrary constant of integration, and the equations are satisfied identically (no compatibility equations are required). Therefore, this is the general solution of the system. This result is generic since it does not require specifying the equation of the state parameter of matter. Therefore, Equation \eqref{solution} provides a family of one-parameter solutions that are complete and independent of matter content.

{The results in Table \ref{tab:bestfits} are for the minimally coupled case ($\xi=0$). The test against cosmological data of the non-minimally coupled case has not yet been performed. The analysis is complicated and out of the scope of the present research. }

 \subsubsection{Dynamical Systems Analysis}

This section presents new findings and an alternative approach to the dynamical system previously developed in \citep{Garcia-Aspeitia:2022uxz, Micolta-Riascos:2023mqo, Gonzalez:2023who}. Through analyzing dynamical systems, we can identify the model's asymptotic states and explore the phase space for different fractional derivative orders and matter models \cite{TWE, wainwrightellis1997, perko, Coley:2003mj, Smale, wiggins2006introduction}. This allows us to classify the equilibrium points and determine the fractional derivative order to produce a power-law-accelerated late-time solution for the scale factor.
 
With this objective in mind, we introduce the new variables, as follows: 
\begin{align}
   x_1= \frac{t(\tau ) \phi '(\tau )}{\sqrt{6} \alpha (\tau )},\quad x_2= t(\tau ) \phi (\tau ),\quad x_3= \alpha (\tau ),
\end{align}
that satisfy\vspace{-9pt}
\begin{equation}
\rho+ \Lambda= -\frac{3 x_2^2 x_3 \left(-\mu \xi +\xi +x_3 \left(\xi +x_1^2+2 \sqrt{6} \xi x_1\right)\right)}{t(\tau
   )^4}+\frac{3 x_3 (-\mu +x_3+1)}{t(\tau )^2},
\end{equation}
and define the new time variable, as follows:
\begin{equation}
    f^{\prime} \equiv \alpha^2 \frac{d f}{d \tau},
\end{equation}
and, thus, we obtain the dynamical system, as follows:
\begin{align}
  & x_1^{\prime} = -2 \mu ^2 x_1+\mu x_1 (x_2+4 x_3+6)-x_1 (3 x_2 x_3+x_2+2 (8-3 x_3) x_3+4), \label{3syst1}\\
  & x_2^{\prime} = x_3 \left(\sqrt{6} x_1 x_3^2+x_2\right),\label{3syst2}\\
  & x_3^{\prime} = x_3 \left(\mu ^2-3 \mu -2 \mu x_3-3 (x_3-3) x_3+2\right). \label{3syst3}
\end{align}

The equilibrium points of \eqref{3syst1}--\eqref{3syst3} are as follows:
\begin{enumerate}
    \item $P_1: (x_1, x_2, x_3) := \left(0, x_{2c}, 0\right)$. The eigenvalues are \newline  $\{0, (\mu -1) (-2 \mu +x_{2c}+4), (\mu -2) (\mu -1)\}$. It is normally hyperbolic with a stable 2D manifold for $1<\mu <2, x_{2c}<2 \mu -4$, an unstable 2D manifold for $\mu <1, x_{2c}<2 \mu -4$ , or $\mu >2, x_{2c}>2 \mu -4$. This point is a saddle for $\mu <1, x_{2c}>2 \mu -4$, or $1<\mu <2, x_{2c}>2 \mu -4$, or $\mu >2, x_{ 2c}<2 \mu -4$.
    \item $P_2: (x_1, x_2, x_3) := \left(x_{1c}, 2 (\mu -2), 0\right)$.    
    The eigenvalues are  $\{0, 0, (-2 + \mu) (-1 + \mu)\}$, indicating that the system is non-hyperbolic.
     
    \item $P_3: (x_1, x_2, x_3) := \left(0, 0, \frac{1}{6} \left(-2\mu -r+9\right)\right)$, with $ r = \sqrt{8 \mu (2 \mu -9)+105}$. The eigenvalues are $\Big\{\frac{1}{3} (-2 \mu -r+9),  \frac{1}{6} (-2 \mu -r+9), \frac {1}{6} (-2 \mu (8 \mu +r-36)+9 r-105)\Big\}.$ The conditions are as follows: 
   \begin{enumerate}
       \item Source for $1 < \mu < 2$.
       \item Sink for $\mu < 1$ or $\mu > 2$.
   \end{enumerate}
   \item 
  \begin{small}
   $P_4: (x_1, x_2, x_3) := \left(\frac{\mu \left(-12 \mu + 3r +65\right)-2 \left(5r + 51\right)}{4 \sqrt {6} (\mu -2) (\mu -1)^2}, \frac{1}{12} \left(4 \mu -r-3\right), \frac{1}{6} \left(-2 \mu -r+9\right)\right).$
   \end{small}
   \newline
   The eigenvalues are $\Big\{\frac{1}{6} (-2 \mu (8 \mu +r-36)+9 r-105), \newline \frac{1}{12} \left( -2 \mu -3 \sqrt{2} \sqrt{2 \mu (5 \mu -27)+(2 \mu -9) r+93}-r+9\right), \newline \frac{1 }{12} \left(-2 \mu +3 \sqrt{2} \sqrt{2 \mu (5 \mu -27)+(2 \mu -9) r+93}-r+9\right) \Big\}.$
   This point is a saddle for all $\mu \neq 1$ and $\mu \neq 2$.
   
   \item $P_5: (x_1, x_2, x_3) := \left(0, 0, \frac{1}{6} \left(-2 \mu +r+9\right)\right).$
   The eigenvalues  are \newline $\Big\{\frac{1}{6} (-2 \mu +r+9), \frac{1}{3} (-2 \mu +r+9),\frac{1 }{6} (2 \mu (-8 \mu +r+36)-3 (3 r+35))\Big\}$. This point is a saddle for all $\mu \in \mathbb{R}$.
   
   \item
   \begin{small}
   $P_6: (x_1, x_2, x_3) := \left(\frac{\mu \left(-12 \mu -3r +65\right)+2 \left(5r-51\right)}{4 \sqrt {6} (\mu -2) (\mu -1)^2}, \frac{1}{12} \left(4 \mu +r-3\right), \frac{1}{6} \left(-2 \mu +r +9\right)\right)$.
   \end{small}
   \newline The eigenvalues are $\Big\{\frac{1}{6} (2 \mu (-8 \mu +r+36)-3 (3 r+35)), \newline \frac{1}{ 12} \left(-2 \mu -3 \sqrt{20 \mu ^2-4 \mu (r+27)+6 (3 r+31)}+r+9\right), \newline \frac{ 1}{12} \left(-2 \mu +3 \sqrt{20 \mu ^2-4 \mu (r+27)+6 (3 r+31)}+r+9\right)\Big\}$.
   
   This point is a saddle for all $\mu \neq 1$ and $\mu \neq 2$.
\end{enumerate}

\subsection{Non-Minimal Coupling}

In this section, we present the new results of the thesis for the most general case $\xi\neq 0$ corresponding to non-minimal coupling. For this analysis, we will use the age parameter given by $\alpha = t H$, and we will use the rules
\begin{equation}
    \frac{d}{d t}= H \frac{d}{d\tau }, \quad \frac{d^2}{d t^2}= H^2 \left(\frac{d^2}{ d\tau ^2} -(1+ q) \frac{d}{d\tau }\right), \quad q:= -1-\frac{\dot{H}}{H^2},
\end{equation} to introduce a new derivative.

Thus, systems \eqref{NRay} and \eqref{NKG} are as follows: \vspace{-10pt}

 \begin{align}
\alpha '(\tau ) & =\frac{(\mu -2) (\mu -1) \left(\xi  \phi ^2-1\right)}{\alpha  \left(\xi  (12 \xi +1) \phi^2-1\right)}+\frac{\alpha  \left(3-3 \xi  (8 \xi +1) \phi ^2\right)}{\xi  (12 \xi +1) \phi ^2-1}  +\frac{\xi  (-2 \mu +12 \xi +9)
   \phi ^2+2 \mu -9}{\xi  (12 \xi +1) \phi ^2-1} \nonumber \\
   & +\frac{2 \xi  t^2 \phi ^2 {\phi^{\prime}}^2}{\alpha  \left(\xi  (12 \xi +1) \phi ^2-1\right)}   +{\phi^{\prime}} \left(-\frac{2 (\mu -1) \xi  t \phi ^2}{\alpha  \left(\xi  (12 \xi +1) \phi ^2-1\right)}-\frac{2
   \xi  t \phi ^2}{\xi  (12 \xi +1) \phi ^2-1}\right), \label{2NRay} 
   \end{align}
   \begin{align}
{\phi^{\prime \prime}}(\tau ) &=\frac{12 (\mu -4) \xi  \phi  \left(\xi  \phi ^2-1\right)}{\alpha  \left(\xi  (12 \xi +1) \phi ^2-1\right)}-\frac{6 (\mu -2) (\mu -1) \xi  \phi  \left(\xi  \phi ^2-1\right)}{\alpha ^2 \left(\xi  (12 \xi +1) \phi ^2-1\right)}+\frac{6 \xi  \phi  \left(\xi  \phi ^2-1\right)}{\xi  (12
   \xi +1) \phi ^2-1}  \nonumber \\
   & +{\phi^{\prime}}^2 \left(\frac{\frac{2 (\mu -1) \xi  t \phi ^2}{\xi  (12 \xi +1) \phi ^2-1}-\frac{12 \xi ^2 t^2
   \phi ^3}{\xi  (12 \xi +1) \phi ^2-1}}{\alpha ^2}+\frac{2 \xi  t \phi ^2}{\alpha  \left(\xi  (12 \xi +1) \phi^2-1\right)}\right)  -\frac{2 \xi  t^2 \phi ^2 {\phi^{\prime}}^3}{\alpha ^2 \left(\xi  (12 \xi +1) \phi ^2-1\right)} \nonumber \\
   & +{\phi^{\prime}}
   \Bigg(\frac{3 \left(\xi  (8 \xi +1) \phi ^2-1\right)}{\xi  (12 \xi +1) \phi ^2-1}+\frac{\frac{(\mu -1) t \phi  \left(\xi  (24 \xi
   +1) \phi ^2-1\right)}{\xi  (12 \xi +1) \phi ^2-1}-\frac{(\mu -2) (\mu -1) \left(\xi  \phi ^2-1\right)}{\xi  (12 \xi +1) \phi^2-1}}{\alpha ^2}   \nonumber \\
   & +\frac{\frac{2 (\mu -4) \left(\xi  \phi ^2-1\right)}{\xi  (12 \xi +1) \phi ^2-1}+\frac{t \left(3 \phi -3
   \xi  (8 \xi +1) \phi ^3\right)}{\xi  (12 \xi +1) \phi ^2-1}}{\alpha }\Bigg), \label{2eq(135)}
\\
t'(\tau)& ={t(\tau)}/{\alpha(\tau)}, \\
\alpha(0) & =t_0 H_0, \quad \phi(0)= \phi_0, \quad {\phi}^{\prime}(0)={{\phi}^{\prime}}_0, \quad t(0)=t_0, \label{2eq(136)}
 \end{align}

and for the initial conditions for the scalar field and the speed of the scalar field, we consider a model with dust-like matter ($p=0$), and we evaluate at $\tau=0$ (today) the expressions \eqref{NewFried} and \eqref{NewP} to obtain the following:
 
\begin{small}
\begin{align}
   & -\frac{\phi_0^2 {{\phi}^{\prime}_0}^2}{6 H_0^2}-\frac{\mu +\xi  \phi_0^2 \left(-\mu +2 t_0 {\phi}^{\prime}_0+1\right)-1}{H_0 t_0} -\xi 
   \phi_0^2  -\Omega_{\Lambda 0}-\Omega_{m0}+1=0,
\\
   & -6 H_0^2 t_0^2 \Omega_{\Lambda 0} \left(\xi  (12 \xi +1) \phi_0^2-1\right)+6 H_0^2 t_0^2 \left(\xi  \phi_0^2-1\right)^2 \nonumber \\
   & +12 H_0 t_0 \left(\xi  \phi_0^2-1\right) \left(\mu  \left(\xi 
   \phi_0^2-1\right)+\xi  \phi_0^2 \left(12 \xi +t_0 {\phi}^{\prime}_0-3\right)+3\right) -6 (\mu -2) (\mu -1) \left(\xi  \phi_0^2-1\right)^2 \nonumber \\
   & +t_0^2 \phi_0^2 \left(\xi  \left(\phi_0^2+12\right)-1\right) {{\phi}^{\prime}_0}^2  +12 (\mu -1) \xi  t_0 \phi_0^2 \left(\xi \phi_0^2-1\right) {\phi}^{\prime}_0=0. 
\end{align}
\end{small}

\subsubsection{Dynamical systems analysis}

In \citep{Rami:2015kha}, the assumption $H=H(\phi, \dot \phi)= \varepsilon \dot{\phi}/\phi$ was explored, corresponding to $\phi'(\tau)/$\linebreak$\phi(\tau)= \varepsilon$, and $\alpha= t \varepsilon \dot{\phi}/\phi$.

In this paper, we consider a more general case where the new variables are introduced as follows: 
\begin{align}
   x_1= \frac{t(\tau ) \phi '(\tau )}{\sqrt{6} \alpha (\tau )},\quad x_2= t(\tau ) \phi (\tau ),\quad  x_3= \alpha (\tau ), \quad x_4= \frac{1}{t(\tau )^2 \left(\xi  (12 \xi +1) \phi (\tau )^2-1\right)},
\end{align}
that satisfy \vspace{-8pt}
\begin{equation}
\rho+    \Lambda= -\frac{3 x_3 x_4 \left(-\mu +x_2^2 x_4 \left(x_1^2 x_3+2 \sqrt{6} \xi  x_1 x_3-12 \xi ^2 (-\mu
   +x_3+1)\right)+x_3+1\right)}{\left(\xi  (12 \xi +1) x_2^2 x_4-1\right)^2},
\end{equation}
and the new time, as follows: 
\begin{equation}
    f^{\prime} \equiv \alpha^3 \frac{d f}{d \tau}, 
\end{equation}
which preserves the arrow of time ($\alpha\geq 0$), obtaining the dynamical system, as follows: \vspace{-10pt}
 
\begin{align}
  & x_1^{\prime}= -24 \xi  x_1^3 x_2^2 x_3^3 x_4+4 \sqrt{6} \xi  x_1^2 x_2^2 x_3^2 x_4 (\mu -3 \xi  x_2+x_3-1) \nonumber \\
   & +x_1 x_3 \Big[-2 \mu ^2+6 \mu +12 \xi ^2 x_2^2 x_4 \left(2 \mu ^2-6 \mu +x_2 (\mu
   +x_3-1)-4 \mu  x_3-2 (x_3-8) x_3+4\right) \nonumber \\
   & +\mu  x_2-3 x_2 x_3-x_2+6 x_3^2+4 \mu  x_3-16 x_3-4\Big] \nonumber \\
   & -\sqrt{6} \xi x_2 \left(-(\mu -2) (\mu -1)+x_3^2+2 (\mu -4) x_3\right) \left(12 \xi ^2 x_2^2 x_4-1\right), \label{EQ1}
\\
  & x_2^{\prime}=  x_3^2 \left(\sqrt{6} x_1
   x_3^2+x_2\right), \label{EQ2}
\\
  & x_3^{\prime}= x_3^2 \Big[\mu ^2-3 \mu +2 \xi  x_1 x_2^2 x_3 x_4 \left(6 x_1 x_3-\sqrt{6} (\mu +x_3-1)\right)\nonumber \\ 
  & +12 \xi ^2 x_2^2 x_4 \left(-(\mu -2) (\mu -1)+x_3^2+2 (\mu -4) x_3\right)-3 x_3^2-2 \mu  x_3+9 x_3+2\Big], \label{EQ3}
\\
  & x_4^{\prime}= -2 x_3^2 x_4 \left(\sqrt{6} \xi  (12 \xi +1) x_1 x_2 x_3^2 x_4+1\right). \label{EQ4}
\end{align}
 
\begin{figure}[h]
 \centering
    \includegraphics[scale=0.8]{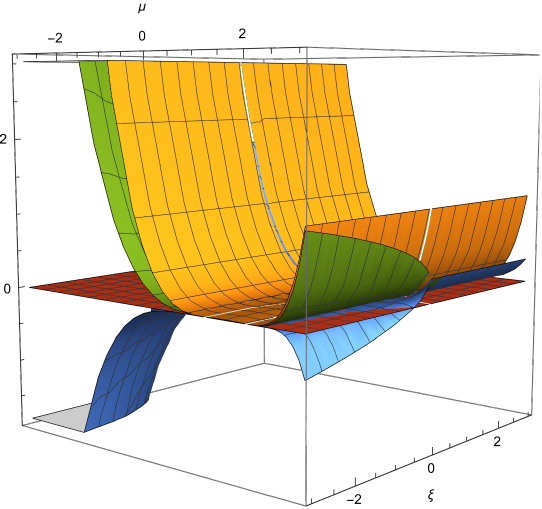}
    \includegraphics[scale=0.83]{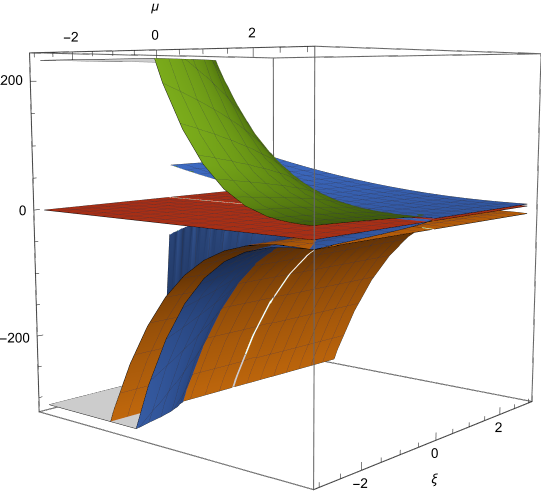}
    \caption{Real parts of the eigenvalues of $Q_2$ ({\bf left}) and $Q_3$ ({\bf right}). $Q_2$ can exhibit source or saddle behavior and $Q_3$ is a saddle. For reference, the red plane corresponds to the value zero, and the other colors denote the three real parts of the eigenvalues.}
    \label{stability2}
\end{figure}

\begin{figure}[h]
 \centering
    \includegraphics[scale=0.8]{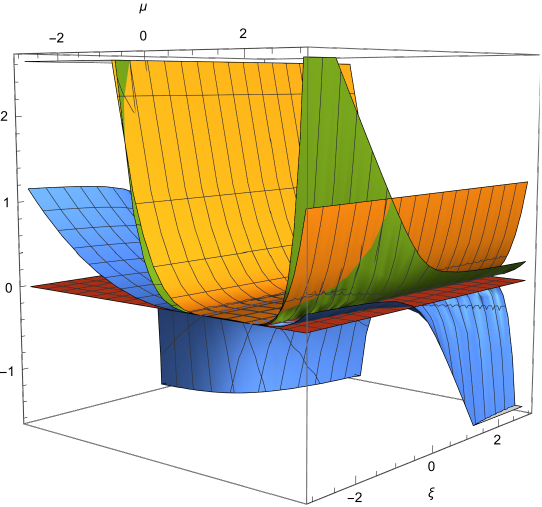}
    \includegraphics[scale=0.83]{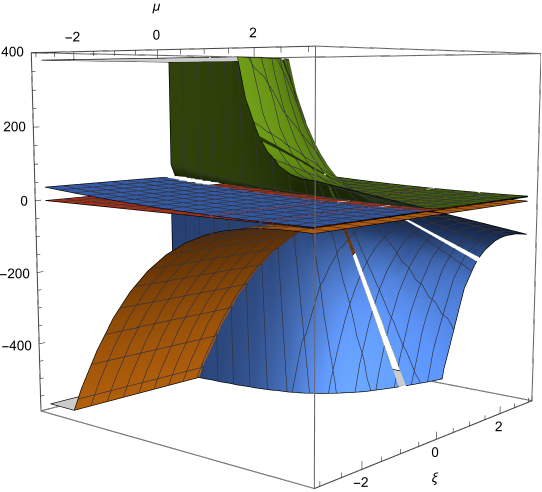}
    \caption{Real parts of the eigenvalues of $Q_4$ ({\bf left}) and $Q_5$ ({\bf right}). $Q_4$ can exhibit either source or saddle behavior and $Q_5$ is a saddle. For reference, the red plane corresponds to the value zero, and the other colors denote the three real parts of the eigenvalues.}
    \label{stability4}
\end{figure}

\begin{enumerate}
    \item 
The set $x_3=0$ is an invariant set of \eqref{EQ1}--\eqref{EQ4} when
\begin{equation}
    \sqrt{6} (\mu -2) (\mu -1) \xi  x_2 \left(12 \xi ^2 x_2^2 x_4-1\right)=0.
\end{equation}
That is, when
\begin{enumerate}
    \item $\mu\in \{1,2\}$ o  
    \item $\xi=0$ o  
    \item $x_2=0$ o 
    \item $x_4 x_2^2= \frac{1}{12 \xi ^2 }$.
\end{enumerate}

\item Taking $x_4=0$, we obtain the dynamical system, as follows:
 \begin{align}
  & x_1^{\prime}=   x_1 x_3 (6 \mu +x_2 (\mu -3 x_3-1)+2 (3 x_3-\mu ) (\mu
   +x_3)-16 x_3-4) \nonumber \\
   & +\sqrt{6} \xi  x_2 \left(-(\mu -2) (\mu -1)+x_3^2+2 (\mu -4) x_3\right), \label{XT1}\\
  & x_2^{\prime}=x_3^2 \left(\sqrt{6} x_1  x_3^2+x_2\right), \label{XT2}\\
  & x_3^{\prime}= x_3^2 \left(\mu ^2-3 \mu -2 \mu  x_3-3 (x_3-3) x_3+2\right), \label{XT3}
 \end{align}
Systems \eqref{XT1}--\eqref{XT3} support the following equilibrium points:
\begin{enumerate}
    \item $Q_1 : (x_1,x_2,x_3) := (x_{1c},0,0)$, which is a non-hyperbolic critical point curve.
    \item $Q_2 : (x_1,x_2,x_3) := \Big(0,0,\frac{1}{6} \left(-r -2 \mu +9\right)\Big)$, where $ r=\sqrt{8 \mu (2 \mu -9)+105}$ with eigenvalues $\{\frac{f_1(\mu,\xi)}{36},\frac{f_2(\mu,\xi)}{108},\frac{f_3(\mu,\xi)}{108}\}$, where $f_i$ are complicated expressions that depend on $\mu$ and $\xi$. Since $x_3\geq0$, this point only exists for $1\leq \mu \leq 2.$ In Figure \ref{stability2}, the point can show source or saddle behavior.
    \item \textls[-25]{$Q_3 : (x_1,x_2,x_3) := \Big(0,0,\frac{1}{6} \left(r -2 \mu +9\right)\Big)$ with eigenvalues $ \{\frac{g_1(\mu,\xi)}{36},\frac{g_2(\mu,\xi)}{108},\frac{g_3(\mu,\xi)}{108}\}$, where $g_i$ are complicated expressions that depend on $\mu$ and $\xi$. In Figure \ref{stability2}, the point is a saddle.}
    \item {$Q_4 :$} 
 $ (x_1,x_2,x_3) := \Big(\frac{\xi \left((\mu -2) (\mu -1) (2 \mu -5) r +(2-\mu ) (\mu -1)
   \left(8 \mu ^2-26 \mu +39\right)\right)}{4 \sqrt{6} (\mu -2) (\mu -1)^2} \newline 
   +\frac{3
   \mu r -10 r -12 \mu ^2+65 \mu -102}{4 \sqrt{6} (\mu -2) (\mu
   -1)^2},\newline\frac{1}{36} \xi \left((-\mu (2 \mu -15)-57) r +\mu \left(8 \mu
   ^2+30 \mu -279\right)+549\right)+\frac{1}{36} \left(-3 r +12 \mu
   -9\right), \newline \frac{1}{6} \left(-r -2 \mu +9\right)\Big)$, with eigenvalues $\{\frac{h_1(\mu,\xi)} {36},\frac{h_2(\mu,\xi)}{108 (\mu -2) (\mu -1)^2},\frac{h_3(\mu,\xi)}{108 (\mu -2) (\mu -1)^2}\}$, \newline where, again, $h_i$ are complicated expressions that depend on $\mu$ and $\xi$. Since $x_3\geq0$, this point only exists for $1\leq \mu \leq 2.$ In Figure \ref{stability4}, the point can show source or saddle behavior.
   \item {$Q_5 :$}
 $ (x_1,x_2,x_3) := \Big(\frac{\xi \left(-\mu (\mu (2 \mu -11)+19) r +10 r +(2-\mu)
   (\mu -1) \left(8 \mu ^2-26 \mu +39\right)\right)}{4 \sqrt{6} (\mu -2) (\mu
   -1)^2} \newline 
   +\frac{-3 \mu r +10 r -12 \mu ^2+65 \mu -102}{4 \sqrt{6} (\mu
   -2) (\mu -1)^2},\newline\frac{1}{36} \xi \left((\mu (2 \mu -15)+57) r +\mu
   \left(8 \mu ^2+30 \mu -279\right)+549\right)+\frac{1}{36} \left(3 r +12
   \mu -9\right), \newline \frac{1}{6} \left(r -2 \mu +9\right)\Big)$, with eigenvalues  $\{\frac{l_1(\mu,\xi) }{36},\frac{l_2(\mu,\xi)}{108 (\mu -2) (\mu -1)^2},\frac{l_3(\mu,\xi)}{108 ( \mu -2) (\mu -1)^2}\}$, where $l_i$ are complicated expressions that depend on $\mu$ and $\xi$. In Figure \ref{stability4}, the point is a saddle.
\end{enumerate}
\end{enumerate}
\vspace{-10pt}
\vspace{-10pt}

\begin{figure}[h]
 \centering
    \includegraphics[scale=0.8]{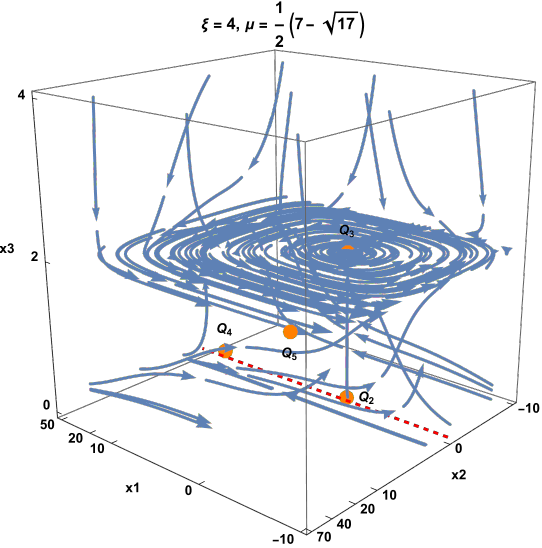}
    \includegraphics[scale=0.8]{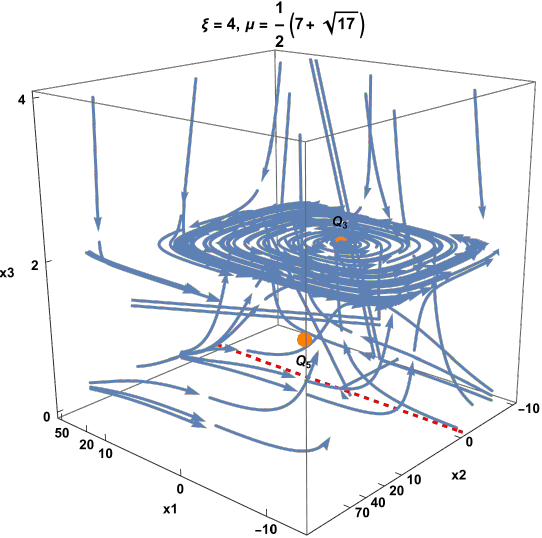}
    \caption{Phase space for systems \eqref{XT1}--\eqref{XT3} for $\xi=4$ and $\mu=\frac{1}{2}(7\pm \sqrt{17 }).$ The red line denotes the non-hyperbolic set of points, $Q_1.$ Since $x_3\geq0$, only two equilibrium points appear in the right-hand graph.
}
    \label{estabilidadOrbita}
\end{figure}

In Figure \ref{estabilidadOrbita} we present some phase space orbits for systems \eqref{XT1}--\eqref{XT3} for $\xi=4$ and $\mu=\frac{1}{2}(7\pm \sqrt{17 }).$ The red line denotes the non-hyperbolic set of points, $Q_1.$ Since $x_3\geq0$, only two equilibrium points appear in the right-hand graph.

\begin{figure}[h]
    \centering 
    \includegraphics[scale=0.8]{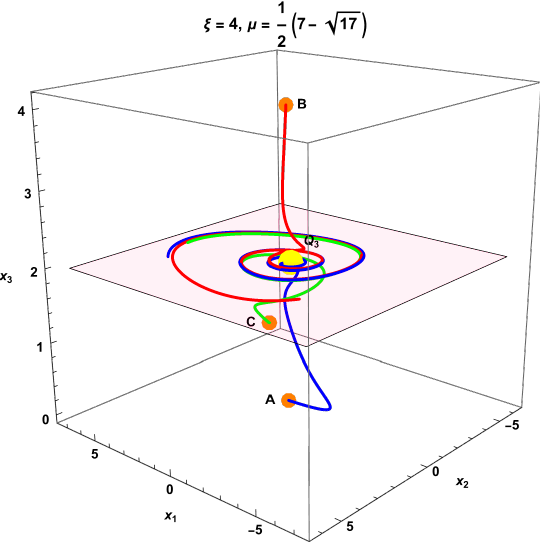}
    \includegraphics[scale=0.8]{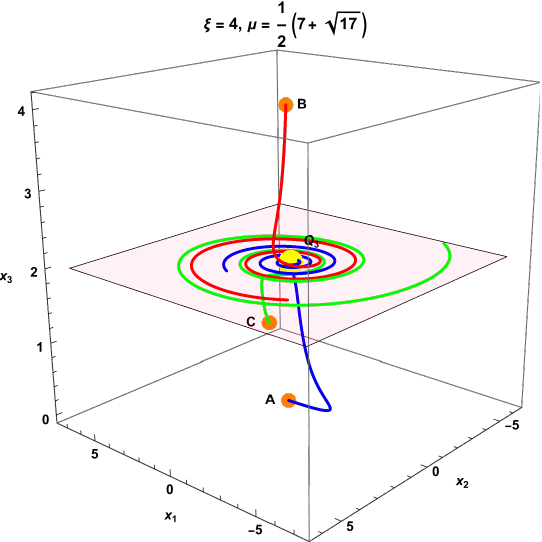}
    \caption{Projection of the flow of \eqref{XT1}--\eqref{XT3} in the plane $x_3=2$ for $\mu=\frac{1}{2}(7\pm \sqrt{ 17})$ and $\xi=4.$ the yellow sphere represents the projection of the point $Q_3=(0,0,\frac{1}{6}(r-2\mu+9))$, which lies on the plane $(x_1,x_2)=(0,0)$. As depicted in the figure, $Q_3$ is an unstable focus. The spiral is constructed from the initial conditions $A=(0.1,0.1)$, $B=(0.1,0.1,4)$, and $C=(1.9,-0.5,1)$.}
    \label{estabilidadOrbita-2d}
\end{figure}

The Figure \ref{estabilidadOrbita-2d} displays the projection of the flow of equations \eqref{XT1}--\eqref{XT3} for $\xi=4$ and $\mu=\frac{1}{2}(7\pm \sqrt{17})$ on the plane $x_3=2$, highlighted in pink. In the plot, the yellow sphere represents the projection of the point $Q_3=(0,0,\frac{1}{6}(r-2\mu+9))$, which lies on the plane $(x_1,x_2)=(0,0)$. As depicted in the figure, $Q_3$ is an unstable focus. The spiral is constructed from the initial conditions $A=(0.1,0.1)$, $B=(0.1,0.1,4)$, and $C=(1.9,-0.5,1)$.

    \vspace{-12pt}

\subsubsection{Alternative Formulation of the Dynamical System} 

\label{SECTAlternative_Formulation}

We define the following variables: 
\begin{equation}
    x= \frac{\dot{\phi}(t)}{\sqrt{6} H(t)}, \quad y=\phi (t), \quad \Omega_\Lambda= \frac{\Lambda } {3H^2(t)},
    \label{altform}
\end{equation}
that satisfy\vspace{-9pt}
\begin{equation}
(x + \sqrt{6} \xi y)^2 + \xi y^2(1-6 \xi)+ \frac{(\mu -1) \left(1-\xi y^2\right) }{\alpha } +\Omega_\Lambda + \Omega_{m}=1,
\end{equation}
where
\begin{equation}
    \Omega_m = \frac{\rho_m}{3 H^2},
\end{equation}
which is the dimensionless energy density of matter, and $\Omega_\Lambda$ is interpreted as the energy density of the cosmological constant.
The equation of state parameter of matter can be expressed in terms of these variables as follows: \vspace{-10pt}

\begin{align}\label{dobleuve}
 w & =  \left\{\alpha  \left(\xi  (12 \xi +1) y^2-1\right) \left(-\alpha +\mu +\alpha  x^2+\xi  y
   \left(2 \sqrt{6} \alpha  x+y (\alpha -\mu +1)\right)+\alpha  \Omega_\Lambda-1\right)\right\}^{-1} \nonumber \\
   & \times \Bigg\{\alpha ^2 \left[x^2 \left(\xi  \left(y^2+12\right)-1\right)+2 \sqrt{6} \xi  x y \left(\xi  y^2-1\right)+\xi  y^2 \left(-(12 \xi +1) \Omega_\Lambda+\xi 
   y^2-2\right)+\Omega_\Lambda+1\right] \nonumber \\
   & +2 \alpha  \left(\xi  y^2-1\right) \left[\mu  \left(\xi  y \left(\sqrt{6} x+y\right)-1\right)+\xi  y \left(-\sqrt{6} x+12 \xi 
   y-3 y\right)+3\right]  \nonumber\\
   & -(\mu -2) (\mu -1) \left(\xi  y^2-1\right)^2\Bigg\}. 
\end{align}

Introducing the new time derivative, i.e., 
\begin{equation}
    \frac{d f}{d \tau} \equiv \frac{1}{H} \frac{d f}{d t}, \label{log-time}
\end{equation}
we obtain the four-dimensional dynamical system, as follows: \vspace{-12pt}
 
\begin{align}\label{eqinit}
 \alpha^2 G   \frac{d x}{d \tau} & = 
  -12 \alpha ^2 \xi  x^3+2 \sqrt{6} \alpha  \xi  x^2 y (-6 \alpha  \xi +\alpha +\mu -1) \nonumber \\
   &   +x \left(3 \alpha  \left(-\mu +8 (\mu -1) \xi ^2
   y^2+(\mu -3) \xi  y^2+3\right)-(\mu -2) (\mu -1) \left(\xi  y^2-1\right)\right) \nonumber \\ 
   &  +\sqrt{6} \xi  y \left(\alpha ^2+2 \alpha  (\mu -4)-(\mu -2) (\mu
   -1)\right) \left(\xi  y^2-1\right),
\\
    \frac{d y}{d \tau} & =\sqrt{6}  x, 
\\
 \alpha^2 G   \frac{d \Omega_\Lambda}{d \tau} & =2 \Omega_\Lambda  \Bigg\{\alpha ^2 \left(-12 \xi  x^2+2 \sqrt{6}
   \xi  x y+3 \xi  (8 \xi +1) y^2-3\right) \nonumber \\ 
   &  \quad\quad\quad\quad  +2 \alpha  \left(\sqrt{6} (\mu -1) \xi  x y+(\mu -4) \left(\xi  y^2-1\right)\right)  -(\mu -2) (\mu -1) \left(\xi 
   y^2-1\right)\Bigg\},
\\
 \label{eqfin}\alpha  G  \frac{d \alpha}{d \tau} & =   3 \alpha ^2+2 \alpha  \mu -9 \alpha -\mu ^2+3 \mu   +\xi  \Big(12 \alpha ^2 x^2-2
   \sqrt{6} \alpha  x y (\alpha +\mu -1) \nonumber \\
   & \quad \quad      +y^2 \left(-2 \alpha  \mu -3 (\alpha -3) \alpha +\mu ^2-3 \mu +2\right)\Big)+12 (1-2 \alpha ) \alpha  \xi ^2
   y^2-2,
\end{align}
 
where
\begin{equation}
    G:= (\xi (12 \xi +1) y^2-1),
\end{equation}
is defined in the phase space, i.e., 
\begin{equation}
(x + \sqrt{6} \xi y)^2 + \xi y^2(1-6 \xi)+ \frac{(\mu -1) \left(1-\xi y^2\right) }{\alpha } +\Omega_\Lambda \leq 1
\end{equation}
which corresponds to $\Omega_{m}\geq 0$.

We introduce temporal rescaling:  \vspace{-9pt}
\begin{align}
f^{\prime}:= \alpha^2 \frac{df}{d \tau}, \label{time-rescaling}
\end{align}
this preserves the arrow of time. Sometimes, we use the time variable $\tau=\ln (a/a_0)$, which is more convenient.

We obtain the equilibrium points with coordinates $(x, y, \Omega_\Lambda, \alpha)$, as follows:
\begin{enumerate}
\item The line of equilibrium points $P_1(y_c)$ with coordinates $\left(-\sqrt{6} \xi y_c, y_c,0,0\right)$ parameterized by $y_c \in \mathbb{R} $. The equation of state parameter is complex infinity.
The eigenvalues are as follows:
\begin{small} 
$\left\{0,-\frac{2 (\mu -2) (\mu -1) \left(\xi y_c^2-1\right)}{\xi (12 \xi +1) y_c^ 2-1},-\frac{(\mu -2) (\mu-1) \left(\xi y_c^2-1\right)}{\xi (12 \xi +1) y_c^2-1 },\frac{(\mu -2) (\mu -1) \left(\xi y_c^2-1\right)}{\xi (12 \xi +1) y_c^2-1}\right\}$.
\end{small}
 The curve is a saddle because it has at least two eigenvalues with different signs.
 
\item The equilibrium point, $P_2: \left(0,0,0,\frac{1}{6} (9- 2\mu -r)\right)$. The equation of state parameter is $w=\frac{r+7}{4-4 \mu }$. The eigenvalues are as follows:
  \newline
\textls[-25]{$\Big\{\frac{1}{3} (9- 2\mu -r),\frac{1}{6} (-2 \mu (8 \mu +r-36)+9 r-105 ), \frac{1}{36} \left(9 (21-4 \mu ) \mu -9 (\mu -3) r-279 - s_{-}\right)$,\newline $\frac{ 1}{36} \left(9 (21-4 \mu ) \mu -9 (\mu -3) r-279 + s_{-}\right)\Big\}$}

where $r=\sqrt{8 \mu(2 \mu -9)+105}$, and $s_{-}$ is the positive square root of
    $s_{-}^2=81 (\mu (4 \mu -21)+(\mu -3) r+31)^2-48 \xi (\mu (\mu (2 \mu (82 \mu -933)+8355)-$17,361$)+(\mu (5 \mu (8 \mu
   -75)+1209)-1413) r+$14,481$)$.
 This is  a saddle for the following: 
\begin{small}
\textls[-15]{{$\Psi:=$} 
 $\left\{\mu \leq 0.463848,  0<\xi \leq \frac{\mu  \left(32 \mu ^3-8 \mu ^2 (r+24)+6 \mu  (5 r+83)-51 r-567\right)+18 (r+17)}{48 (\mu -2) (\mu -1) (4 (\mu
   -4) \mu +23)}\right\}\cup $\newline $\left\{0.463848<\mu <1,  0<\xi \leq \frac{\mu  (2 \mu  (4 \mu  (4 \mu +r-24)-15 r+249)+51 r-567)-18 (r-17)}{48 (\mu -2) (\mu
   -1) (4 (\mu -4) \mu +23)}\right\}\cup $\newline $\left\{\mu >2,  0<\xi \leq \frac{\mu  (2 \mu  (4 \mu  (4 \mu +r-24)-15 r+249)+51 r-567)-18 (r-17)}{48 (\mu -2)
   (\mu -1) (4 (\mu -4) \mu +23)}\right\}$.}
\end{small}
This is a source for the following: 

   $\frac{1}{48} \left(-\frac{\sqrt{8 \mu  (2 \mu -9)+105} \left(\mu  \left(8 \mu ^2-30 \mu +51\right)-18\right)}{(\mu -2) (\mu -1) (4 (\mu -4) \mu
   +23)}+\frac{38-20 \mu }{4 (\mu -4) \mu +23}+\frac{20}{\mu -2}-\frac{7}{\mu -1}+8\right)\leq \xi <0$.
  
  This case is discarded under the assumption $\xi\geq 0$. Figure \ref{fig:P2} presents the real parts of the eigenvalues associated with $P_2$. This shows that the point is well for $(\mu,\xi)\in \Psi$, or a saddle (assuming $\xi\geq 0$).
     \begin{figure}[h]
       \centering
       \includegraphics[scale=0.4]{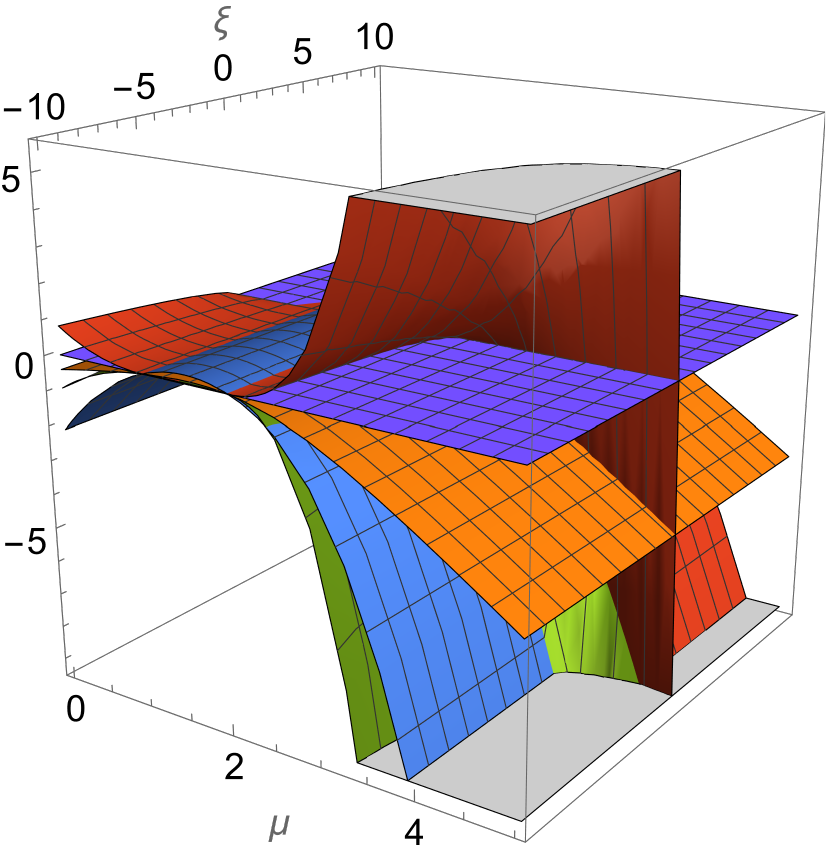}
       \caption{Real parts of the eigenvalues associated with $P_2$. This shows that the point is a sink if $(\mu,\xi)\in \Psi$, or a saddle (assuming $\xi\geq 0$). For reference, the magenta-colored plane corresponds to the value zero, and the other colors denote the real parts of the four eigenvalues.}
       \label{fig:P2}
   \end{figure}
   
\item The equilibrium point $P_3: \left(0,0,0,\frac{1}{6} (9-2 \mu +r)\right)$. The equation of state parameter is $w=\frac{r-7}{4 (\mu -1)}$. The eigenvalues are as follows: \newline

\textls[-35]{$\Big\{\frac{1}{3} (9-2 \mu +r), \frac{1}{6} (2 \mu (-8 \mu +r+36)-3 (3 r +35)), \frac{1}{36}
   \left(9 (21-4 \mu ) \mu +9 (\mu -3) r-279 - s_{+}\right)$,\newline $\frac{1}{36} \left(9 ( 21-4 \mu ) \mu +9 (\mu -3) r-279 + s_{+}\right)\Big\}$
 where $r=\sqrt{8 \mu(2 \mu -9)+105}$, and $s_{+}$ is the positive square root of}

$s_{+}^2= 81 ((21-4 \mu )
   \mu +(\mu -3) r-31)^2 + 48 \xi (\mu (\mu (2 (933-82 \mu ) \mu -8355)+17,361)+(\mu (5 \mu (8 \mu -75)+1209)-1413) r-14,481)$.
The point is always saddle, as shown in Figure \ref{fig:P3}.
   \begin{figure}[h]
       \centering
       \includegraphics[scale=0.4]{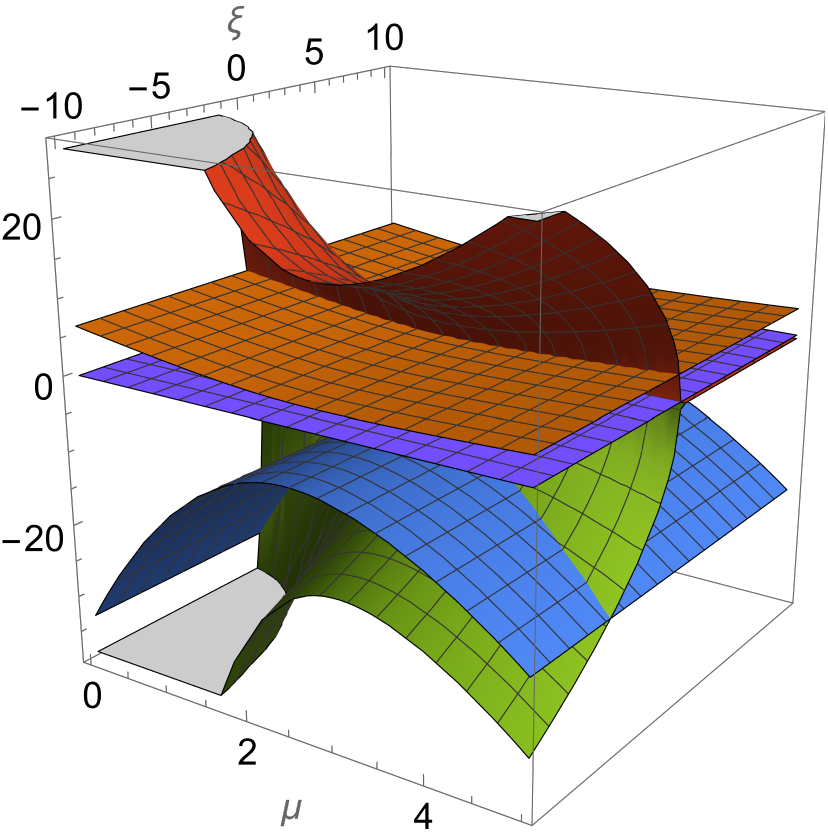}
       \caption{Real parts of the eigenvalues associated with $P_3$. Hence, the point is a saddle. For reference, the magenta-colored plane corresponds to the value zero, and the other colors denote the real parts of the four eigenvalues.}
       \label{fig:P3}
   \end{figure}

\item The equilibrium point $P_4: \left(0,-\frac{1}{\sqrt{\xi }},0,\frac{1}{2}\right)$. The equation of state parameter is indeterminate. The eigenvalues are \newline  $\left\{-\frac{1}{2},1,\frac{1}{4} \left(2 \mu -\sqrt{8 \mu -19}-2\right), \frac{1}{4} \left(2 \mu +\sqrt{8 \mu -19}-2\right)\right\}$. Eigenvalue $1$ corresponds to the coordinate $\Omega_\Lambda$. This point is always a saddle when $\Omega_\Lambda\neq 0$. In the invariant set $\Omega_\Lambda=0$, they can be attractors if $\mu < 1$. For $\mu=1$, the eigenvalues reduce to $-\frac{1}{2}, \pm i \frac{\sqrt{11}}{4}$, and as shown in the lower panel of Figure \ref{fig:enter-label1}, we see spiral attractors. The analysis of this invariant set is presented in Section \ref{invariantset-0-CC}.

\item The equilibrium point $P_5: \left(0,\frac{1}{\sqrt{\xi }},0,\frac{1}{2}\right)$. The equation of state parameter is indeterminate. The eigenvalues are \newline  $\left\{-\frac{1}{2},1,\frac{1}{4} \left(2 \mu -\sqrt{8 \mu -19}-2\right), \frac{1}{4} \left(2 \mu +\sqrt{8 \mu -19}-2\right)\right\}$. Eigenvalue $1$ corresponds to the coordinate $\Omega_\Lambda$. This point is always a saddle when $\Omega_\Lambda\neq 0$. In the invariant set where $\Omega_\Lambda=0$, they can be attractors if $\mu < 1$. For $\mu=1$, the eigenvalues reduce to $-\frac{1}{2}, \pm i \frac{\sqrt{11}}{4}$, and as shown in the lower panel in Figure \ref{fig:enter-label1}, are spiral attractors. The analysis of this invariant set is presented in Section \ref{invariantset-0-CC}.

\begin{figure}[h]
\centering
    \includegraphics[width=0.45\textwidth]{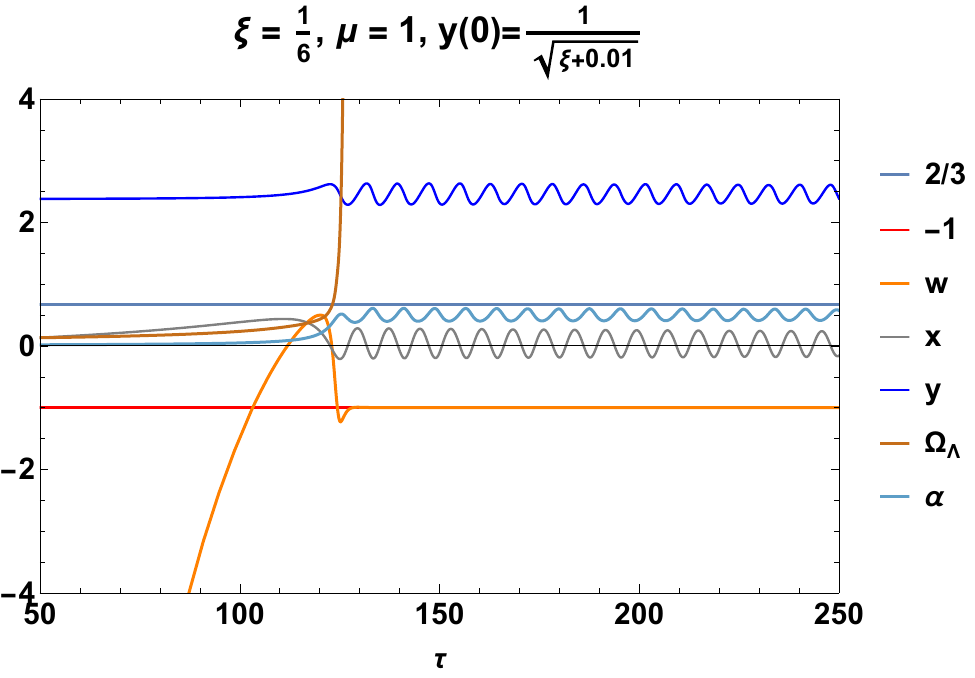}
    \includegraphics[width=0.45\textwidth]{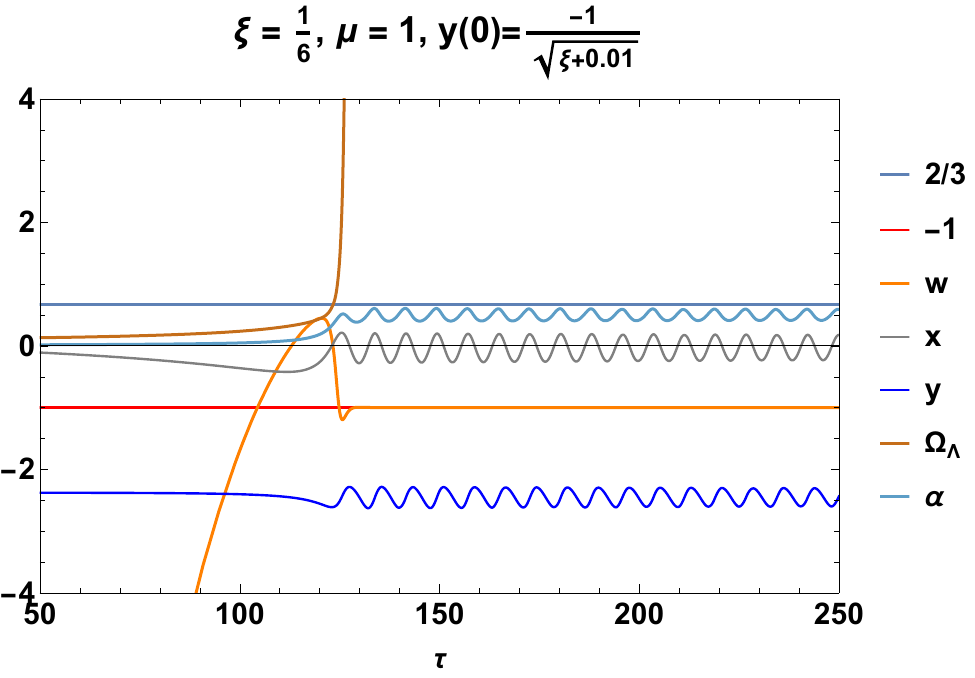}
    \includegraphics[width=0.6\textwidth]{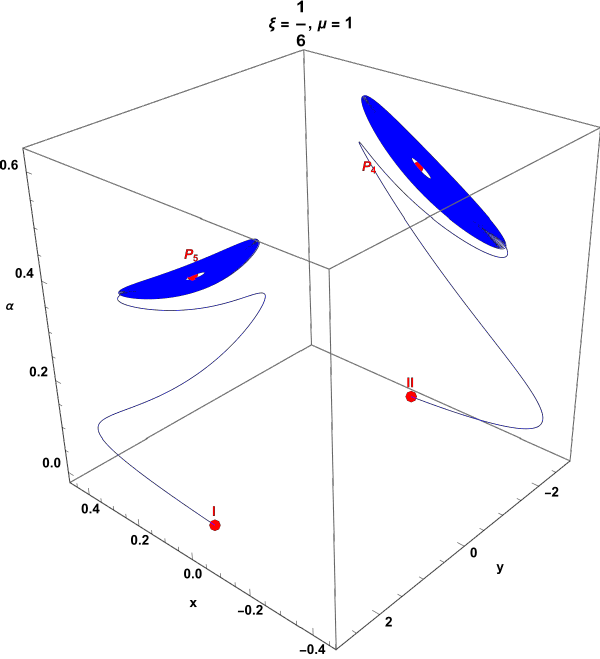}
    \caption{The upper panels show the numerical solutions $x(\tau)$, $y(\tau)$, $\Omega_{\Lambda}(\tau)$, and $\alpha(\tau)$ of systems \eqref{eqinit}--\eqref{eqfin}, as well as the effective equation of the state of matter $w(\tau)$ defined in \eqref{dobleuve} for $\mu=1$.
    Two spiral orbits of the system, starting at the initial conditions labeled $I$ and $II$ in Table \ref{Tab1}, asymptotically tend to the equilibrium points $P_4$ and $P_5$.}
    \label{fig:enter-label1}
\end{figure}
There are additional points that cancel out the numerator and denominator, such as the following:

\item The set $P_6(x_c): \left(x_c, -\frac{1}{\sqrt{\xi }},\Omega_{\Lambda 0}, 0\right)$.

\item The line $P_7(x_c): \left(x_c, -\frac{1}{\sqrt{\xi }}, 0, 0\right)$. In both cases, the stability analysis of these point curves will be left for future research since it cannot be implemented with the techniques developed in this paper.

\item The line $P_8: \left(0, -\frac{1}{\sqrt{\xi }}, \Omega_{\Lambda 0}, 0\right)$. The eigenvalues are \newline  $\left\{-2,4,\frac{\mu -\sqrt{\mu -1}-1}{\alpha }+o(1),\frac{\mu +\sqrt {\mu -1}-1}{\alpha }+o(1)\right\}$ when $\alpha\rightarrow 0$. Therefore, this point is a saddle.

\item The point $P_9: \left(0, -\frac{1}{\sqrt{\xi }}, 0, 0\right)$. The eigenvalues are \newline  \newline $\left\{-2,4,\frac{\mu -\sqrt{\mu -1}-1}{\alpha }+o(1),\frac{\mu +\sqrt {\mu -1}-1}{\alpha }+o(1)\right\}$ \textls[-10]{when $\alpha\rightarrow 0$. Therefore, this point is a saddle.}

\item The set $P_{10}(x_c): \left(x_c, \frac{1}{\sqrt{\xi }}, \Omega_{\Lambda 0}, 0\right)$.

\item The line $P_{11}(x_c): \left(x_c, \frac{1}{\sqrt{\xi }}, 0, 0\right)$. In both cases, the stability analysis of these point curves will be left for future research since it cannot be implemented with the techniques developed in this paper.

\item The line $P_{12}: \left(0, \frac{1}{\sqrt{\xi }}, \Omega_{\Lambda 0}, 0\right)$. The eigenvalues are \newline  \newline $\left\{-2,4,\frac{\mu -\sqrt{\mu -1}-1}{\alpha }+o(1),\frac{\mu +\sqrt {\mu -1}-1}{\alpha }+o(1)\right\}$ when $\alpha\rightarrow 0$. \textls[-10]{Therefore, this point is a saddle.}

\item The point $P_{13}:\left(0, \frac{1}{\sqrt{\xi }}, 0, 0\right)$. The eigenvalues are \newline  \newline $\left\{-2,4,\frac{\mu -\sqrt{\mu -1}-1}{\alpha }+o(1),\frac{\mu +\sqrt {\mu -1}-1}{\alpha }+o(1)\right\}$ when $\alpha\rightarrow 0$. Therefore, this point is a saddle. All have indeterminate parameters of the equation of the state of matter.

\item The line $\left(0, -\frac{1}{\sqrt{\xi (12 \xi +1)}}, \Omega_{\Lambda 0}, -\mu -\sqrt{\mu ( 2 \mu -11)+18}+4\right)$.

\item The line $\left(0, -\frac{1}{\sqrt{\xi (12 \xi +1)}}, \Omega_{\Lambda 0}, -\mu + \sqrt{\mu ( 2 \mu -11)+18}+4\right)$.

\item The line $\left(0, \frac{1}{\sqrt{\xi (12 \xi +1)}}, \Omega_{\Lambda 0}, -\mu -\sqrt{\mu (2 \mu -11)+18}+4\right)$.

\item The line $\left(0, \frac{1}{\sqrt{\xi (12 \xi +1)}}, \Omega_{\Lambda 0}, -\mu + \sqrt{\mu (2 \mu -11)+18}+4\right)$. Stability can be determined numerically since the Jacobian matrix has infinite entries.
    \end{enumerate}
    
In these four cases, the stability analysis of these curves/sets of points will be left for future research since it cannot be implemented with the techniques developed in this paper.
They all have infinitely complex parameters of the equation of the state of matter.

In Figure \ref{fig:enter-label1}, the top panels show the numerical solutions $x(\tau)$, $y(\tau)$, $\Omega_{\Lambda}(\tau)$, and $ \alpha(\tau)$ of systems \eqref{eqinit}--\eqref{eqfin}, as well as the effective equation of the state of matter $w(\tau)$ defined in \eqref{dobleuve}. The values of the parameters $\xi$ and $\mu$ are shown in the figure.
In both cases, the equation of state $w(\tau)$ begins in a phantom regime ($w<-1$), crosses the matter-dominated region ($w=0$), and tends asymptotically toward a de Sitter regime ($w=-1$).
Two spiral orbits of the system starting at the initial conditions labeled $I$ and $II$ in Table \ref{Tab1} asymptotically tend to the equilibrium points $P_4$ and $P_5$.

\begin{figure}[h]
 \centering
    \includegraphics[width=0.45\textwidth]{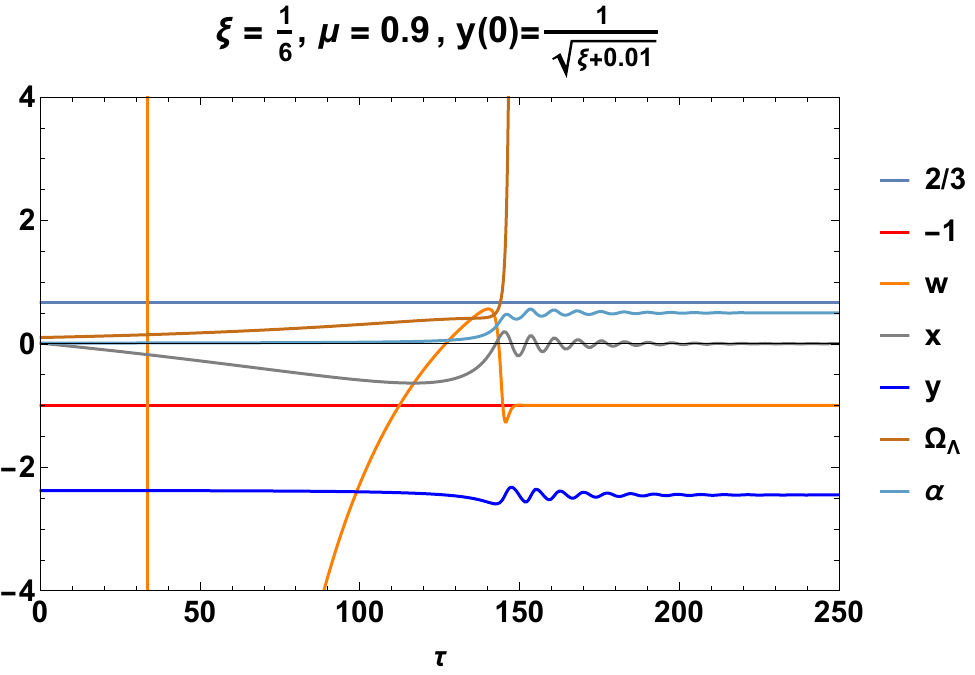}
    \includegraphics[width=0.45\textwidth]{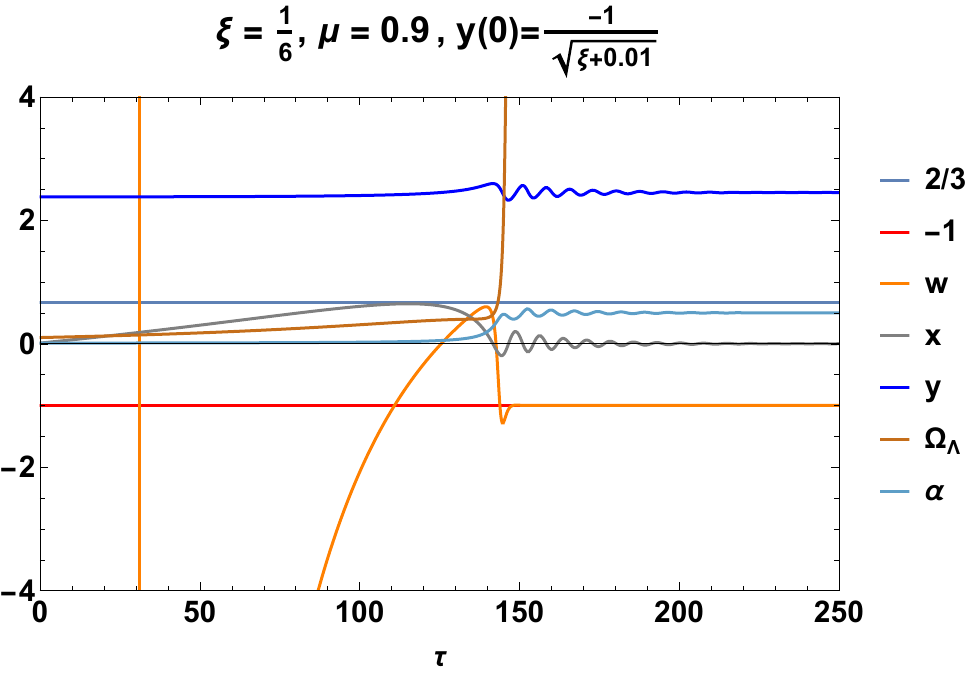}
    \includegraphics[width=0.6\textwidth]{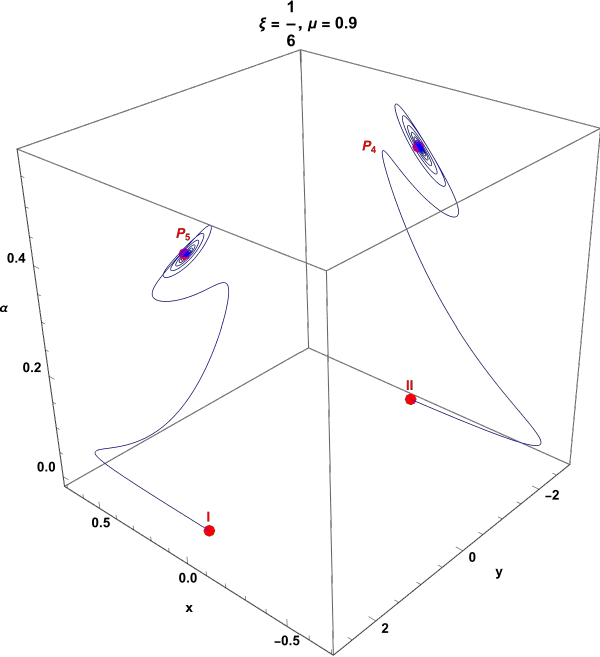}
    \caption{The upper panels show the numerical solution $x(\tau)$, $y(\tau)$, $\Omega_{\Lambda}(\tau)$ and $\alpha(\tau)$ of systems \eqref{eqinit}--\eqref{eqfin}, as well as the effective equation of the state of matter $w(\tau)$ defined in \eqref{dobleuve} for $\mu<1$.
    Two spiral orbits of the system, starting at the initial conditions labeled $I$ and $II$ in Table \ref{Tab1}, asymptotically tend the equilibrium points $P_4$ and $P_5$.}
    \label{fig:enter-label0}
\end{figure}

The upper panels of {Figure} 
 \ref{fig:enter-label0} show the numerical solutions $x(\tau)$, $y(\tau)$, $\Omega_{\Lambda}(\tau)$, and $\alpha(\tau)$ of systems \eqref{eqinit}--\eqref{eqfin}, as well as the effective equation of the state of matter $w(\tau)$ defined in \eqref{dobleuve} for $\mu<1$. Two spiral orbits of the system, starting at the initial conditions labeled $I$ and $II$ in Table \ref{Tab1}, asymptotically tend the equilibrium points, $P_4$ and $P_5$.

\begin{table}[h]
\caption{\label{Tab1} List of initial conditions used in the numerical integration of systems \eqref{eqinit}--\eqref{eqfin}. The values of $\xi$ and $\mu$ considered are the following: (a) $\xi=1/6$ and $\mu=1.0$ in Figure \ref{fig:enter-label1}, (b) $\xi=1/6$ and $\mu=0.9$ in Figure \ref{fig:enter-label0}, (c)   $\xi=1/6$ and $\mu=1.1$ in Figure \ref{fig:enter-label2} and (d) $\xi=4$ and $\mu=1.1$ in Figure \ref{fig:ciclo-3}.}
\footnotesize\setlength{\tabcolsep}{5pt}
\centering
    \begin{tabular}{lccccccc}\hline
Sol.  & \multicolumn{1}{c}{$x(0)$} & \multicolumn{1}{c}{$y(0)$} & \multicolumn{1}{c}{$\Omega_{\Lambda}(0)$} & \multicolumn{1}{c}{$\alpha(0)$}\\\hline
        I &  $0.01$ & $\frac{1}{\sqrt{\xi+0.01}}$ & $0.1$ & $0.01$\\
        II &  $0.01$ & $\frac{-1}{\sqrt{\xi+0.01}}$ & $0.1$ & $0.01$\\\hline
    \end{tabular}
\end{table}

\vspace{-10pt}
\begin{figure}[h]
 \centering
    \includegraphics[width=0.45\textwidth]{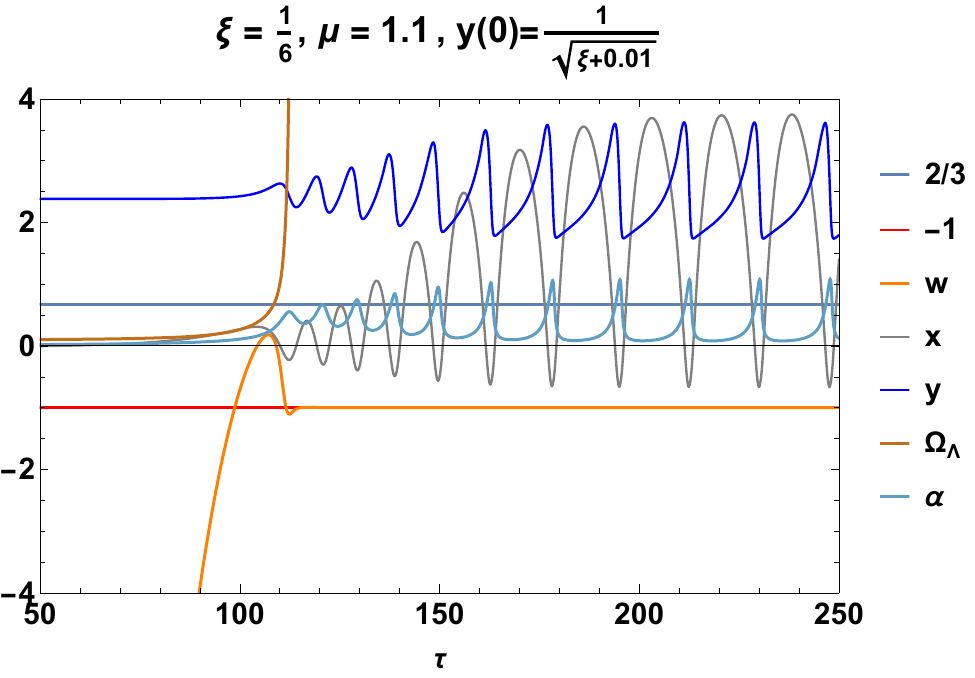}
    \includegraphics[width=0.45\textwidth]{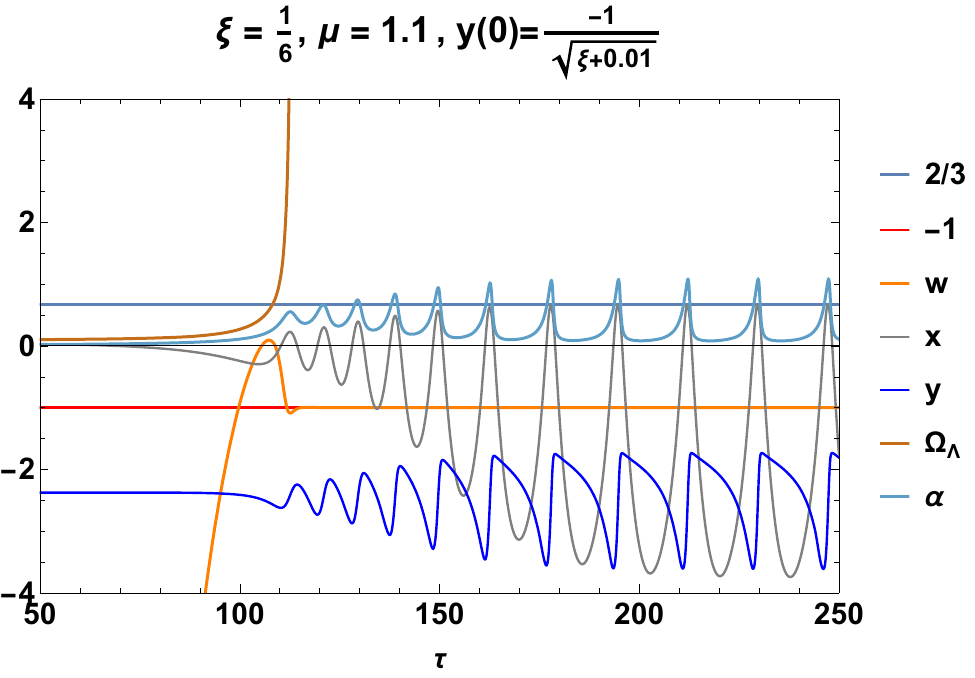}
    \includegraphics[width=0.6\textwidth]{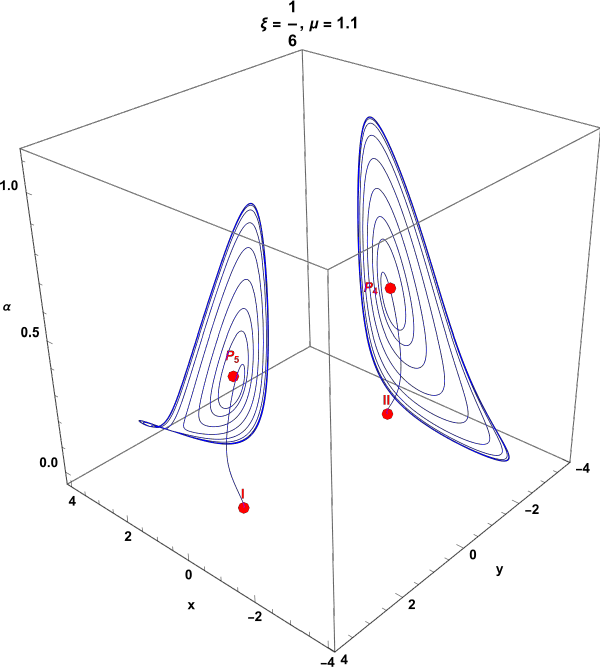}
    \caption{The upper panels show the numerical solutions $x(\tau)$, $y(\tau)$, $\Omega_{\Lambda}(\tau)$, and $\alpha(\tau)$ of systems \eqref{eqinit}--\eqref{eqfin}, as well as the effective state equation $w(\tau)$ defined in \eqref{dobleuve} for $\mu>1$. Two orbits starting at the initial conditions labeled $I$ and $II$ in Table \ref{Tab1} asymptotically tend to a limit cycle.}
    \label{fig:enter-label2}
\end{figure}

In Figure \ref{fig:enter-label2}, the top panels show the numerical solutions $x(\tau)$, $y(\tau)$, $\Omega_{\Lambda}(\tau)$, and $\alpha(\tau)$ of systems \eqref{eqinit}--\eqref{eqfin}, as well as the effective equation of state $w(\tau)$ defined in \eqref{dobleuve}. The values of the parameters $\xi$ and $\mu$ are shown in the figure.
In both cases, the equation of state $w(\tau)$ begins in a phantom regime ($w<-1$), crosses the matter-dominated region ($w=0$), and tends asymptotically toward a de Sitter regime ($w=-1$). The oscillating behavior of the functions $x(\tau)$, $y(\tau)$, $\Omega_{\Lambda}(\tau)$ and $\alpha(\tau)$ is still present but exhibits a greater amplitude compared to that shown in Figure \ref{fig:enter-label1}. Two orbits starting at the initial conditions labeled $I$ and $II$ in Table \ref{Tab1} tend asymptotically to a limit cycle.

\begin{figure}[h]
    \centering
    \includegraphics[width=0.6\textwidth]{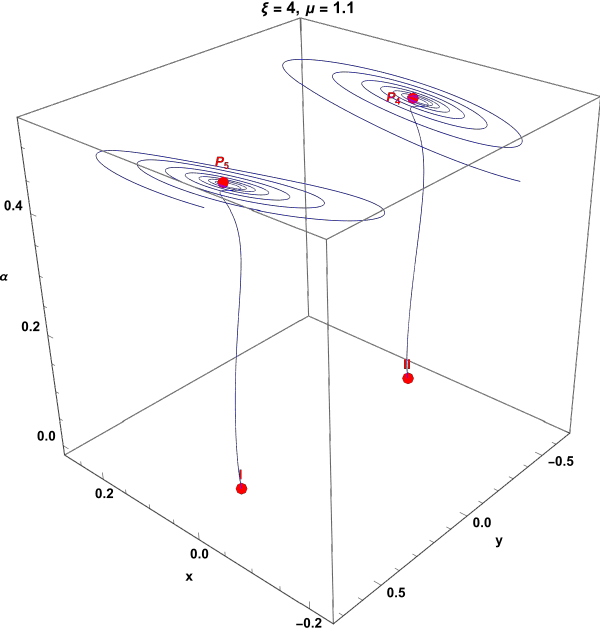}
    \caption{Dynamics  in the invariant set $\Omega_\Lambda=0$ of systems \eqref{eqinit}--\eqref{eqfin} for the values of the parameters $\xi=4$ and $\mu=1.1$. Two orbits that begin at the initial conditions labeled $I$ and $II$ in Table~\ref{Tab1} tend to the saddle points $P_4$ and $P_5$ and then form an unstable spiral centered on said points.}
    \label{fig:ciclo-3}
\end{figure}

Figure \ref{fig:ciclo-3} presents the dynamics in the invariant set $\Omega_\Lambda=0$ of systems \eqref{eqinit}--\eqref{eqfin} for the values of the parameters $\xi =4$ and $\mu=1.1$. Two orbits that begin at the initial conditions labeled $I$ and $II$ in Table \ref{Tab1} tend to the saddle points $P_4$ and $P_5$ and then form an unstable spiral centered on said points.

\begin{figure}[h]
 \centering
    \includegraphics[width=0.3\textwidth]{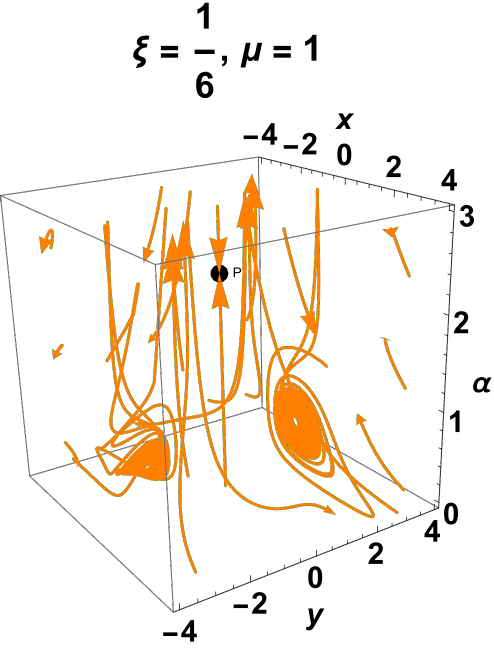}
    \includegraphics[width=0.3\textwidth]{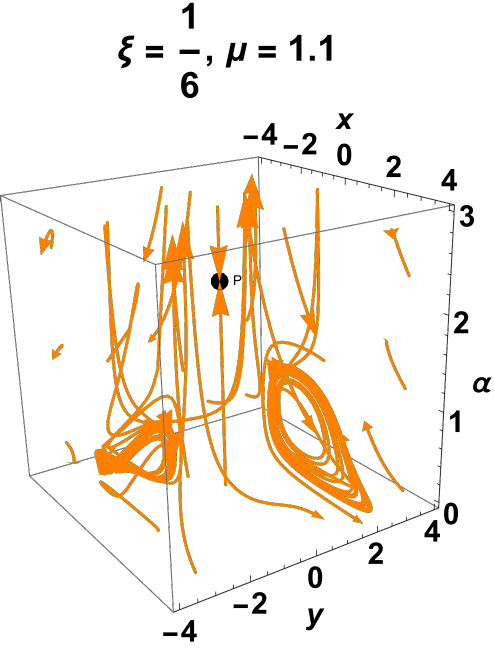}
    \includegraphics[width=0.3\textwidth]{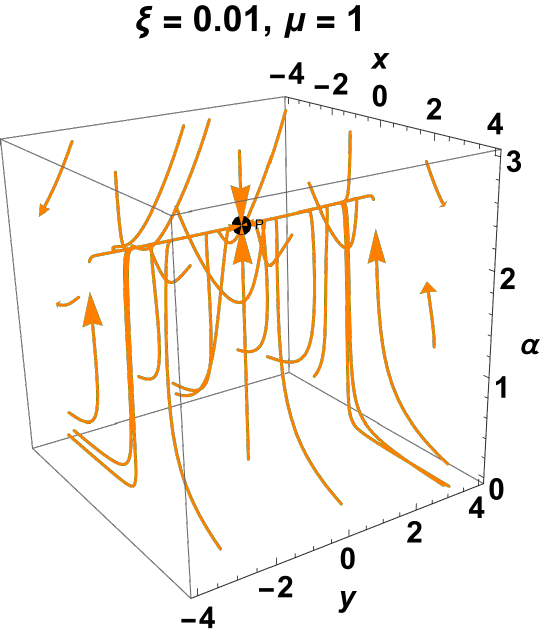}
    \caption{Phase diagrams for systems \eqref{eqinit}--\eqref{eqfin} for different values of the parameters $\xi$ and $\mu.$}
    \label{fig:enter-label}
\end{figure}

Finally, in Figure \ref{fig:enter-label}, phase diagrams are presented for the \eqref{eqinit}--\eqref{eqfin} systems for different values of the parameters $\xi$ and $\mu.$

In this section, we numerically examined the invariant set $\Omega_\Lambda=0$ of the\linebreak \eqref{eqinit}--\eqref{eqfin} systems. However, to characterize such an invariant set analytically, we will present a comprehensive analysis of dynamical systems in Section \ref{invariantset-0-CC}. 
\vspace{-10pt}

\subsubsection{Invariant Set $\Omega_\Lambda=0$}
\label{invariantset-0-CC}
Replacing $\Omega_\Lambda=0$ in \eqref{altform}, and defining the new variables, we have the following: 
\begin{equation}
    x= \frac{\dot{\phi}(t)}{\sqrt{6} H(t)}, \quad y=\phi (t),
\end{equation}
which satisfies \vspace{-9pt}
\begin{equation}
(x + \sqrt{6} \xi y)^2  + \xi  y^2(1-6 \xi)+ \frac{(\mu -1) \left(1-\xi  y^2\right)}{\alpha}+ \Omega_{m}=1,
\end{equation}
where
\begin{equation}
    \Omega_m = \frac{\rho_m}{3 H^2}
\end{equation}
is the dimensionless energy density of matter; we obtain an alternative formulation, which describes the dynamics on the invariant set $\Omega_\Lambda=0$. 
The equation of state parameter of matter can be expressed in terms of the new variables as follows:

\begin{align}\label{dobleuve2}
 w & =  \left\{\alpha  \left(\xi  (12 \xi +1) y^2-1\right) \left(-\alpha +\mu +\alpha  x^2+\xi  y
   \left(2 \sqrt{6} \alpha  x+y (\alpha -\mu +1)\right) -1\right)\right\}^{-1} \nonumber \\
   & \times \Bigg\{\alpha ^2 \left[x^2 \left(\xi  \left(y^2+12\right)-1\right)+2 \sqrt{6} \xi  x y \left(\xi  y^2-1\right)+\xi  y^2 \left(\xi 
   y^2-2\right) +1\right] \nonumber \\
   & +2 \alpha  \left(\xi  y^2-1\right) \left[\mu  \left(\xi  y \left(\sqrt{6} x+y\right)-1\right)+\xi  y \left(-\sqrt{6} x+12 \xi 
   y-3 y\right)+3\right] \nonumber \\
   & -(\mu -2) (\mu -1) \left(\xi  y^2-1\right)^2\Bigg\}. 
\end{align}

Introducing the derivative \eqref{log-time}, we obtain the three-dimensional dynamical system as follows: 
\begin{small}
\begin{align}
 \alpha^2 G   \frac{d x}{d \tau} & = 
  -12 \alpha ^2 \xi  x^3+2 \sqrt{6} \alpha  \xi  x^2 y (-6 \alpha  \xi +\alpha +\mu -1) \nonumber \\
   &   +x \left(3 \alpha  \left(-\mu +8 (\mu -1) \xi ^2
   y^2+(\mu -3) \xi  y^2+3\right)-(\mu -2) (\mu -1) \left(\xi  y^2-1\right)\right) \nonumber \\ 
   &  +\sqrt{6} \xi  y \left(\alpha ^2+2 \alpha  (\mu -4)-(\mu -2) (\mu
   -1)\right) \left(\xi  y^2-1\right),\\
    \frac{d y}{d \tau} & =\sqrt{6}  x, 
   \\
\alpha  G  \frac{d \alpha}{d \tau} & =   3 \alpha ^2+2 \alpha  \mu -9 \alpha -\mu ^2+3 \mu  +\xi  \Big(12 \alpha ^2 x^2-2
   \sqrt{6} \alpha  x y (\alpha +\mu -1) 
   \nonumber \\ 
   &  \quad\quad  +y^2 \left(-2 \alpha  \mu -3 (\alpha -3) \alpha +\mu ^2-3 \mu +2\right)\Big)+12 (1-2 \alpha ) \alpha  \xi ^2
   y^2-2,  \label{eqfin2}
\end{align}
\end{small}
where 
{\small\begin{equation}
    G:= (\xi  (12 \xi +1) y^2-1), 
\end{equation}}
is defined on the phase space
{\small\begin{equation}
(x + \sqrt{6} \xi y)^2  + \xi  y^2(1-6 \xi)+ \frac{(\mu -1) \left(1-\xi  y^2\right)}{\alpha } \leq 1
\end{equation}}
which corresponds to $\Omega_{m}\geq 0$. 

Introducing the time-rescaling \eqref{time-rescaling}, 
which preserves the arrow of time, we obtain the~system, as follows:  

\begin{small}
\begin{align}
  x^{\prime} & =  \frac{1}{\xi  (12 \xi +1) y^2-1}\Bigg[-12 \alpha ^2 \xi  x^3+2 \sqrt{6} \alpha  \xi  x^2 y (-6 \alpha  \xi +\alpha +\mu -1)
  \nonumber \\ 
  &+x \left(3 \alpha \left(-\mu +8 (\mu -1) \xi ^2 y^2+(\mu -3) \xi  y^2+3\right)-(\mu -2) (\mu -1) \left(\xi y^2-1\right)\right) \nonumber \\
  & +\sqrt{6} \xi  y \left(\alpha ^2+2 \alpha  (\mu -4)-(\mu -2) (\mu -1)\right) \left(\xi y^2-1\right)\Bigg], \label{eq4.126}\\
   y ^{\prime} & = \sqrt{6} \alpha ^2 x,  \label{eq4.127}
\\
 \alpha^{\prime} &=\frac{\alpha  }{\xi  (12 \xi +1) y^2-1} \Bigg[3 \alpha ^2+2 \alpha  \mu -9 \alpha -\mu ^2+3 \mu +12 (1-2 \alpha ) \alpha  \xi ^2 y^2-2\nonumber \\
  &  +\xi  \left(12 \alpha ^2 x^2-2 \sqrt{6} \alpha  x y (\alpha +\mu -1)+y^2 \left(-2 \alpha 
   \mu -3 (\alpha -3) \alpha +\mu ^2-3 \mu +2\right)\right)\Bigg].  \label{eq4.128}
\end{align}
\end{small}
sometimes it is more useful to use the time variable $\tau=\ln (a/a_0)$.

The equilibrium points of systems \eqref{eq4.126}, \eqref{eq4.127}, and 
  \eqref{eq4.128} in the coordinates $(x, y, \alpha)$ are as follows: 
\begin{enumerate}
\item The equilibrium point curve $P_{1}: \left(- \sqrt{6} \xi y_{c}, y_{c}, 0 \right)$ parameterized by $y_c \in \mathbb{R}$. The equation of state parameter is complex infinity.
The eigenvalues are \newline
$\left\{0,-\frac{(\mu -2) (\mu -1) \left(\xi y^2-1\right)}{\xi (12 \xi +1) y^2 -1},\frac{(\mu -2) (\mu -1) \left(\xi
    y^2-1\right)}{\xi (12 \xi +1) y^2-1}\right\}$. The curve is a saddle because it has at least two eigenvalues with different signs.

\item The equilibrium point $P_2: \left(0,0,\frac{1} {6} (9- 2\mu -r)\right)$. There exists ($\alpha>0$) for $1\leq \mu\leq 2$. The eigenvalues are $\Big\{\frac{1}{6} \left(-2 \mu \left(8 \mu +r-36\right)+9 r-105\right)$, \newline $ \frac{1}{72} \left(-72 \mu ^2-u_{-}-18 r\mu +378 \mu +54 r-558\right)$, \newline $ \frac{1}{72 } \left(-72 \mu ^2+u_{-}-18 r\mu +378 \mu +54 r-558\right)\Big\}$, where $r=\sqrt{8 \mu(2 \mu -9)+105}$ and $u_{-}$ are the positive square roots of \newline 
$u_{-}^2 = 324 \left(\mu \left(4 \mu +r-21\right)-3 r+31\right)^2 \newline -192 \left(\mu \left(3 \left (403 r-5787\right)+\mu \left(2 \mu \left(82 \mu +20 r-933\right)-375 r+8355\right)\right)-1413 r+14,481\right) \xi$.

The critical point is a saddle in its existence interval, as shown in Figure \ref{fig:P2-Eigenvalues.png}.\vspace{-3pt}
\begin{figure}[h]
 \centering
    \includegraphics[width=0.45\textwidth]{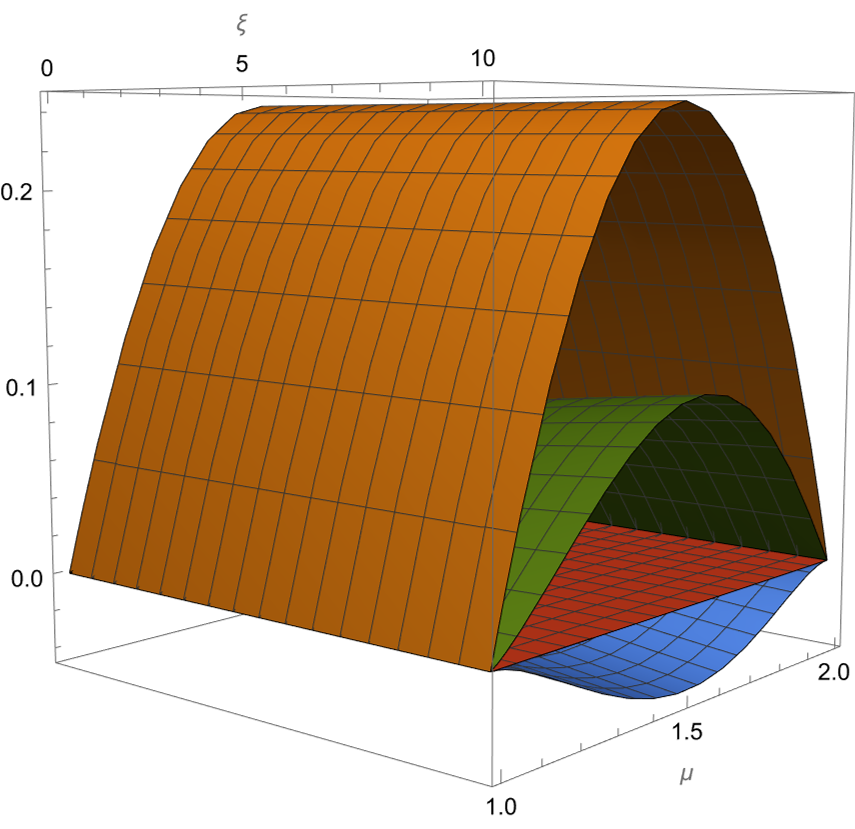}
    \caption{Eigenvalues of $P_2$ show that the point is a saddle. For reference, the red plane corresponds to the value zero, and the other colors denote the three eigenvalues; two are non-negative, and one is non-positive.}
    \label{fig:P2-Eigenvalues.png}
\end{figure}
The equation of state parameter is $w = \frac{-2 \mu  \left(r+9\right)+3 r+35}{4 \mu  (2 \mu -3)-4}$.

\item The equilibrium point $P_3: \left(0,0,\frac{1}{6} (9- 2\mu +r)\right)$. The eigenvalues are \newline  $\Big\{\frac{1}{6} \left(2 \mu  \left(-8 \mu +r+36\right)-3 \left(3 r+35\right)\right)$, \newline $\frac{1}{72} \left(-u_{+}-72 \mu ^2+18r \mu +378 \mu -54 r-558\right)$, \newline $ \frac{1}{72} \left(u_{+}-72 \mu ^2+18r \mu +378 \mu -54 r-558\right) \Big\}$, where $r=\sqrt{8 \mu(2 \mu -9)+105}$ and $u_{+}$ are the positive square roots of the following: 
\begin{small}
$u_{+}^2 = 324 \left(4 \mu ^2-\left(r+21\right) \mu +3 r+31\right)^2 \newline +192 \left(\mu  \left(3 \left(403 r+5787\right)+\mu  \left(2\mu  \left(-82 \mu +20 r+933\right)-15 \left(25 r+557\right)\right)\right)-9 \left(157 r+1609\right)\right) \xi$.
\end{small}
$P_3$ is a sink for

$\left\{\mu >2,  0<\xi \leq \frac{1}{48}\right\}\cup \left\{\mu =1,  0<\xi \leq \frac{27}{154}\right\}\cup
   \left\{\mu =2,  0<\xi \leq \frac{27}{280}\right\} \newline \cup \left\{\mu +\frac{1}{4}
   \left(-5-\frac{9}{\sqrt[3]{58+\sqrt{4093}}}+\sqrt[3]{58+\sqrt{4093}}\right)=0,  0<\xi \leq -\frac{27 (\mu
   -3)}{4 \left(2 \mu ^3-3 \mu ^2-12 \mu +90\right)}\right\} \newline \cup \left\{0<\mu <\frac{1}{4}
   \left(5-\frac{9}{\sqrt[3]{\sqrt{4093}-58}}+\sqrt[3]{\sqrt{4093}-58}\right),  0<\xi \leq
   \frac{1}{48}\right\} \newline \cup \left\{\frac{1}{4}
   \left(5-\frac{9}{\sqrt[3]{\sqrt{4093}-58}}+\sqrt[3]{\sqrt{4093}-58}\right)<\mu <1,  0<\xi \leq
   \frac{1}{48}\right\}$.

It is a saddle otherwise.

The equation of state parameter is $w = \frac{2 \mu  \left(r-9\right)-3 r+35}{4 \mu  (2 \mu -3)-4}$.

\item The equilibrium point $P_4: \left(0,-\frac{1}{\sqrt{\xi}},\frac{1}{2}\right)$. The eigenvalues are \newline $\Big\{-\frac{1}{2},\frac{1}{4} \left(2 \mu -\sqrt{8 \mu -19}-2\right), \frac{1}{4} \left(2 \mu +\sqrt{8 \mu -19}-2\right)\Big\}$. The equation of state parameter is $w=\frac{1}{24 \left(\xi ^{3/2}+\xi \right)}$. The point is a sink for $0<\mu<1$ and $\xi > 0$ and becomes a saddle for $\mu>1$, as shown in Figure \ref{fig:ciclo-3}.

\item The equilibrium point $P_5: \left(0,\frac{1}{\sqrt{\xi}},\frac{1}{2}\right)$. The eigenvalues are \newline $\Big\{-\frac{1}{2}, \frac{1}{4} \left(2 \mu -\sqrt{8 \mu -19}-2\right), \frac{1}{4} \left(2 \mu +\sqrt{8 \mu -19}-2\right)\Big\}$. The equation of state parameter is $w = \frac{1}{24 \left(\xi ^{3/2}+\xi \right)}$. The point is a sink for $0<\mu < 1$ and $\xi > 0$ and is a saddle for $\mu>1$, as shown in Figure \ref{fig:ciclo-3}.

\item The equilibrium point $P_{7}: \left(x_{c}, -\frac{1}{\sqrt{\xi}}, 0\right)$. 

\item The equilibrium point $P_9: \left(0,-\frac{1}{\sqrt{\xi}},0\right)$. The equation of state parameter is $w=0$. 

\item The equilibrium point $P_{11}: \left(x_{c}, \frac{1}{\sqrt{\xi}}, 0\right)$. 

\item The equilibrium point $P_{13}: \left(0,\frac{1}{\sqrt{\xi}},0\right)$. The equation of state parameter is $w=0$. 

Due to the complexity of the stability analysis of these four points, it will be left for future research since it cannot be implemented with the techniques developed in this~paper.

\item The equilibrium points $\left(0,-\frac{1}{\sqrt{\xi  (12 \xi +1)}},-\mu -\sqrt{\mu  (2 \mu -11)+18}+4\right)$. 

\item The equilibrium points $\left(0,-\frac{1}{\sqrt{\xi  (12 \xi +1)}},-\mu +\sqrt{\mu  (2 \mu -11)+18}+4\right)$. 

\item The equilibrium points $\left(0,\frac{1}{\sqrt{\xi  (12 \xi +1)}},-\mu -\sqrt{\mu  (2 \mu -11)+18}+4\right)$.

\item The equilibrium points $\left(0,\frac{1}{\sqrt{\xi  (12 \xi +1)}},-\mu +\sqrt{\mu  (2 \mu -11)+18}+4\right)$. 
\end{enumerate}

In all four cases, the equation of state parameter is complex and infinite. Due to the complexity of the stability analysis of this set of equilibrium points, numerical solution methods can be used. Since the stability analysis of these four points is very complex, it will be left for future research since it cannot be implemented with the techniques developed in this paper. 

\subsubsection{Asymptotic Expansions for $\varepsilon:= \mu-1 \rightarrow 0$}

Next, we will study the perturbation problem related to the study of the phase portrait of a system of differential equations, as follows:
\begin{equation}
\dot x=X(x;\varepsilon), \quad x\in \mathbb{R}^3, \quad \varepsilon \sim 0
\end{equation}
in a neighborhood of the origin, where the unperturbed vector field is as follows: $X(x;0)$ \citep{fenichel, Kevorkian1, Fusco, dumortier, holmes, Verhulst, Berglund}.

To obtain asymptotic expansions for $\mu \rightarrow 1$, the reparametrization $\varepsilon:= \mu-1$ is introduced and the limit $\varepsilon \rightarrow 0 $ is analyzed using the asymptotic expansion, as follows: \vspace{-10pt}
 
\begin{align*}
    \rho = \rho_0 + \varepsilon \rho_1 + \mathcal{O}(\varepsilon^2), \;
    p = p_0 + \varepsilon p_1 + \mathcal{O}(\varepsilon^2), \;
    H = H_0 + \varepsilon H_1 + \mathcal{O}(\varepsilon^2), \;
    \phi = \phi_0 + \varepsilon \phi_1 + \mathcal{O}(\varepsilon^2).
\end{align*} 

After making the substitution in the field equations and comparing the coefficients with equal powers of $\varepsilon$, the following is obtained. 

At \textbf{{order} 
$0$}, we have the following equations: 
\begin{small}
\begin{align}
   & \dot{H_0} = \frac{2 t \xi  \dot{\phi_0}^2-2 t \xi 
   H_0 \phi_0 \dot{\phi_0}+3 H_0 \left(-\xi  ((8 \xi  t+t) H_0-2) \phi_0^2+t
   H_0-2\right)}{t \left(\xi  (12 \xi +1) \phi_0^2-1\right)},\\
   & \ddot{\phi_{0}}= \frac{-12 t \xi ^2 \phi_0 \dot{\phi_0}^2-3 t H_0 \left(\xi  (8 \xi +1) \phi_0^2-1\right) \dot{\phi_0}+6 \xi  H_0 (t H_0-6) \phi_0 \left(\xi  \phi_0^2-1\right)}{t \left(\xi  (12 \xi +1) \phi_0^2-1\right)}, 
\end{align}
\end{small}
and the auxiliary equation is as follows:
\begin{small}
\begin{align}
& \dot{\rho_{0}}= \frac{3 H_0 }{t \left(\xi  (12 \xi +1) \phi_0^2-1\right)}  \Bigg[t \left(\xi  (6 \xi +1)
   \phi_0^2+6 \xi -1\right) \dot{\phi_0}^2+12 t \xi  H_0 \phi_0 \left(\xi  (6 \xi +1) \phi_0^2-1\right)
   \dot{\phi_0} \nonumber \\
   & +6 H_0 \left(\xi  \phi_0^2-1\right) \left(\xi  (12 \xi +(6 \xi  t+t) H_0-2) \phi_0^2-t H_0+2\right)\Bigg],
\end{align}
\end{small}

and the constraints are as follows: 

\begin{small}
\begin{align}
   & \rho_{0}= 3\left(1- \xi  \phi_0^2\right) H_0^2-6 \xi  \phi_0 \dot{\phi_0}
   H_0-\frac{1}{2} \dot{\phi_0}^2-\Lambda, \\
   & p_0= -\frac{1}{2 t \left(\xi  (12 \xi +1) \phi_0^2-1\right)} \Bigg[6 t H_0^2 \left(\xi  \phi_0^2-1\right)^2+12
   H_0 \left(\xi  \phi_0 \left(2 (6 \xi -1) \phi_0+t \dot{\phi_0}\right)+2\right) \left(\xi  \phi_0^2-1\right) \nonumber \\
   & +t \left(\xi  \phi_0^2+12 \xi -1\right) \dot{\phi_0}^2+2 t \Lambda  \left(1-\xi  (12 \xi +1) \phi_0^2\right)\Bigg]. 
\end{align}
\end{small}

At \textbf{order $1$}, we have the following equations:\vspace{-10pt}

\begin{small}
\begin{align}
    & \dot{H_1}= \frac{1}{t^2 \left(\xi  (12 \xi +1) \phi_0^2-1\right)^2}  \Bigg[-24 \xi ^2 H_0^2 \phi_0 \phi_1 t^2+4 \xi 
   \left(\xi  (12 \xi +1) \phi_0^2-1\right) \dot{\phi_0} \dot{\phi_1} t^2 \nonumber \\
   & -2 H_1 \left(\xi  (12 \xi +1) \phi_0^2-1\right) \left(\xi  \phi_0 \left(t \dot{\phi_0}-3 \phi_0\right)+3\right) t \nonumber \\
   & +2 H_0 \Bigg(3 t H_1 \left(\xi  \phi_0^2 \left(-\xi  (8 \xi +1) (12 \xi +1) \phi_0^2+20 \xi +2\right)-1\right) \nonumber \\
   & +\xi  \Big\{ -\xi  (12 \xi +1) \phi_0^4-t \xi  (12 \xi +1) \dot{\phi_1} \phi_0^3+\left(t (12 \xi +1) \phi_1 \dot{\phi_0} \xi +12 \xi
   +2\right) \phi_0^2 \nonumber \\
   & +\left(72 \xi  \phi_1+t \dot{\phi_1}\right) \phi_0+t \phi_1 \dot{\phi_{0}}\Big\}-1\Bigg) t \nonumber \\
   & +2 \xi  (6 \xi +1) \phi_0^2+\xi  \phi_0 \left(2 t \dot{\phi_0}-\xi  (12 \xi +1)
   \left(\phi_0^3+2 t \dot{\phi_0} \phi_0^2+4 t^2 \phi_1 \dot{\phi_0}^2\right)\right)-1\Bigg], 
\end{align}
\begin{align}
    & \ddot{\phi_{1}} = \frac{1}{t^2 \left(\xi   (12 \xi +1) \phi_0^2-1\right)^2}  \Bigg[6 \xi ^3 (12 \xi +1) (2 t (H_0+(t H_0-3) H_1)+1) \phi_0^5 \nonumber \\
    & +t \xi ^2 (12 \xi +1) \left(6 t \xi  \phi_1 H_0^2-36 \xi  \phi_1 H_0-3 t (8 \xi +1) \dot{\phi_1} H_0+(24 \xi -3 t (8 \xi +1) H_1+1)
   \dot{\phi_0}\right) \phi_0^4 \nonumber \\
   & +12 \xi ^2 \left(-2 \xi  (12 \xi +1) \dot{\phi_0} \dot{\phi_1} t^2-(6 \xi +1) (2 t
   (H_0+(t H_0-3) H_1)+1)\right) \phi_0^3 \nonumber \\
   & +2 t \xi  \Big\{6 t \xi  (6 \xi -1) \phi_1 H_0^2+3 \left(12 (1-6 \xi ) \xi  \phi_1+t (10 \xi +1) \dot{\phi_1}\right) H_0 \nonumber \\
   & +\dot{\phi_0} \left(6 t (12 \xi +1) \phi_1 \dot{\phi_0} \xi ^2-18 \xi +3 t (10 \xi +1) H_1-1\right)\Big\} \phi_0^2 \nonumber \\
   & +6 \xi  \left(2 t \left(-3 H_1+H_0 \left(t H_1-2 t \xi  \phi_1 \dot{\phi_0}+1\right)+2 t \xi  \dot{\phi_0} \dot{\phi_1}\right)+1\right) \phi_0 \nonumber \\
   & +t \left(6 t \xi  \phi_1 H_0^2-3 \left(12 \xi  \phi_1+t \dot{\phi_1}\right) H_0+\dot{\phi_0} \left(12 t \phi_1 \dot{\phi_0} \xi ^2-3 t H_1+1\right)\right)\Bigg], 
\end{align}
\end{small}
 
and the auxiliary equation is as follows: \vspace{-13pt}

\begin{small}
\begin{align}
& \dot{\rho_1}= \frac{1}{t^2 \left(\xi  (12 \xi +1) \phi_0^2-1\right)^2}  \Bigg[36 t^2 \xi  \phi_0
   \left(\xi  (6 \xi +1) \left(\xi  (12 \xi +1) \phi_0^2-2\right) \phi_0^2-6 \xi +1\right) \phi_1 H_0^3 \nonumber \\
   & +3 t \Bigg(18 t H_1 \left(\xi  \phi_0^2-1\right) \left(\xi  (6 \xi +1) \phi_0^2-1\right) \left(\xi  (12 \xi +1) \phi_{0}^2-1\right) \nonumber \\
   & +\xi  \Bigg\{\xi ^2 (1-36 \xi  (16 \xi +1)) \phi_0^6+12 \xi ^2 (12 \xi +1) \left(2 (6 \xi -1) \phi_1+t (6 \xi
   +1) \dot{\phi_1}\right) \phi_0^5 \nonumber \\
   & +3 \xi  \left(24 \xi  (8 \xi +1)+4 t \xi  (6 \xi +1) (12 \xi +1) \phi_1 \dot{\phi_0}-1\right) \phi_0^4 \nonumber \\
   & -24 \xi  \left(2 (6 \xi -1) \phi_1+t (9 \xi +1) \dot{\phi_1}\right) \phi_0^3-3
   \left(8 t (3 \xi +1) \phi_1 \dot{\phi_0} \xi +12 \xi -1\right) \phi_0^2 \nonumber \\
   & +12 \left((36 \xi -2) \phi_1+t \dot{\phi_1}\right) \phi_0+12 t \phi_1 \dot{\phi_0}\Bigg\}-1\Bigg) H_0^2 \nonumber \\
   & +3 \Bigg(2 \left(\xi  (6 \xi +1) \phi_0^2+6 \xi -1\right) \left(\xi  (12 \xi +1) \phi_0^2-1\right) \dot{\phi_0} \dot{\phi_1} t^2 \nonumber \\
   & +24 H_1 \left(\xi  (12 \xi +1) \phi_0^2-1\right) \left(\xi  \phi_0 \left(\phi_0 \left(\xi  (6 \xi -1) \phi_0^2-6
   \xi +2\right)+t \left(\xi  (6 \xi +1) \phi_0^2-1\right) \dot{\phi_0}\right)-1\right) t \nonumber \\
   & +\xi  \phi_0 \Bigg\{\xi ^2 (12 (5-24  \xi ) \xi +7) \phi_0^5+3 \xi  (8 \xi  (12 \xi -5)-7) \phi_0^3+3 (20 \xi +7) \phi_0 \nonumber \\
   & -144 t^2 \xi ^2 \phi_1 \dot{\phi_0}^2+2 t \left(\xi  \left(\left(\xi -144 \xi ^3\right) \phi_0^2-2\right) \phi_0^2+1\right) \dot{\phi_0} \Bigg\}-7\Bigg) H_0 \nonumber \\
   & +t (3 t H_1-1) \left(\xi  (6 \xi +1) \phi_0^2+6 \xi -1\right) \left(\xi  (12 \xi +1)
   \phi_0^2-1\right) \dot{\phi_0}^2\Bigg],
\end{align}
\end{small}
 
and the constraints are as follows:  \vspace{-10pt}
 
\begin{small}
\begin{align}
   & \rho_{1}= -6 \xi  \phi_0 \phi_1 H_0^2-\frac{3 (2 t H_1-1) \left(\xi  \phi_0^2-1\right) H_0}{t}-6 \xi  (H_1
   \phi_0+H_0 \phi_1) \dot{\phi_0}-\left(6 \xi  H_0 \phi_0+\dot{\phi_0}\right)
   \dot{\phi_1}, 
\\
   & p_1= -\frac{1}{t^2 \left(\xi  (12 \xi +1) \phi_0^2-1\right)^2}  \Bigg[-144 t^2 \phi_0 \phi_1 \dot{\phi_0}^2 \xi ^3+6 t^2 H_0^2 \phi_0 \left(\xi  \phi_0^2-1\right) \left(\xi  (12 \xi +1) \phi_0^2+12 \xi -1\right) \phi_1 \xi  \nonumber \\
   & +6 t \phi_0 \left(\xi  \phi_0^2-1\right) \left(\xi  (12 \xi +1) \phi_0^2-1\right) \dot{\phi_0} \xi  +3 \left(\xi  \phi_0^2-1\right)^2 \left(\xi  (12 \xi +1) \phi_0^2-1\right)\nonumber \\
   & +6 t H_1 \left(\xi  \phi_0^2-1\right) \left(\xi  (12 \xi
   +1) \phi_0^2-1\right) \left(\xi  \phi_0 \left(2 (6 \xi -1) \phi_0+t \dot{\phi_0}\right)+2\right) \nonumber \\
   & +t^2 \left(\xi  \phi_0^2+12 \xi -1\right) \left(\xi  (12 \xi +1) \phi_0^2-1\right) \dot{\phi_0} \dot{\phi_1}  +6 t H_0 \Big\{t H_1 \left(\xi  (12 \xi +1) \phi_0^2-1\right) \left(\xi  \phi_0^2-1\right)^2 \nonumber \\
   & +\xi  \Bigg(\xi ^2 (12 \xi +1) \phi_0^6+\xi ^2 (12 \xi +1) \left(4 (6 \xi -1) \phi_1+t \dot{\phi_1}\right) \phi_0^5 +\xi  \left(t \xi (12 \xi +1) \phi_1 \dot{\phi_0}-3 (8 \xi +1)\right) \phi_0^4 \nonumber \\
   & -2 \xi  \left(4 (6 \xi -1) \phi_1+t (6 \xi +1) \dot{\phi_1}\right) \phi_0^3+\left(2 t (6 \xi -1) \phi_1 \dot{\phi_0} \xi +12 \xi +3\right) \phi_0^2  \nonumber \\
   &  +\left((72 \xi -4) \phi_1+t \dot{\phi_1}\right) \phi_0+t \phi_1 \dot{\phi_0}\Bigg)-1\Big\}\Bigg].
\end{align}
\end{small}

For simplicity, we will solve the $\xi=0$ case.

At \textbf{order $0$}, we have the following equations: 
\begin{align}
&\dot{H_0}= -\frac{3 H_0 (-2+t H_0)}{t},\\
& \dot{\rho_0}= -\frac{3 H_0\left(-6H_0 (2-t H_0)-t \dot{\phi}^2\right)}{t},\\
& \ddot{\phi_0}= -3H_0  \dot{\phi_0}.
\end{align}
and the constraints are as follows: \vspace{-10pt}
\begin{align}
& \rho_{0}=3 H_0^2-\Lambda -\frac{1}{2} \dot{\phi_0}^2,\\
& p_0=\frac{6 t H_0^2-24 H_0 +2 \Lambda  t-t  \dot{\phi_0}^2}{2 t}.
\end{align}

At \textbf{order $1$}, we have the following equations: 
\begin{small}
\begin{align}
& \dot{H_1}= \frac{-1+6 t H_1+2 t H_0 (-1-3 t H_1)}{t^2},\\
& \dot{\rho_1}= \frac{1}{t^2}\left(3 t H_0^2 (-1-18 t H_1)+t (-1+3 t H_1) \dot{\phi_0}^2+3 H_0 
\left(-7+24 t H_1+2 t^2 \dot{\phi_0} \dot{\phi_1}\right)\right),\\
& \ddot{\phi_1}= \frac{(1-3 t H_1)  \dot{\phi_0} -3 t H_0 \dot{\phi_1}}{t}. 
\end{align}
\end{small}
and the constraints are as follows: \vspace{-6pt}
\begin{align}
& \rho_1=\frac{3 H_0 (2 t H_1-1)}{t}-\dot{\phi_0} \dot{\phi_1},\\
& p_1=-\frac{6 t H_0 (-t H_1-1)+12 t H_1+t^2 \dot{\phi_0} \dot{\phi_1} -3}{t^2}.
\end{align}

Solving the resulting systems, we obtain the following:  
\begin{small}
\begin{align}
& H_0(t)=\frac{7 t^6}{3 t^7+7 c_1}, \label{H0}\\
& H_1(t)=\frac{-51 t^{14}+7 c_2 t^7-980 c_1 \ln (t) t^7+49 c_1{}^2}{7 t \left(3 t^7+7 c_1\right){}^2}, \\
&  \rho_{0}(t) = \frac{147 t^{12}}{\left(3 t^7+7 c_1\right){}^2}-\frac{c_3{}^2}{2 \left(3 t^7+7 c_1\right){}^2}+c_4, \label{rho0}
\end{align}
\end{small}
where $c_1$, $c_2$, $c_3$ 
and $c_4$ are integration constants. 
\begin{small}
\begin{align}
&  \phi_{0}(t)=c_5+\frac{c_3}{7 \sqrt[7]{3} 7^{\frac{6}{7}} c_1{}^{\frac{6}{7}}}  \Bigg[-2 \sin \left(\frac{\pi }{7}\right) \tan ^{-1}\left(\cot \left(\frac{\pi }{7}\right)-\frac{\sqrt[7]{\frac{3}{7}} t \csc
   \left(\frac{\pi }{7}\right)}{\sqrt[7]{c_1}}\right) \nonumber \\
   & +\ln \left(\sqrt[7]{3} 7^{\frac{6}{7}} t+7 \sqrt[7]{c_1}\right)-\cos \left(\frac{\pi }{7}\right) \ln
   \left(3^{\frac{2}{7}} 7^{\frac{5}{7}} t^2-2 \sqrt[7]{3} 7^{\frac{6}{7}} \sqrt[7]{c_1} \cos \left(\frac{\pi }{7}\right) t+7 c_1{}^{\frac{2}{7}}\right) \nonumber \\
   & +\ln \left(3^{\frac{2}{7}} 7^{\frac{5}{7}} t^2+2
   \sqrt[7]{3} 7^{\frac{6}{7}} \sqrt[7]{c_1} \sin \left(\frac{3 \pi }{14}\right) t+7 c_1{}^{\frac{2}{7}}\right) \sin \left(\frac{3 \pi }{14}\right) \nonumber \\
   & -\ln \left(3^{\frac{2}{7}} 7^{\frac{5}{7}} t^2-2 \sqrt[7]{3} 7^{\frac{6}{7}} \sqrt[7]{c_1} \sin \left(\frac{\pi }{14}\right) t+7 c_1{}^{\frac{2}{7}}\right) \sin \left(\frac{\pi }{14}\right) \nonumber \\
   & +2 \tan^{-1}\left(\frac{\sqrt[7]{\frac{3}{7}} \sec \left(\frac{3 \pi }{14}\right) t}{\sqrt[7]{c_1}}+\tan \left(\frac{3 \pi }{14}\right)\right) \cos
   \left(\frac{3 \pi }{14}\right)  \nonumber \\
   & +2 \tan ^{-1}\left(\frac{\sqrt[7]{\frac{3}{7}} t \sec \left(\frac{\pi }{14}\right)}{\sqrt[7]{c_1}}-\tan \left(\frac{\pi
   }{14}\right)\right) \cos \left(\frac{\pi }{14}\right)\Bigg], \label{phi0}
\\
 &  \rho_{1}(t)= -\frac{1}{7 \left(3 t^7+7 c_1\right){}^3} \Bigg[-189 c_7 t^{21}+3465
   t^{19}-1323 c_1 c_7 t^{14}+294 (21 c_1-c_2) t^{12} \nonumber \\
   & +21 \left(c_3 c_6-147 c_1{}^2 c_7\right) t^7+5145 c_1{}^2 t^5+14 c_1 c_3{}^2+c_2 c_3{}^2+49 c_1 c_3
   c_6-2401 c_1{}^3 c_7 \nonumber \\
   &  +4 \left(10290 c_1 t^{12}+18 c_3{}^2 t^7+7 c_1 c_3{}^2\right) \ln (t)\Bigg],
   \end{align}
  \end{small}
where $c_5$, $c_6$ and $c_7$ are integration constants. 

Finally, 
 \begin{tiny}
\begin{align}
 & \phi_{1}(t)=
   -\frac{9 c_3 \, _3F_2\left(\frac{8}{7},\frac{8}{7},2;\frac{15}{7},\frac{15}{7};-\frac{3 t^7}{7 c_1}\right) t^8}{2744 c_1{}^2}+\frac{9 c_3 \,
   _2F_1\left(\frac{8}{7},2;\frac{15}{7};-\frac{3 t^7}{7 c_1}\right) \ln (t) t^8}{343 c_1{}^2}  +\frac{c_2 c_3 t}{343 c_1 \left(3 t^7+7 c_1\right)}+\frac{2 c_3 t}{49 \left(3 t^7+7 c_1\right)}  \nonumber \\
   &-\frac{4 c_3 \, _3F_2\left(\frac{1}{7},\frac{1}{7},2;\frac{8}{7},\frac{8}{7};-\frac{3 t^7}{7 c_1}\right) t}{49 c_1}  +\frac{4 c_3 \, _2F_1\left(\frac{1}{7},2;\frac{8}{7};-\frac{3 t^7}{7 c_1}\right) \ln (t) t}{49 c_1}+c_8+\frac{2 \left(\frac{3}{7}\right)^{\frac{6}{7}} c_2
   c_3 \ln \left(\sqrt[7]{3} 7^{\frac{6}{7}} t+7 \sqrt[7]{c_1}\right)}{2401 c_1{}^{\frac{13}{7}}} \nonumber \\
   &  +\frac{4 \left(\frac{3}{7}\right)^{\frac{6}{7}} c_3 \ln \left(\sqrt[7]{3}
   7^{\frac{6}{7}} t+7 \sqrt[7]{c_1}\right)}{343 c_1{}^{\frac{6}{7}}}+\frac{c_6 \ln \left(\sqrt[7]{3} 7^{\frac{6}{7}} t+7 \sqrt[7]{c_1}\right)}{7 \sqrt[7]{3} 7^{\frac{6}{7}} c_1{}^{\frac{6}{7}}}  -\frac{2 \left(\frac{3}{7}\right)^{\frac{6}{7}} c_2 c_3 \cos ^2\left(\frac{\pi }{14}\right) \ln \left(3^{\frac{2}{7}} 7^{\frac{5}{7}} t^2-2 \sqrt[7]{3} 7^{\frac{6}{7}}
   \sqrt[7]{c_1} \cos \left(\frac{\pi }{7}\right) t+7 c_1{}^{\frac{2}{7}}\right)}{2401 c_1{}^{\frac{13}{7}}} \nonumber \\
   & -\frac{4 \left(\frac{3}{7}\right)^{\frac{6}{7}} c_3 \cos
   ^2\left(\frac{\pi }{14}\right) \ln \left(3^{\frac{2}{7}} 7^{\frac{5}{7}} t^2-2 \sqrt[7]{3} 7^{\frac{6}{7}} \sqrt[7]{c_1} \cos \left(\frac{\pi }{7}\right) t+7
   c_1{}^{\frac{2}{7}}\right)}{343 c_1{}^{\frac{6}{7}}}  -\frac{c_6 \cos ^2\left(\frac{\pi }{14}\right) \ln \left(3^{\frac{2}{7}} 7^{\frac{5}{7}} t^2-2 \sqrt[7]{3} 7^{\frac{6}{7}} \sqrt[7]{c_1}
   \cos \left(\frac{\pi }{7}\right) t+7 c_1{}^{\frac{2}{7}}\right)}{7 \sqrt[7]{3} 7^{\frac{6}{7}} c_1{}^{\frac{6}{7}}} \nonumber \\
   & +\frac{3 \left(\frac{3}{7}\right)^{\frac{6}{7}} c_2 c_3 \csc
   \left(\frac{\pi }{14}\right) \ln \left(3^{\frac{2}{7}} 7^{\frac{5}{7}} t^2+2 \sqrt[7]{3} 7^{\frac{6}{7}} \sqrt[7]{c_1} \sin \left(\frac{3 \pi }{14}\right) t+7
   c_1{}^{\frac{2}{7}}\right) \sin ^2\left(\frac{\pi }{7}\right)}{4802 c_1{}^{\frac{13}{7}}} \nonumber \\
   &  +\frac{3 \left(\frac{3}{7}\right)^{\frac{6}{7}} c_3 \csc \left(\frac{\pi }{14}\right)
   \ln \left(3^{\frac{2}{7}} 7^{\frac{5}{7}} t^2+2 \sqrt[7]{3} 7^{\frac{6}{7}} \sqrt[7]{c_1} \sin \left(\frac{3 \pi }{14}\right) t+7 c_1{}^{\frac{2}{7}}\right) \sin ^2\left(\frac{\pi
   }{7}\right)}{343 c_1{}^{\frac{6}{7}}} \nonumber \\
   & -\frac{4 \left(\frac{3}{7}\right)^{\frac{6}{7}} \tan ^{-1}\left(\cot \left(\frac{\pi }{7}\right)-\frac{\sqrt[7]{\frac{3}{7}} t \csc
   \left(\frac{\pi }{7}\right)}{\sqrt[7]{c_1}}\right) c_2 c_3 \sin \left(\frac{\pi }{7}\right)}{2401 c_1{}^{\frac{13}{7}}}  -\frac{8 \left(\frac{3}{7}\right)^{\frac{6}{7}}
   \tan ^{-1}\left(\cot \left(\frac{\pi }{7}\right)-\frac{\sqrt[7]{\frac{3}{7}} t \csc \left(\frac{\pi }{7}\right)}{\sqrt[7]{c_1}}\right) c_3 \sin
   \left(\frac{\pi }{7}\right)}{343 c_1{}^{\frac{6}{7}}} \nonumber \\
   & -\frac{2 \tan ^{-1}\left(\cot \left(\frac{\pi }{7}\right)-\frac{\sqrt[7]{\frac{3}{7}} t \csc
   \left(\frac{\pi }{7}\right)}{\sqrt[7]{c_1}}\right) c_6 \sin \left(\frac{\pi }{7}\right)}{7 \sqrt[7]{3} 7^{\frac{6}{7}} c_1{}^{\frac{6}{7}}}   -\frac{2
   \left(\frac{3}{7}\right)^{\frac{6}{7}} c_2 c_3 \ln \left(3^{\frac{2}{7}} 7^{\frac{5}{7}} t^2+2 \sqrt[7]{3} 7^{\frac{6}{7}} \sqrt[7]{c_1} \sin \left(\frac{3 \pi }{14}\right) t+7
   c_1{}^{\frac{2}{7}}\right) \sin ^3\left(\frac{\pi }{14}\right)}{2401 c_1{}^{\frac{13}{7}}} \nonumber \\
   & -\frac{4 \left(\frac{3}{7}\right)^{\frac{6}{7}} c_3 \ln \left(3^{\frac{2}{7}} 7^{\frac{5}{7}} t^2+2
   \sqrt[7]{3} 7^{\frac{6}{7}} \sqrt[7]{c_1} \sin \left(\frac{3 \pi }{14}\right) t+7 c_1{}^{\frac{2}{7}}\right) \sin ^3\left(\frac{\pi }{14}\right)}{343
   c_1{}^{\frac{6}{7}}}  -\frac{c_6 \ln \left(3^{\frac{2}{7}} 7^{\frac{5}{7}} t^2+2 \sqrt[7]{3} 7^{\frac{6}{7}} \sqrt[7]{c_1} \sin \left(\frac{3 \pi }{14}\right) t+7 c_1{}^{\frac{2}{7}}\right)
   \sin ^3\left(\frac{\pi }{14}\right)}{7 \sqrt[7]{3} 7^{\frac{6}{7}} c_1{}^{\frac{6}{7}}} \nonumber \\
   &  +\frac{2 \left(\frac{3}{7}\right)^{\frac{6}{7}} c_2 c_3 \ln \left(3^{\frac{2}{7}} 7^{\frac{5}{7}} t^2-2
   \sqrt[7]{3} 7^{\frac{6}{7}} \sqrt[7]{c_1} \cos \left(\frac{\pi }{7}\right) t+7 c_1{}^{\frac{2}{7}}\right) \sin ^2\left(\frac{\pi }{14}\right)}{2401
   c_1{}^{\frac{13}{7}}}  \nonumber \\
   & +\frac{4 \left(\frac{3}{7}\right)^{\frac{6}{7}} c_3 \ln \left(3^{\frac{2}{7}} 7^{\frac{5}{7}} t^2-2 \sqrt[7]{3} 7^{\frac{6}{7}} \sqrt[7]{c_1} \cos \left(\frac{\pi
   }{7}\right) t+7 c_1{}^{\frac{2}{7}}\right) \sin ^2\left(\frac{\pi }{14}\right)}{343 c_1{}^{\frac{6}{7}}}  +\frac{c_6 \ln \left(3^{\frac{2}{7}} 7^{\frac{5}{7}} t^2-2 \sqrt[7]{3} 7^{\frac{6}{7}}
   \sqrt[7]{c_1} \cos \left(\frac{\pi }{7}\right) t+7 c_1{}^{\frac{2}{7}}\right) \sin ^2\left(\frac{\pi }{14}\right)}{7 \sqrt[7]{3} 7^{\frac{6}{7}} c_1{}^{\frac{6}{7}}}  \nonumber \\
   & -\frac{12   \left(\frac{3}{7}\right)^{\frac{6}{7}} \tan ^{-1}\left(\frac{\sqrt[7]{\frac{3}{7}} \sec \left(\frac{3 \pi }{14}\right) t}{\sqrt[7]{c_1}}+\tan \left(\frac{3 \pi
   }{14}\right)\right) c_2 c_3 \cos \left(\frac{\pi }{14}\right) \sin ^2\left(\frac{\pi }{14}\right)}{2401 c_1{}^{\frac{13}{7}}}    -\frac{24 \left(\frac{3}{7}\right)^{\frac{6}{7}} \tan ^{-1}\left(\frac{\sqrt[7]{\frac{3}{7}} \sec \left(\frac{3 \pi }{14}\right) t}{\sqrt[7]{c_1}}+\tan \left(\frac{3 \pi
   }{14}\right)\right) c_3 \cos \left(\frac{\pi }{14}\right) \sin ^2\left(\frac{\pi }{14}\right)}{343 c_1{}^{\frac{6}{7}}}  \nonumber \\
   &  -\frac{2 \left(\frac{3}{7}\right)^{\frac{6}{7}}
   \tan ^{-1}\left(\frac{\sqrt[7]{\frac{3}{7}} \sec \left(\frac{3 \pi }{14}\right) t}{\sqrt[7]{c_1}}+\tan \left(\frac{3 \pi }{14}\right)\right) c_6 \cos
   \left(\frac{\pi }{14}\right) \sin ^2\left(\frac{\pi }{14}\right)}{7 c_1{}^{\frac{6}{7}}}    -\frac{2 \left(\frac{3}{7}\right)^{\frac{6}{7}} c_2 c_3 \ln \left(3^{\frac{2}{7}}
   7^{\frac{5}{7}} t^2-2 \sqrt[7]{3} 7^{\frac{6}{7}} \sqrt[7]{c_1} \sin \left(\frac{\pi }{14}\right) t+7 c_1{}^{\frac{2}{7}}\right) \sin \left(\frac{\pi }{14}\right)}{2401
   c_1{}^{\frac{13}{7}}}  \nonumber \\
   &  -\frac{4 \left(\frac{3}{7}\right)^{\frac{6}{7}} c_3 \ln \left(3^{\frac{2}{7}} 7^{\frac{5}{7}} t^2-2 \sqrt[7]{3} 7^{\frac{6}{7}} \sqrt[7]{c_1} \sin \left(\frac{\pi
   }{14}\right) t+7 c_1{}^{\frac{2}{7}}\right) \sin \left(\frac{\pi }{14}\right)}{343 c_1{}^{\frac{6}{7}}}  -\frac{c_6 \ln \left(3^{\frac{2}{7}} 7^{\frac{5}{7}} t^2-2 \sqrt[7]{3} 7^{\frac{6}{7}}
   \sqrt[7]{c_1} \sin \left(\frac{\pi }{14}\right) t+7 c_1{}^{\frac{2}{7}}\right) \sin \left(\frac{\pi }{14}\right)}{7 \sqrt[7]{3} 7^{\frac{6}{7}} c_1{}^{\frac{6}{7}}} \nonumber \\
   & +\frac{\left(\frac{3}{7}\right)^{\frac{6}{7}} c_6 \cos ^2\left(\frac{\pi }{14}\right) \ln \left(3^{\frac{2}{7}} 7^{\frac{5}{7}} t^2+2 \sqrt[7]{3} 7^{\frac{6}{7}}
   \sqrt[7]{c_1} \sin \left(\frac{3 \pi }{14}\right) t+7 c_1{}^{\frac{2}{7}}\right) \sin \left(\frac{\pi }{14}\right)}{7 c_1{}^{\frac{6}{7}}} \nonumber \\
   & +\frac{4 \left(\frac{3}{7}\right)^{\frac{6}{7}} \tan ^{-1}\left(\frac{\sqrt[7]{\frac{3}{7}} \sec \left(\frac{3 \pi }{14}\right) t}{\sqrt[7]{c_1}}+\tan \left(\frac{3 \pi
   }{14}\right)\right) c_2 c_3 \cos ^3\left(\frac{\pi }{14}\right)}{2401 c_1{}^{\frac{13}{7}}}+\frac{8 \left(\frac{3}{7}\right)^{\frac{6}{7}} \tan
   ^{-1}\left(\frac{\sqrt[7]{\frac{3}{7}} \sec \left(\frac{3 \pi }{14}\right) t}{\sqrt[7]{c_1}}+\tan \left(\frac{3 \pi }{14}\right)\right) c_3 \cos
   ^3\left(\frac{\pi }{14}\right)}{343 c_1{}^{\frac{6}{7}}} \nonumber \\
   & +\frac{2 \tan ^{-1}\left(\frac{\sqrt[7]{\frac{3}{7}} \sec \left(\frac{3 \pi }{14}\right)
   t}{\sqrt[7]{c_1}}+\tan \left(\frac{3 \pi }{14}\right)\right) c_6 \cos ^3\left(\frac{\pi }{14}\right)}{7 \sqrt[7]{3} 7^{\frac{6}{7}} c_1{}^{\frac{6}{7}}}+\frac{4
   \left(\frac{3}{7}\right)^{\frac{6}{7}} \tan ^{-1}\left(\frac{\sqrt[7]{\frac{3}{7}} t \sec \left(\frac{\pi }{14}\right)}{\sqrt[7]{c_1}}-\tan \left(\frac{\pi
   }{14}\right)\right) c_2 c_3 \cos \left(\frac{\pi }{14}\right)}{2401 c_1{}^{\frac{13}{7}}} \nonumber \\
   & +\frac{8 \left(\frac{3}{7}\right)^{\frac{6}{7}} \tan
   ^{-1}\left(\frac{\sqrt[7]{\frac{3}{7}} t \sec \left(\frac{\pi }{14}\right)}{\sqrt[7]{c_1}}-\tan \left(\frac{\pi }{14}\right)\right) c_3 \cos
   \left(\frac{\pi }{14}\right)}{343 c_1{}^{\frac{6}{7}}}+\frac{2 \tan ^{-1}\left(\frac{\sqrt[7]{\frac{3}{7}} t \sec \left(\frac{\pi
   }{14}\right)}{\sqrt[7]{c_1}}-\tan \left(\frac{\pi }{14}\right)\right) c_6 \cos \left(\frac{\pi }{14}\right)}{7 \sqrt[7]{3} 7^{\frac{6}{7}}
   c_1{}^{\frac{6}{7}}},
   \end{align}
   \end{tiny}
   where $c_8$ is an integration constant.

In Figure \ref{fig:Asymptotic}, $H(t)$, $\phi(t)$, and $\rho$ are presented vs. the asymptotic expansions $H_0(t)$, $\phi_0(t)$, and $\rho_0$, defined, respectively, by \eqref{H0}, \eqref{phi0}, and \eqref{rho0} for the initial conditions $\rho_{m}(0.1)=1, H(0.1)=1,\dot{ \phi}(0.1)=1, \phi (0.1)=1$ (arbitrarily chosen), and the parameter values $c_1= 9.571429 \times{10}^{-7}$, $c_3= 7\times { 10}^{-6}$, $c_4=-1.5$, $c_5= 0.881801$, $\mu =1.1, \xi =\frac{1}{6},V(\phi)=\Lambda =1$ for $\varepsilon=0.1$. The accuracy of the asymptotic approximation when $\varepsilon\rightarrow 0$ is illustrated as follows: 
   \begin{figure}[h]
    \centering
       \includegraphics[scale=1.0]{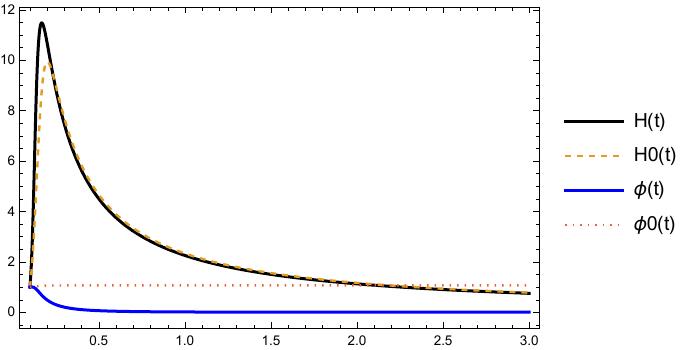}
       \includegraphics[scale=1.0]{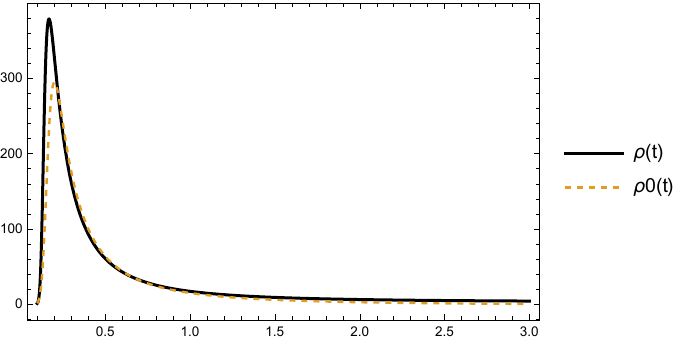}
       \caption{Numerical solutions $H(t)$, $\phi(t)$, and $\rho$ vs. the asymptotic expansions $H_0(t)$, $\phi_0(t)$, and $\rho_0(t)$, defined, respectively, by \eqref{H0}, \eqref{phi0}, and \eqref{rho0} (valid for $\xi=0$) for the initial conditions $\rho_{m}(0.1)=1, H(0.1)=1, \dot{\phi}(0.1)=1, \phi (0.1)=1$, and the parameter values $c_1= 9.571429 \times{10}^{-7}$, $c_3= 7\times {10 }^{-6}$, $c_4=-1.5$, $c_5= 0.881801$, $\mu =1.1, \xi =\frac{1}{6},V(\phi)=\Lambda=1$ for $\varepsilon=0.1$. The accuracy of the asymptotic approximation is illustrated when $\varepsilon\rightarrow 0$, even though the analytical solutions are calculated for $\xi=0$.}
       \label{fig:Asymptotic}
   \end{figure}
  
\subsubsection{Asymptotic Expansions for $\xi \to 0$}
We consider the following expansions: 
\begin{align*}
    \rho = \rho_0 + \xi \rho_1 + \mathcal{O}(\xi^2), \;
    H = H_0 + \xi H_1 + \mathcal{O}(\xi^2), \;
    \phi = \phi_0 + \xi \phi_1 + \mathcal{O}(\xi^2),
\end{align*}
where we assume that the non-minimal coupling parameter is a perturbation parameter, $0<\xi \ll 1$.

Replacing them in \eqref{NRay}--\eqref{NFried}, we have the following: 
\begin{small}
\begin{align}
    & -\frac{1}{t^3}\Big[-3 t^3 H_0 \dot{\phi_0}^2 +18 t^3 H_0^3 + 3 \mu t^2 H_0^2 -39 t^2 H_0^2 - 3(\mu - 1)(4\mu -11) t H_0  \nonumber \\
    & + 3(\mu - 2)(\mu -1)^2 + t^3 \dot{\rho_0} + \mu t^2 \dot{\phi_0}^2 - t^2 \dot{\phi_0}^2\Big]  \nonumber \\
    & - \frac{\xi}{t^3}\Big[54 t^3 H_0^2 H_1 + 6(\mu -13)^2 H_0 H_1 -6 t^3 H_0 \dot{\phi_0} \dot{\phi_1} + 18 t^3 H_0 \dot{\phi_0}^2 \nonumber \\
    & - 18 t^3 H_0^3 \phi_0^2 - 3 t^2 H_0^2 \phi_0 \left((\mu-13)\phi_0 + 12 t \dot{\phi_0}\right)  \nonumber \\
    & - 6 t^2 H_0 \phi_0 \left((\mu -1)\dot{\phi_0} + 3 t V'(\phi_0)\right) \nonumber \\
    & + 3(\mu - 1)(4\mu - 11) t H_0 \phi_0^2 - 3 t H_1 \left((\mu -1)(4\mu - 11) + t^2 \dot{\phi_0}^2\right)  \nonumber \\
    & + t^3 \dot{\rho_1} + 2(\mu -1) t^2 \dot{\phi_0} \dot{\phi_1} - 6 \mu t^2 \dot{\phi_0}^2 + 6 \mu \phi_0 V'(\phi_0) \nonumber \\
    & - 6 t^2 \phi_0 V(\phi_0) + 6 t^2 \dot{\phi_0}^2 - 3 \mu^3 \phi_0^2 + 6 \mu^2 t \phi_0 \dot{\phi_0} +12 \mu^2 \phi_0^2 \nonumber \\
    & - 12 \mu t \phi_0 \dot{\phi_0} - 15 \mu \phi_0^2 + 6 t \phi_0 \dot{\phi_0} + 6 \phi_0^2 \Big] = 0, 
\\
    & \frac{1}{t^2} \Big[ -t^2 \dot{H_0} - 3 t^2 H_0^2- 2 \mu  t H_0 + 8 t H_0 + (\mu -2)(\mu -1) \Big] \nonumber \\
    & +\frac{\xi}{t^2} \Big[-6 t^2 H_0 H_1 + 2 t^2 H_0 \phi_0 \dot{\phi_0} \nonumber \\
    & - t^2 \dot{H_1} - 2 (\mu -4) t H_1 + 2 t^2 \phi_0 V'(\phi_0) - 2 t^2 \dot{\phi_0}^2+2 (\mu -1) t \phi_0 \dot{\phi_0} \Big] = 0, 
\\
    & -\frac{1}{t^2} \Big[3 t^2 H_0 \dot{\phi_0} + t^2 V'(\phi_0) + t^2 \ddot{\phi_0} - \mu  t \dot{\phi_0}+ t \dot{\phi_0}\Big]  \nonumber \\
    & -\frac{\xi}{t^2} \Big[3 t^2 H_0 \dot{\phi_1} + 6 \phi_0 ((\mu -2) (\mu -1) - t H_0 (t H_0 +2 \mu -8)) \nonumber \\
    &   + 3 t^2 H_1 \dot{\phi_0} + t^2 \phi_1 V''(\phi_0)  + t^2 \ddot{\phi_1} - \mu  t \dot{\phi_1} + t \dot{\phi_1} \Big] = 0.
\end{align}
\end{small}
 
Assuming a constant potential $V(\phi)=\Lambda$, the zeroth and first-order equations are as follows:

\textbf{Zero order}, $\mathcal{O}(\xi^0)${, as follows:} 

 \begin{align}
   & \dot{\rho_0}=-\frac{(-3 t H_0+\mu -1) \left(t \left(t \dot{\phi_{0}}^2-3 H_0 (2 t H_0+\mu -5)\right)+3 (\mu -2)(\mu -1)\right)}{t^3}, \\
  &  \dot{H_0}=\frac{t H_0 (-3 t H_0-2 \mu +8)+(\mu -2) (\mu -1)}{t^2}, \\
  &  \ddot{\phi _0}=\frac{(-3 t H_0+\mu -1) \dot{\phi_0}}{t},
 \end{align}

with the constraint to order zero specified as follows: 
 \begin{equation}
     -\frac{(\mu -1) H_0}{t}+H_0^2+\frac{1}{6} \left(-2 (\Lambda + \rho_{0})-\dot{\phi_0}^2\right)=0.
 \end{equation}

The last two equations have the general solution, as follows: 
\begin{small}
\begin{align}
   & H_0(t)=\frac{-2 \mu -\frac{2 c_1 r t_0^r}{t^r+c_1 t_0^r}+r+9}{6 t}, \label{eq94}\\
   & \phi_{0}(t)= \frac{2 c_2 \left(\frac{t}{t_0}\right)^{\frac{1}{2} (4 \mu +r-9)} \,
   _2F_1\left(1,\frac{r+4 \mu -9}{2 r};\frac{3 r+4 \mu -9}{2 r};-\frac{\left(\frac{t}{t_0}\right)^r}{c_1}\right)}{c_1 (4 \mu +r-9)}+\phi_c, \label{eq95}
\end{align}
\end{small}
where $c_1$, $c_2$, and $\phi_c$ are integration constants. 
Therefore, the first equation reduces to the following: 
 
\begin{small}
\begin{align}
\dot{\rho_0}= & -\frac{\left(2 \mu +r \left(-\frac{1}{2}+\frac{c_1 t_0^r}{t^r+c_1 t_0^r}\right)-\frac{11}{2}\right) }{6 t^3} \nonumber \\
&  \times \Bigg(20 \mu ^2-75 \mu
   -\frac{r^2 \left(t^r-c_1 t_0^r\right){}^2}{\left(t^r+c_1 t_0^r\right){}^2}+\frac{(\mu -3) r \left(t^r-c_1 t_0^r\right)}{t^r+c_1
   t_0^r} +\frac{6 c_2{}^2 t^{4 \mu +r-9} t_0^{-4 \mu +r+9}}{\left(t^r+c_1 t_0^r\right){}^2}+90\Bigg),
\end{align} 
\end{small}
which is integrable, giving us the following solution: 
\begin{small}
\begin{align}
\rho_0(t)= & \frac{r^3}{24 t^2}+\frac{r^2 \left(-5 \mu -\frac{8 c_1 t^r t_0^r}{\left(t^r+c_1 t_0^r\right){}^2}+14\right)}{24 t^2}-\frac{c_2{}^2 t^{4 \mu
   +r-11} t_0^{-4 \mu +r+9}}{2 \left(t^r+c_1 t_0^r\right){}^2} \nonumber\\
   & -\frac{r \left(16 \mu ^2-92 \mu +\frac{8 (5 \mu -12) t^r}{t^r+c_1
   t_0^r}+153\right)}{24 t^2}  +\frac{5 (4 \mu -11) (\mu  (4 \mu -15)+18)}{24 t^2} -\Lambda.
\end{align}
\end{small}
where $c_3 = - \Lambda$ is an integration constant. 

\textbf{{First order,} $\mathcal{O}(\xi)${:} }\vspace{-4pt}
\begin{small}
 \begin{align}
& \dot{\rho_1}=\frac{1}{t^3} \Bigg[3 t H_1 \left(t \left(t \dot{\phi_0}^2-2 H_0 (9 t H_0(t)+\mu -13)\right)+(\mu -1) (4 \mu -11)\right) \nonumber \\
   & +(-3 t H_0+\mu -1) \Big(-6 t \phi_0 (2 t H_0+\mu -1) \dot{\phi_0} \nonumber \\
   & +3 \phi_0^2 ((\mu -2) (\mu -1)-t H_0 (2 t H_0+\mu -5))  +2 t^2 \dot{\phi_0} \left(3 \dot{\phi_0}-\dot{\phi_1}\right)\Big)\Bigg], \label{eq98}
   \end{align}
   \begin{align}
& \dot{H_1}= -\frac{2\left(H_1 (3 t H_0+\mu -4)-\phi_0 (t H_0+\mu -1) \dot{\phi_0}+t \dot{\phi_0}^2\right)}{t}, \\
& \ddot{\phi_1}= -\frac{t \left((3 t H_0-\mu +1)\dot{\phi_1}+3 t H_1 \dot{\phi_0}\right)+6 \phi_0 ((\mu -2) (\mu -1)-t H_0 (t H_0+2 \mu -8))}{t^2},
 \end{align}
\end{small}
with the restriction in the first order, as follows: 
\begin{small}
 \begin{align}
  \rho_{1}=\frac{H_1(6 t H_0-3 \mu +3)-3 H_0 \dot{\phi_0}\left( \phi_0 (t H_0-\mu +1)+2 t \dot{\phi_0}\right)-t \dot{\phi_0}\dot{\phi_1}}{t}. 
 \end{align}
\end{small}

This expression can be used to define $\rho_1$, such that Equation \eqref{eq98} is a compatibility condition that is satisfied for all $t$. Hence, we have a reduced system, as follows:  
 \begin{small}
\begin{align}
    \ddot{\phi_1}= & -\frac{6 c_2 H_1 t^{\frac{1}{2} (4 \mu +r-9)} t_0^{\frac{1}{2} (-4 \mu +r+9)}+\dot{\phi_1} \left((-4 \mu +r+11) t^r-c_1 (4 \mu +r-11)
   t_0^r\right)}{2 t \left(t^r+c_1 t_0^r\right)} \nonumber \\
   & +\frac{c_2 t^{\frac{1}{2} (4 \mu +r-9)-2} t_0^{-2 \mu -\frac{r}{2}}  \, _2F_1\left(1,\frac{r+4 \mu -9}{2 r};\frac{3 r+4 \mu -9}{2 r};-\frac{t^r t_0^{-r}}{c_1}\right)}{3 c_1 (4 \mu +r-9) \left(t^r+c_1
   t_0^r\right){}^2} \nonumber \\
   & \times\Bigg(t_0^{9/2} \left(4 (69-14 \mu ) \mu +r^2+(8
   \mu -30) r-423\right) t^{2 r} \nonumber\\
   & -2 c_1 \left(4 \mu  (14 \mu -69)+r^2+423\right) t^r t_0^{r+\frac{9}{2}}  +c_1{}^2 \left(4 (69-14 \mu ) \mu +r^2+(30-8 \mu ) r-423\right) t_0^{2
   r+\frac{9}{2}}\Bigg) \nonumber\\
   & +\frac{\Lambda  t_0^2 \left(56 \mu ^2-276 \mu -\frac{4 r^2 t^{2 r}}{\left(t^r+c_1 t_0^r\right){}^2}-r^2+8 \mu  r+\frac{4 r (-4 \mu +r+15) t^r}{t^r+c_1
   t_0^r}-30 r+423\right)}{6 t^2},
   \end{align}
\begin{align}
   \dot{H_1} & =\frac{H_1 \left(-1+r \left(-1+\frac{2 c_1 t_0^r}{t^r+c_1 t_0^r}\right)\right)}{t} \nonumber \\
   & +\frac{2 c_2{}^2 t_0^{9-4
   \mu } t^{4 \mu +r-11} \left((4 \mu +r+3) t^r-c_1 (-4 \mu +r-3) t_0^r\right) \, _2F_1\left(1,\frac{r+4 \mu -9}{2 r};\frac{3 r+4 \mu -9}{2 r};-\frac{t^r
   t_0^{-r}}{c_1}\right)}{3 c_1 (4 \mu +r-9) \left(t^r+c_1 t_0^r\right){}^2} \nonumber \\
   & -\frac{c_2 \Lambda  t^{\frac{1}{2} (4 \mu +r)-\frac{13}{2}} t_0^{\frac{1}{2} (-4 \mu +r+13)}
   \left((4 \mu +r+3) t^r-c_1 (-4 \mu +r-3) t_0^r\right)}{3 \left(t^r+c_1 t_0^r\right){}^2}  -\frac{2 c_2{}^2 t^{4 \mu +r-11} t_0^{-4 \mu +r+9}}{\left(t^r+c_1
   t_0^r\right){}^2}.
\end{align}
 \end{small}

The second equation is integrable, giving us the following: \vspace{-11pt}
 
\begin{small}
\begin{align}
    H_1(t)= & -\frac{c_2{}^2 (-4 \mu +r-3) (4 \mu +r-9) \Gamma \left(\frac{4 \mu -9}{r}\right) t^{4 \mu +r-10} t_0^{-4 \mu -r+9} \, _2\tilde{F}_1\left(1,\frac{4 \mu
   -9}{r};\frac{r+4 \mu -9}{r};-\frac{\left(\frac{t}{t_0}\right)^r}{c_1}\right)}{36 r \left(\left(\frac{t}{t_0}\right)^r+c_1\right){}^2} \nonumber \\
   & -\frac{c_2{}^2 (-4 \mu +r+11) \Gamma
   \left(\frac{r+4 \mu -9}{r}\right) t^{2 (2 \mu +r)-10} t_0^{-4 \mu -2 r+9} \, _2\tilde{F}_1\left(1,\frac{r+4 \mu -9}{r};\frac{2 r+4 \mu
   -9}{r};-\frac{\left(\frac{t}{t_0}\right)^r}{c_1}\right)}{3 c_1 r \left(\left(\frac{t}{t_0}\right)^r+c_1\right){}^2} \nonumber \\
   & +\frac{c_2 t^{\frac{1}{2} (4 \mu +r)-10} t_0^{-2 (2
   \mu +r)}}{36
   \left(\left(\frac{t}{t_0}\right)^r+c_1\right){}^2} \Bigg[\frac{2 c_2 t_0^9 t^{\frac{1}{2} (4 \mu +r)} \left(12 (-4 \mu +r+11) t^r+c_1 (-4 \mu +r-3) (4 \mu +r-9) t_0^r\right)}{c_1 (4 \mu +r-9)} 
   \nonumber \\
   & \times\, _2F_1\left(1,\frac{r+4 \mu -9}{2
   r};\frac{3 r+4 \mu -9}{2 r};-\frac{\left(\frac{t}{t_0}\right)^r}{c_1}\right) -12 \Lambda  (-4 \mu +r+11) t^{r+\frac{9}{2}} t_0^{\frac{1}{2} (4 \mu
   +r+13)} \nonumber \\
   &+12 c_1 \Lambda  t^{9/2} (4 \mu +r-11) t_0^{\frac{1}{2} (4 \mu +3 r+13)}  +\frac{72 c_2 t_0^{r+9} t^{\frac{1}{2} (4 \mu +r)}}{9-4 \mu }\Bigg]  \nonumber\\
   &+c_4 \left(1+\frac{\left(\frac{t}{t_0}\right)^r}{c_1}\right){}^{-\frac{r+1}{r}} \left(1+c_1
   \left(\frac{t_0}{t}\right)^r\right){}^{\frac{1}{r}-1}, \label{eq104}
\end{align} 
\end{small}
where $c_4$ is a constant of integration.

The equation for $\phi_1$ can be solved in quadratures, as follows: 

\begin{small}
\begin{align}
   \phi_{1}(t)= & \int _1^t \; e^{\left(\int _1^{\mu_3}\left(\frac{\mu -1}{\mu_1}-3 H_0(\mu_1)\right)d\mu_1\right)}\times \Bigg[c_5+\int _1^{\mu_3}\frac{e^{\left(-\int
   _1^{\mu_2}\left(\frac{\mu -1}{\mu_1}-3 H_0(\mu_1)\right)d\mu_1\right)}}{\mu_2^2} \Bigg(-3 H_1(\mu_2) \dot{\phi_0}(\mu_2) \mu_2^2 \nonumber \\
   & -6 ((\mu -2) (\mu -1)-H_0(\mu_2) \mu_2 (2 \mu
   +H_0(\mu_2) \mu_2-8)) \phi_{0}(\mu_2)\Bigg) d\mu_2\Bigg] d\mu_3+c_6,
\end{align} 
\end{small}
where $c_5$ and $c_6$ are integration constants and $H_0(\mu_1)$ is calculated using \eqref{eq94}; $\dot{\phi_0}(\mu_2)$ is calculated by taking the derivative with respect to $t $ in \eqref{eq95}, and $\phi_{0}(\mu_2)$ is evaluated using~\eqref{eq95}. Finally, the expression for $H_1(\mu_2) $ is \eqref{eq104}.

	\section{Conclusions}
\label{conclusions}
In this paper, we studied corrections to the Friedman and Klein--Gordon equations based on the formalism of fractional calculus using Caputo's derivative. Thus, we presented fractional calculus as a viable and attractive option for applications to the theory of gravity. Based on the diverse applications of fractional calculus in physics and engineering as shown in the paper, this work presents a didactic exposition of the main aspects of fractional calculus, showing the different approaches that exist and definitions of fractional derivatives.

This paper analyzes the fractional theory in cosmological models with a scalar field with a coupling constant, $\xi$. We qualitatively analyze the case of $\xi = 0$ and the general case of $\xi \neq 0$. Using different formulations of dynamical systems, we examine their equilibrium points and determine how the fractional term affects their stability---exposing the complex behavior presented by specific equilibrium points in the system's phase space and numerical solutions to different values of $\xi$ and $\mu$. We use methods from perturbation theory to analyze the cosmological behavior, obtaining solutions through two asymptotic expansions proposed for the system functions around the parameters $\mu=1$ and $\xi=0$. In these cases, the model recovers standard cosmology and the theory of minimal coupling to gravity, with an excellent approximation to the numerical solutions of the system.

In the paper's introductory Section \ref{intr}, we discuss the challenges of traditional calculations in modeling power-law phenomena. We then present the various applications of fractional calculus, highlighting its effectiveness in recent studies in cosmology. This makes it a viable option to explore gravity theory further. We also introduce the research questions, pose the fractional formulation of the gravity problem, and outline the paper's general and specific objectives. We then describe the methodology used to develop the paper. Finally, we emphasize the scientific novelty of this research. Specifically, we mention that our analysis of cosmology with a scalar field with conformal and non-minimal coupling to gravity in the fractional context seeks to generalize previous results in the framework of the fractional formulation of gravity, making it a novel contribution.

Section \ref{cap3} summarizes the main results of fractional calculus, mentioning some approaches to possible fractional derivatives, such as Gr\"unwald–Letnikov, Riemann--Liouville and Caputo.  We review the known rules of differentiation in the fractional context under the different approaches and show the fractional derivative of known functions. Emphasis is placed on viewing integration as the inverse operation of differentiation, which facilitates the definition of fractional integration. This is achieved by generalizing Cauchy's formula for iterated integrals, thereby enabling us to define the fractional derivative. The Liouville, Riemann, Liouville--Caputo, and Caputo derivatives are defined. Finally, we discuss what fractional differential equations are, using the Wolfram Language 13.3 tools \citep{Mathematica}, and we solve the classical problem of the harmonic oscillator. {We also explore its quantum version and the relation between fractional calculus and $q$-deformed Lie algebras. We solve a physical problem of a point particle that moves under the effect of a friction force in the fractional context from different approaches using Wolfram Language 13.3 \citep{Mathematica}.} 

It is important to emphasize that fractional differential equations, which involve fractional derivatives, generalize ordinary differential equations. Fractional differential equations have been widely used in engineering, physics, chemistry, biology, and other fields involving relaxation and oscillation models. Recently, they have been used in cosmology. For this reason, since the version of Wolfram Language 13.1, two basic operators for fractional calculation were implemented, the functions \textbf{{FractionalD} 
}and \textbf{{CaputoD}}. The algorithms of the \textbf{{MittagLefflerE}} functions have been updated, as they are of crucial importance in the theory of fractional calculus, and the powerful \textbf{{DSolve}} function was heavily updated in version 13.1 to support FDE. This paper solves several fractional differential equations using the \textbf{{DSolve}} function of Wolfram Language 13.3 \citep{Mathematica}. The solutions to such equations are generally given in terms of the function \textbf{{MittagLefflerE}} and the primary function for fractional calculus applications. Its role in fractional differential equation solutions is similar in importance to the \textbf{{Exp}} function for ODE solutions: any ODE with constant coefficients can be solved in terms of Mittag-Leffler functions.

According to the analysis carried out in this paper, the behavior of the fractional harmonic oscillator is very similar to the behavior of the ordinary damped harmonic oscillator. These examples demonstrate that the order of a fractional differential equation can be used as a control parameter to model some complicated systems.

In Section \ref{cap4}, cosmological models are introduced within the fractional formulation of the gravity framework, and previous results are discussed where the coupling is minimal, $\xi=0.$ The paper's main result is the application of the tools defined in Section \ref{cap3} to fractional cosmologies with non-minimal coupling $\xi\neq0.$ Different dynamical systems and relevant cases, such as the invariant set $\Omega_{\Lambda}= 0$, are studied. Some numerical integration methods from the specialized Mathematica software (Version 13.3) \cite{Mathematica}
 are used to study the behavior of the solutions of the systems of differential equations coming from the cosmological model, and the stability of the critical points is studied using appropriate tools.

\textls[-15]{In Sections \ref{SECTAlternative_Formulation} and \ref{invariantset-0-CC}, we found a great technical difficulty when analyzing the stability of the curves and critical points $P_6(x_c): \left(x_c, -\frac{1} {\sqrt{\xi }},\Omega_{\Lambda 0}, 0\right)$,
$P_7(x_c): \left(x_c, -\frac{1}{\sqrt{\xi }}, 0, 0\right)$,
$P_8: \left(0, -\frac{1}{\sqrt{\xi }}, \Omega_{\Lambda 0}, 0\right)$,
$P_9: \left(0, -\frac{1}{\sqrt{\xi }}, 0, 0\right)$,
$P_{10}(x_c): \left(x_c, \frac{1}{\sqrt{\xi }}, \Omega_{\Lambda 0}, 0\right)$,
$P_{11}(x_c): \left(x_c, \frac{1}{\sqrt{\xi }}, 0, 0\right)$,
$P_{12}: \left(0, \frac{1}{\sqrt{\xi }}, \Omega_{\Lambda 0}, 0\right)$,
$P_{13}:\left(0, \frac{1}{\sqrt{\xi }}, 0, 0\right)$. Due to the complexity of the stability analysis of these sets of equilibrium points, numerical solution methods have to be used. However, partial stability information can be obtained by computing the eigenvalues using the time variable $\tau=\ln (a/a_0)$ and taking the limit when $\alpha \rightarrow 0$.}

In the study of curves,  

$\left(0, -\frac{1}{\sqrt{\xi (12 \xi +1)}}, \Omega_{\Lambda 0}, -\mu \mp \sqrt{\mu (2 \mu -11)+18}+4\right)$

and

$\left(0, \frac{1}{\sqrt{\xi (12 \xi +1)}}, \Omega_{\Lambda 0}, -\mu \mp \sqrt{\mu (2 \mu - 11)+18}+4\right)$

The technical difficulties lie in the parameter of the equation for the state of matter, which is complex and infinite. Furthermore, stability has to be determined numerically since the Jacobian matrix has infinite entries. These situations are due to the non-differentiability of the flow when $\alpha\rightarrow 0$ and \begin{small}
$ G:= (\xi (12 \xi +1) y^2-1) \rightarrow 0.$
\end{small} The last case is even more complex since, in a neighborhood of $G\equiv 0$, the direction of the flow changes, entering the context of non-smooth mechanics. In all these cases, the stability analysis of these curves/sets will be left for future research since they cannot be implemented with the techniques developed in this paper.

Non-smooth mechanics is a modeling approach that does not require the temporal evolution of positions and velocities to be smooth functions. Due to potential impacts, the speeds of the mechanical system may experience sudden jumps at certain moments to comply with kinematic restrictions. Consider, for example, a rigid model of a ball falling to the ground. Just before the impact between the ball and the ground, the ball has a pre-impact velocity that does not disappear. At the instant of impact, the velocity must jump to a post-impact velocity of at least zero, or penetration will occur. Non-smooth mechanical models are often used in contact dynamics. 

A review of research on non-smooth dynamical systems from the point of view of applications is provided in reference \citep{POPP2000765}. Two types of lack of smoothness are considered: impact and friction. A list of basic models is provided for each type, application areas are specified, and research results are reviewed. The literature on non-smooth dynamical systems will be reviewed in future work to resolve the difficulties above.

Research hypotheses are implemented in this research, and the paper's general and specific objectives are satisfactorily addressed. It is possible to obtain relevant information on the flow properties associated with several autonomous systems of ordinary differential equations from a cosmological context through qualitative techniques from the dynamical systems theory. The review of the literature on dynamical systems and fractional calculus is exhaustive. The work tools used in the paper were appropriately selected, identifying the use of the Caputo fractional derivative and its application in formulating a fractional cosmological action and its respective variational calculation. The main results of fractional calculus for the qualitative analysis of cosmological models are presented in a didactic manner. 

The analysis of cosmologies featuring a scalar field with conformal and non-minimal coupling to gravity within a fractional context is novel, generalizing previous results within the framework of the fractional formulation of gravity. From a mathematical point of view, this approach describes the phase space of the models, obtaining qualitative results and explicit and approximated exact solutions with high numerical precision. This represents a different and attractive scenario due to its high level of applicability in describing natural phenomena.

\vspace{6pt} 

\section*{Author contributions}
Conceptualization, K.M., A.D.M., and G.L.; methodology, G.L.; software,  K.M., A.D.M., and G.L.; validation,  K.M., A.D.M., and G.L.; formal analysis, K.M., A.D.M., G.L., C.M., and A.P.; investigation, K.M., A.D.M., G.L., C.M., and A.P.; resources, G.L., C.M., and A.P.;  writing---original draft preparation, G.L.; writing---review and editing, G.L., C.M., and A.P.; visualization,  K.M., A.D.M., and G.L.; supervision, G.L. and {A.D.M.}
; project administration, G.L. and K.M.; funding acquisition, K.M., A.D.M., G.L., C.M., and A.P. All authors have read and agreed to the published version of the manuscript.

\section*{Funding}
K.M. and G.L. were funded by Vicerrectoría de Investigación y Desarrollo Tecnológico (VRIDT) at Universidad Católica del Norte (UCN) through Concurso Fondo de Tesis 2023 UCN, Resolución VRIDT N°063/2023. VRIDT-UCN funded G.L. through Resolución VRIDT No. 026/2023, Resolución VRIDT No. 027/2023, Proyecto de Investigación Pro Fondecyt Regular 2023 (Resolución VRIDT N°076/2023) and Resolución VRIDT N°09/2024. A.D.M. was supported by Agencia Nacional de Investigación y Desarrollo---ANID  Subdirección de Capital Humano/Doctorado Nacional/año 2020 folio 21200837, Gastos operacionales proyecto de Tesis/2022 folio 242220121 and VRIDT-UCN. C. M. was supported by ANID Subdirección de Capital Humano/Doctorado Nacional/año 2021- folio 21211604.  A.P. acknowledges the funding of VRIDT-UCN  through Resoluci\'{o}n VRIDT No. 096/2022 and Resoluci\'{o}n VRIDT No. 098/2022. G.L., A.D.M., and A.P. have the financial support of ANID through Proyecto Fondecyt Regular 2024,  Folio 1240514, Etapa 2024.

\section*{Data availability}

No new data were created or analyzed in this study. Data sharing is not applicable. 

\acknowledgments{We are thankful for the support of Núcleo de Investigación Geometría Diferencial y Aplicaciones, Resolución VRIDT No. 096/2022.  A.P. thanks Nikolaos
 Dimakis and the Universidad de La Frontera for the hospitality provided while part of this work was carried out. We thank Bayron Micolta-Riascos, Antonella Cid, and Juan Magaña for their helpful comments and suggestions. G.L. would like to express his gratitude toward faculty member Alan Coley and staff members Anna Maria Davis, Nora Amaro, Jeanne Clyburne, and Mark Monk for their warm hospitality during the implementation of the final details of the research in the Department of Mathematics and Statistics at Dalhousie University.}

\section*{Conflicts of interest}
We declare no conflicts of interest. The funders had no role in the design of the study; in the collection, analyses, or interpretation of data; in the writing of the manuscript; or in the decision to publish the results.

\appendix
\section[\appendixname~\thesection]{Special Functions}
\label{app1}
This appendix presents several special functions that appear recurrently in these topics.

\subsection{Gamma Function}

The Gamma function extends the factorial function; that is, it extends a function of integers to a function of real or complex numbers. Namely,
for natural numbers, the factorial of $n$ is defined as follows: \vspace{-11pt}
\begin{equation}
    n!=1\times 2\times 3\times \cdots \times n=\prod_{j=1}^{n}j.
    \label{1.1.1.1}
\end{equation}\vspace{-5pt}

On the other hand, the Gamma function can be written as follows: 
\begin{equation}
    \Gamma (z)=\int_{0}^{\infty}t^{z-1}e^{-t}\,dt,
    \label{1.1.1.2}
\end{equation}
which works for values of the complex number $z$ with $\Re(z)>0$. The following can be observed: 
\begin{equation}
    \Gamma (z+1)=z\Gamma (z),
    \label{1.1.1.3}
\end{equation}
and also
\begin{equation}
    \Gamma (n+1)=n!.
    \label{1.1.1.4}
\end{equation}

Additionally, one can write the binomial coefficient in terms of the Gamma function, as follows:
\begin{equation}
\begin{pmatrix}
z\\
v
\end{pmatrix}
=\frac{z!}{v!(z-v)!}=\frac{\Gamma (z+1)}{\Gamma (v+1)\Gamma (z+1-v)}.
\label{1.1.1.6}
\end{equation}

This is added to the reflection formula, as follows: \vspace{-4pt}
\begin{equation}
    \Gamma (z)\Gamma (1-z)=\frac{\pi}{\sin{(\pi z})},
    \label{1.1.1.7}
\end{equation}
and to the incomplete gamma function, as follows: \vspace{-6pt}
\begin{equation}
    \Gamma (a,z)=\int_{z}^{\infty}t^{a-1}e^{-t}dt,
    \label{1.1.1.8}
\end{equation}
where
\begin{equation}
    \Gamma (a,0)=\Gamma (a).
    \label{1.1.1.9}
\end{equation}
\subsection{Mittag-Leffler Functions}
From the Maclaurin series expansion of the exponential, we have the following: 
\begin{equation}
    e^z=\sum_{n=0}^{\infty}\frac{z^n}{n!},
    \label{1.1.2.1}
\end{equation}

We can replace the factorial with the Gamma function, as follows:
\begin{equation}
    e^z=\sum _{n=0}^{\infty}\frac{z^n}{\Gamma (n+1)}.
    \label{1.1.2.2}
\end{equation}

This can then be extended as follows:\vspace{-6pt}
\begin{equation}
    E(\alpha,z)=\sum_{n=0}^{\infty}\frac{z^n}{\Gamma (\alpha n+1)},
    \label{1.1.2.3}
\end{equation}
where $\alpha$ is an arbitrary real number that is greater than zero.

In fractional calculus, this function is of similar importance to the exponential function in standard calculus. For some values of $\alpha$ and functions of $z$, already-known functions can be obtained, as follows:
\begin{equation}
    E(2,-z^2)=\cos{z}, \hspace{20px} E(1/2,z^{1/2})=e^z\left[1+\text{erf}( z^{1/2})\right],
    \label{1.1.2.4}
\end{equation}
where the error function, $\text{erf}(z)$, is given by
\begin{equation}
    \text{erf}(z)=\frac{2}{\sqrt{\pi}}\int_{0}^{z}e^{-t^2}dt.
    \label{1.1.2.5}
\end{equation}

The Mittag-Leffler function can also be extended as follows: \vspace{-4pt}
\begin{equation}
    E(\alpha,\beta,z)=\sum _{n=0}^{\infty}\frac{z^n}{\Gamma (\alpha n+\beta)},
    \label{1.1.2.6}
\end{equation}
which is known as the generalized Mittag-Leffler function and has several special cases, such as the following: 
\begin{equation}
    E(1,2,z)=(e^z-1)/z, \hspace{20px} E(2,2,z^2)=\sinh{(z)}/z.
    \label{1.1.2.7}
\end{equation}
\subsection{Hypergeometric Functions}
These functions contain many particular cases of special functions and are widely used in fractional calculus. From the following definition: 
\begin{equation}
    {}^{}_{p}{F_q}(\{a_i\};\{b_j\};z)=\frac{\prod_{j=1}^{q}\Gamma (b_j)}{\prod_{i=1}^{p}\Gamma (a_i)}\sum _{n=0}^{\infty}\frac{\prod_{i=1}^{p}\Gamma (a_i+n)z ^n}{\prod_{j=1}^{q}\Gamma (b_j+n)n!},
    \label{1.1.3.1}
\end{equation}

It follows that the derivative with respect to $z$ is as follows:
\begin{equation}
    \frac{d}{dz}\left[\hspace{0px}{}^{}_{p}{F_q}(\{a_i\};\{b_j\};z)\right]=\frac{ \prod_{i=1}^{p}\Gamma (a_i)}{\prod_{j=1}^{q}\Gamma (b_j)}\hspace{0px}\left[{}^{}_{ p}{F_q}(\{a_i+1\},\{b_j+1\};z)\right].
    \label{1.1.3.2}
    \end{equation}

\subsection{Euler Polynomials: Incomplete Riemann or Hurwitz Zeta Function}

The Euler polynomials, $E_n$, are generated by (see (23.1.1), \cite{abramowitz1965handbook}), as follows: 
\begin{equation}
    E_n(z) = \lim_{w\rightarrow 0}D_w^{n} \left(\frac{2 e^{2 zw }}{1+e^{w}}\right),\label{A.268}
\end{equation}
where the first few are given as follows: \vspace{-9pt}
\begin{align}
  E_0 (z) &  = 1,  \label{A.269} \\
  E_1 (z) &  = z - 1/2, \label{A.270}\\
  E_2 (z) & = z^2 - z. \label{A.271}
\end{align}

The Euler polynomials fulfill the following relation: 
\begin{equation}
    E_n (1 + z) + E_n (z) = 2z^n. \label{A.272}
\end{equation}

Extending the definition of the Euler polynomials from $n\in\mathbb{N}$ to the fractional case, $\alpha\in \mathbb{R}$, can be conducted easily,  interpreting the generating function of the Euler polynomials as the sum of a geometric series, as follows: \vspace{-4pt}
\begin{equation}
    \sum_{n=0}^{\infty} q^n = \frac{1}{1-q}, \qquad |q|<1,
\end{equation}
to obtain 
\begin{equation}
E_n(z)= \lim_{w\rightarrow 0} 2 \sum_{n=0}^{\infty} (-1)^{n} D_w^{n} e^{(z + n) w} , \qquad w < 0, \quad z \geq 0. \label{A.275}
\end{equation}

This rule may now easily be extended to the non-integer case by replacing $n \mapsto \alpha$. The ad hoc extension of the Euler polynomials to the fractional case is given by the following: 
\begin{equation}
        E_\alpha (z)= 2^{1+\alpha} \left(\zeta(-\alpha, z/2) - \zeta(-\alpha,(1 + z)/2)\right). \label{A19}
\end{equation} 
where the Hurwitz $\zeta$ function is given by the following: \vspace{-4pt}
\begin{equation}
    \zeta(s, z)= \sum_{n=0}^{\infty} (n+z)^{-s}. \label{A18}
\end{equation}

This ad hoc extension of the Euler polynomials to the fractional case corresponds to the use of the Liouville fractional derivative. Furthermore, the following relation has been confirmed:  
\begin{equation}
    E_{\alpha} (1 + z) + E_{\alpha} (z) = 2z^{\alpha} \label{A.272b}
\end{equation} independently of the fractional derivative.


\begin{thebibliography}{999}

\bibitem[Bandyopadhyay and Kamal, 2014]{bandyopadhyay2014stabilization}
Bandyopadhyay, B.; Kamal, S.
\newblock {\em Stabilization and Control of Fractional Order Systems: A Sliding
Mode Approach};
\newblock Lecture Notes in Electrical Engineering; Springer International
Publishing: Berlin/Heidelberg, Germany, 2014.

\bibitem[Herrmann, 2014]{herrmann2014fractional}
Herrmann, R. 
\newblock {\em Fractional Calculus: An Introduction for Physicists}, 2nd ed.;
\newblock World Scientific Publishing Company: Singapore, 2014.

\bibitem[Klafter et~al., 2012]{klafter2012fractional}
Klafter, J.; Lim, S.C.; Metzler, R. 
\newblock {\em Fractional Dynamics: Recent Advances};
\newblock World Scientific: Singapore, 2012.


\bibitem[Lorenzo and Hartley, 2016]{lorenzo2016fractional}
Lorenzo, C.F.; Hartley, T.T. 
\newblock {\em The Fractional Trigonometry: With Applications to Fractional
Differential Equations and Science};
\newblock Wiley: Hoboken, NJ, USA, 2016.

\bibitem[Malinowska et~al., 2015]{malinowska2015advanced}
Malinowska, A.B.; Odzijewicz, T.; Torres, D.F.M.
\newblock {\em Advanced Methods in the Fractional Calculus of Variations};
\newblock Springer briefs in applied sciences and technology; Springer
International Publishing: Berlin/Heidelberg, Germany, 2015.

\bibitem[Monje et~al., 2010]{monje2010fractional}
Monje, C.A.; Chen, Y.Q.; Vinagre, B.M.; Xue, D.; Feliu-Batlle, V.
\newblock {\em Fractional-Order Systems and Controls: Fundamentals and
Applications};
\newblock Advances in Industrial Control; Springer: London, UK, 2010.

\bibitem[Padula and Visioli, 2014]{padula2014advances}
Padula, F.; Visioli, A.
\newblock {\em Advances in Robust Fractional Control};
\newblock Springer International Publishing: Berlin/Heidelberg, Germany, 2014.

\bibitem[Tarasov, 2013]{Tarasov2013}
Tarasov, V.E.
\newblock Review of some promising fractional physical models.
\newblock {\em Int. J. Mod. Phys. B} {\bf 2013}, {\em 27}, {1330005}.

\bibitem[Tarasov, 2019]{tarasov2019applications}
Tarasov, V.E.
\newblock {\em Applications in Physics, Part A};
\newblock De Gruyter Reference; De Gruyter: Berlin, Germany, 2019.



\bibitem[Kilbas et~al., 2006]{book1:2006}
Kilbas, A.; Srivastava, H.; Trujillo, J.
\newblock Theory and applications of fractional differential equations.
\newblock In {\em North Holland Mathematical Studies}; Elsevier: Amsterdam, The Netherlands, 2006; Volume 204.





\bibitem[Podlubny, 1998]{book2:1999}
Podlubny, I.
\newblock {\em Fractional Differential Equations};
\newblock Elsevier: Amsterdam, The Netherlands, 1998; Volume 198



\bibitem[Uchaikin, 2013]{Uch:2013}
Uchaikin, V.V.
\newblock {\em Fractional Derivatives for Physicists and Engineers};
\newblock Higher Education Press: Beijing, China, 2013.
\bibitem[Wheatcraft and Meerschaert, 2008]{Wheatcraft-Meerschaert-2008} Wheatcraft, S.W.; Meerschaert, M.M. Fractional conservation of mass. {\em Adv. Water Resour.} {\bf 2008}, {\em 31}, 1377--1381.

\bibitem[Oldham, 1972]{Oldham-1972} Oldham, K.B. Signal-independent electroanalytical method. {\em Anal. Chem.} {\bf 1972}, {\em 44}, 196--198.

\bibitem[Pospíšil, Hromadová, Sokolová, \& Lanza, 2019]{Pospíšil2019}
Pospíšil, L.; Hromadová, M.; Sokolová, R.; Lanza, C. Kinetics of radical dimerization. Simple evaluation of rate constant from convolution voltammetry and faradaic phase angle data. {\em Electrochim. Acta} {\bf 2019}, {\em 300}, 284--289.

\bibitem[Atangana and Bildik, 2013]{Atangana-Bildik-2013}
Atangana, A.; Bildik, N. The use of fractional order derivative to predict the groundwater flow. {\em Math. Probl. Eng.} {\bf 2013}, {\em 2013},~543026.

\bibitem[Atangana and Vermeulen, 2014]{Atangana-Vermeulen-2014}
Atangana, A.; Vermeulen, P.D. Analytical solutions of a space-time fractional derivative of groundwater flow equation. In {\em Abstract and Applied Analysis}; Hindawi Limited: London, UK, 2014; Volume 2014, pp. 1--11.

\bibitem[Benson, Wheatcraft, \& Meerschaert, 2000a]{Benson-Wheatcraft-Meerschaert-2000a}
Benson, D.A.; Wheatcraft, S.W.; Meerschaert, M.M. Application of a fractional advection‐dispersion equation. {\em Water Resour. Res.} {\bf 2000}, {\em 36}, 1403--1412. 
\bibitem[Benson, Wheatcraft, \& Meerschaert, 2000b]{Benson-Wheatcraft-Meerschaert-2000b}
Benson, D.A.; Wheatcraft, S.W.; Meerschaert, M.M. The fractional‐order governing equation of Lévy motion. {\em Water Resour. Res.} {\bf 2000}, {\em 36}, 1413--1423.

\bibitem[Benson, Schumer, Meerschaert \& Wheatcraft, 2001]{Benson-Schumer-Meerschaert-Wheatcraft-2001}
Benson, D.A.; Schumer, R.; Meerschaert, M.M.; Wheatcraft, S.W. Fractional dispersion, Lévy motion, and the MADE tracer tests. {\em Transp. Porous Media} {\bf 2001}, {\em 42}, 211--240.


\bibitem[Metzler and Klafter, 2000]{Metzler-Klafter-2000} Metzler, R.; Klafter, J. The random walk's guide to anomalous diffusion: A fractional dynamics approach. {\em Phys. Rep.} {\bf 2000}, {\em 339}, 1--77.

\bibitem[Mainardi, Luchko \& Pagnini, 2007]{Mainardi-Luchko-Pagnini-2007}
Mainardi, F.; Luchko, Y.; Pagnini, G. The fundamental solution of the space-time fractional diffusion equation. {\em arXiv} {\bf 2007}, {arXiv:cond-mat/0702419}.

\bibitem[Atangana \& Kilicman, 2014]{Atangana-Kilicman-2014}
Atangana, A.; Kilicman, A. On the generalized mass transport equation to the concept of variable fractional derivative. {\em Math. Probl. Eng.} {\bf 2014}, {\em 2014}, {542809}. 

\bibitem[Gorenflo \& Mainardi, 2003]{Gorenflo-Mainardi-2003} 
Gorenflo, R.; Mainardi, F. Fractional diffusion processes: Probability distributions and continuous time random walk. In {\em Processes with Long-Range Correlations: Theory and Applications}; Springer: Berlin/Heidelberg, Germany, 2003; pp. 148--166. 

\bibitem[Colbrook, Ma, Hopkins, \& Squire, 2017]{Colbrook-Ma-Hopkins-Squire-2017}
Colbrook, M.J.; Ma, X.; Hopkins, P.F.; Squire, J. Scaling laws of passive-scalar diffusion in the interstellar medium. {\em Mon. Not. R. Astron. Soc.} {\bf 2017}, {\em 467}, 2421--2429.

\bibitem[Mainardi, 2022]{Mainardi-2022}
Mainardi, F. {\em Fractional Calculus and Waves in Linear Viscoelasticity: An Introduction to Mathematical Models}; World Scientific: Singapore, 2022. 

\bibitem[Laskin, 2000]{Laskin-2000}
Laskin, N. Fractional quantum mechanics. {\em Phys. Rev. E} {\bf 2000}, {\em 62}, 3135.

\bibitem[Laskin, 2002]{Laskin-2002}
Laskin, N. Fractional schrödinger equation. {\em Phys. Rev. E} {\bf 2002}, {\em 66}, 056108.

\bibitem[Bhrawy \& Zaky, 2017]{Bhrawy-Zaky-2017}
Bhrawy, A.H.; Zaky, M. An improved collocation method for multi-dimensional space–time variable-order fractional Schrödinger equations. {\em Appl. Numer. Math.} {\bf 2017}, {\em 111}, 197--218.

\bibitem[Lim, 2006]{Lim:2006hp}
Lim, S.C. 
\newblock {Fractional derivative quantum fields at positive temperature}.
\newblock {\em Phys. A} {\bf 2006}, {\em 363}, 269--281.

\bibitem[Lim and Eab, 2019]{LimEab+2019+237+256}
Lim, S.C.; Eab, C.H.
\newblock {\em Fractional Quantum Fields}; 
\newblock De Gruyter: Berlin, Germany, 2019; pp. 237--256.

\bibitem[Moniz and Jalalzadeh, 2020]{Moniz:2020emn}
Moniz, P.V.; Jalalzadeh, S. 
\newblock {From Fractional Quantum Mechanics to Quantum Cosmology: An Overture}.
\newblock {\em Mathematics} {\bf 2020}, {\em 8}, 313.


\bibitem[Rasouli et~al., 2021]{Rasouli:2021lgy}
Rasouli, S.M.M.; Jalalzadeh, S.; Moniz, P.V.
\newblock {Broadening quantum cosmology with a fractional whirl}.
\newblock {\em Mod. Phys. Lett. A} {\bf 2021}, {\em 36}, 2140005.

\bibitem[V.~Moniz and Jalalzadeh, 2020]{VargasMoniz:2020hve}
Moniz, P.V.; Jalalzadeh, S.
\newblock {\em {Challenging Routes in Quantum Cosmology}};
\newblock World Scientific Publishing: Singapore, 2020.

\bibitem[Calcagni and Kuroyanagi, 2021]{Calcagni:2020tvw}
Calcagni, G.; Kuroyanagi, S.
\newblock {Stochastic gravitational-wave background in quantum gravity}.
\newblock {\em J. Cosmol. Astropart. Phys.} {\bf 2021}, {\em 3}, 019.


\bibitem[El-Nabulsi, 2013a]{El-Nabulsi:2013hsa}
El-Nabulsi, R.A.
\newblock {Fractional derivatives generalization of Einstein`s field
equations}.
\newblock {\em Indian J. Phys.} {\bf 2013}, {\em 87},~195--200.
\newblock {{https://doi.org/10.1007/s12648-012-0201-4}}.

\bibitem[El-Nabulsi, 2013b]{El-Nabulsi:2013mwa}
El-Nabulsi, R.A.
\newblock {Non-minimal coupling in fractional action cosmology}.
\newblock {\em Indian J. Phys.} {\bf 2013}, {\em 87}, 835--840.

\bibitem[Jalalzadeh et~al., 2021]{Jalalzadeh:2021gtq}
Jalalzadeh, S.; da~Silva, F.R.; Moniz, P.V. 
\newblock {Prospecting black hole thermodynamics with fractional quantum
mechanics}.
\newblock {\em Eur. Phys. J. C} {\bf 2021}, {\em 81}, 632.

\bibitem[Vacaru, 2012a]{Vacaru:2010wn}
Vacaru, S.I. 
\newblock {Fractional Dynamics from Einstein Gravity, General Solutions, and
Black Holes}.
\newblock {\em Int. J. Theor. Phys.} {\bf 2012}, {\em 51}, 1338--1359.

\bibitem[Vacaru, 2012b]{Vacaru:2010wj}
Vacaru, S.I.
\newblock {Fractional Nonholonomic Ricci Flows}.
\newblock {\em Chaos Solitons Fractals} {\bf 2012}, {\em 45}, 1266--1276.

\bibitem[Rami, 2009]{rami2009fractional}
Rami, E.N.A.
\newblock Fractional dynamics, fractional weak bosons masses and physics beyond
the standard model.
\newblock {\em Chaos Solitons Fractals} {\bf 2009}, {\em 41},~2262--2270.

\bibitem[Debnath et~al., 2012]{Debnath2012}
Debnath, U.; Jamil, M.; Chattopadhyay, S. 
\newblock {Fractional Action Cosmology: Emergent, Logamediate, Intermediate,
Power Law Scenarios of the Universe and Generalized Second Law of Thermodynamics}.
\newblock {\em Int. J. Theor. Phys.} {\bf 2012}, {\em 51}, 812--837.

\bibitem[Debnath et~al., 2013]{Debnath2013}
Debnath, U.; Chattopadhyay, S.; Jamil, M. 
\newblock {Fractional action cosmology: Some dark energy models in emergent,
logamediate, and intermediate scenarios of the universe}.
\newblock {\em J. Theor. Appl. Phys.} {\bf 2013}, {\em 7}, 25.

\bibitem[El-Nabulsi, 2007]{El-Nabulsi:2007wgc}
El-Nabulsi, R.A. 
\newblock {Cosmology with a fractional action principle}.
\newblock {\em Rom. Rep. Phys.} {\bf 2007}, {\em 59}, 763--771.


\bibitem[El-Nabulsi, 2009]{el2009fractional}
El-Nabulsi, R.A.
\newblock Fractional Lagrangian Formulation of General Relativity and Emergence
of Complex, Spinorial and Noncommutative Gravity.
\newblock {\em Int. J. Geom. Methods Mod. Phys.}
{\bf 2009}, {\em 6},~25--76.


\bibitem[Jamil et~al., 2012]{Jamil:2011uj}
Jamil, M.; Momeni, D.; Rashid, M.A. 
\newblock {Fractional Action Cosmology with Power Law Weight Function}.
\newblock {\em J. Phys. Conf. Ser.} {\bf 2012}, {\em 354},~012008.

\bibitem[Shchigolev, 2016]{Shchigolev:2015rei}
Shchigolev, V.K. 
\newblock {Testing Fractional Action Cosmology}.
\newblock {\em Eur. Phys. J. Plus} {\bf 2016}, {\em 131}, 256.

\bibitem[Giusti, 2020]{Giusti:2020rul}
Giusti, A.
\newblock {MOND-like Fractional Laplacian Theory}.
\newblock {\em Phys. Rev. D} {\bf 2020}, {\em 101}, 124029.

\bibitem[Rami, 2015]{Rami:2015kha}
Rami, E.-N.A.
\newblock {Fractional action oscillating phantom cosmology with conformal
coupling}.
\newblock {\em Eur. Phys. J. Plus} {\bf 2015}, {\em 130}, 102.

\bibitem[Calcagni, 2010a]{Calcagni:2009kc}
Calcagni, G. 
\newblock {Fractal universe and quantum gravity}.
\newblock {\em Phys. Rev. Lett.} {\bf 2010}, {\em 104}, 251301.


\bibitem[Calcagni, 2010b]{Calcagni:2010bj}
Calcagni, G.
\newblock {Quantum field theory, gravity and cosmology in a fractal universe}.
\newblock {\em J. High Energy Phys.} {\bf 2010}, {\em 3}, 120.

\bibitem[Calcagni, 2013]{Calcagni:2013yqa}
Calcagni, G.
\newblock {Multi-scale gravity and cosmology}.
\newblock {\em J. Cosmol. Astropart. Phys.} {\bf 2013}, {\em 12}, 041.


\bibitem[Calcagni, 2021b]{Calcagni_2021}
Calcagni, G. 
\newblock Quantum scalar field theories with fractional operators.
\newblock {\em Class. Quantum Gravity} {\bf 2021}, {\em 38}, 165006.

\bibitem[Calcagni, 2021a]{Calcagni:2021aap}
Calcagni, G.
\newblock {Classical and quantum gravity with fractional operators}.
\newblock {\em Class. Quantum Gravity} {\bf 2021}, {\em 38}, 165005;
\newblock Erratum in {\em Class. Quantum Gravity} {\bf 2021}, {\em 38}, 169601.




\bibitem[Calcagni and De~Felice, 2020]{Calcagni:2020ads}
Calcagni, G.; De~Felice, A.
\newblock {Dark energy in multifractional spacetimes}.
\newblock {\em Phys. Rev. D} {\bf 2020}, {\em 102}, 103529.



\bibitem[Calcagni et~al., 2019]{Calcagni:2019ngc}
Calcagni, G.; Kuroyanagi, S.; Marsat, S.; Sakellariadou, M.; Tamanini, N.; Tasinato, G.
\newblock {Quantum gravity and gravitational-wave astronomy}.
\newblock {\em J. Cosmol. Astropart. Phys.} {\bf 2019}, {\em 10}, 012.

\bibitem[Calcagni et~al., 2016]{Calcagni:2016ofu}
Calcagni, G.; Kuroyanagi, S.; Tsujikawa, S.
\newblock {Cosmic microwave background and inflation in multi-fractional
spacetimes}.
\newblock {\em J. Cosmol. Astropart. Phys.} {\bf 2016}, {\em 8}, 039.

\bibitem[Vacaru, 2010]{Vacaru:2010fb}
Vacaru, S.I.
\newblock {New Classes of Off-Diagonal Cosmological Solutions in Einstein Gravity}.
\newblock {\em Int. J. Theor. Phys.} {\bf 2010}, {\em 49}, 2753--2776.

\bibitem[Roberts, 2014]{Roberts:2009ix}
Roberts, M.D.
\newblock {fractional derivative Cosmology}.
\newblock {\em SOP Trans. Theor. Phys.} {\bf 2014}, {\em 1}, 310.

\bibitem[Shchigolev, 2011]{Shchigolev:2010vh}
Shchigolev, V.K.
\newblock {Cosmological Models with fractional derivatives and Fractional
Action Functional}.
\newblock {\em Commun. Theor. Phys.} {\bf 2011}, {\em 56}, 389--396.


\bibitem[Shchigolev, 2013a]{Shchigolev:2012rp}
Shchigolev, V.K. 
\newblock {Cosmic Evolution in Fractional Action Cosmology}.
\newblock {\em Discontinuity Nonlinearity Complex.} {\bf 2013}, {\em 2}, 115--123.

\bibitem[Shchigolev, 2013b]{Shchigolev:2013jq}
Shchigolev, V.K.
\newblock {Fractional Einstein-Hilbert Action Cosmology}.
\newblock {\em Mod. Phys. Lett. A} {\bf 2013}, {\em 28}, 1350056.



\bibitem[Shchigolev, 2021]{Shchigolev:2021lbm}
Shchigolev, V.K.
\newblock {Fractional-order derivatives in cosmological models of accelerated
expansion}.
\newblock {\em Mod. Phys. Lett. A} {\bf 2021}, {\em 36}, 2130014.

\bibitem[García-Aspeitia et~al., 2022]{Garcia-Aspeitia:2022uxz}
García-Aspeitia, M.A.; Fernandez-Anaya, G.; Hernández-Almada, A.; Leon, G.; Magaña, J.
\newblock {Cosmology under the fractional calculus approach}.
\newblock {\em Mon. Not. R. Astron. Soc.} {\bf 2022}, {\em 517}, 4813--4826.


\bibitem[Jalalzadeh et~al., 2022]{Jalalzadeh:2022uhl}
Jalalzadeh, S.; Costa, E.W.O.; Moniz, P.V. 
\newblock {de Sitter fractional quantum cosmology}.
\newblock {\em Phys. Rev. D} {\bf 2022}, {\em 105}, L121901.

\bibitem[Landim, 2021{\natexlab{a}}]{Landim:2021ial}
Landim, R.G.
\newblock {Fractional dark energy: Phantom behavior and negative absolute
temperature}.
\newblock {\em Phys. Rev. D} {\bf 2021}, {\em 104},~103508.
\newblock {{https://doi.org/10.1103/PhysRevD.104.103508}}.

\bibitem[Landim, 2021{\natexlab{b}}]{Landim:2021www}
Landim, R.G.
\newblock {Fractional dark energy}.
\newblock {\em Phys. Rev. D} {\bf 2021}, {\em 103},~083511.
\newblock {{https://doi.org/10.1103/PhysRevD.103.083511}}.

\bibitem[Micolta-Riascos et~al., 2023]{Micolta-Riascos:2023mqo}
Micolta-Riascos, B.; Millano, A.D.; Leon, G.; Erices, C.; Paliathanasis, A. 
\newblock Revisiting fractional cosmology.
\newblock {\em Fractal Fract.} {\bf 2023}, {\em 7}, 149.

\bibitem[Gonz\'alez et~al., 2023]{Gonzalez:2023who}
Gonz\'alez, E.; Leon, G.; Fernandez-Anaya, G.
\newblock {Exact solutions and cosmological constraints in fractional
cosmology}.
\newblock {\em Fractal Fract.} {\bf 2023}, {\em 7}, 368.

\bibitem[Leon Torres et~al., 2023]{LeonTorres:2023ehd}
Leon, G.; García-Aspeitia, M.A.; Fernandez-Anaya, G.; Hern\'andez-Almada, A.; Maga\~na, J.; Gonz\'alez, E. 
\newblock Cosmology under the fractional calculus approach: A possible $H_0$ tension resolution? 
\newblock {\em PoS } {\bf 2023}, {\em CORFU2022}, 248.
\newblock {https://doi.org/10.22323/1.436.0248}.


\bibitem[Wolfram~Research, 2023]{Mathematica}
Wolfram~Research. 
\newblock {\em Mathematica}; {V}ersion 13.3;
\newblock {Wolfram Research}: Champaign, IL, USA, 2023.


\bibitem[West, 2021]{west2021fractional}
West, B.J. 
\newblock {Fractional Calculus and the Future of Science}. \emph{Entropy} \textbf{2021}, \emph{23}, 1566.


\bibitem[Stanislavsky, 2004]{Stanislavsky-Fractional-Oscillator-2004}
Stanislavsky, A.
\newblock Fractional oscillator.
\newblock {\em Phys. Rev. E} {\bf 2004}, {\em 70}, 051103.


\bibitem[Arraut and~Segovia, 2018]{Arraut:2016dlm}
Arraut, I.; Segovia, C. 
\newblock A q-deformation of the Bogoliubov transformations.
\newblock{\em Phys. Lett. A} \textbf{2018}, {\em 382}, 464--466.
{https://doi.org/10.10 16/j.physleta.2017.12.044}.


\bibitem[Hou and~Xu, 1995]{Hou-Xu1995} Hou, B.Y.; Xu, L.C. \newblock The Hopf algebraic structure of q-deformed Heisenberg algebra when q is a root of unity. 
\newblock{\em Commun. Theor. Phys.} {\bf 1995}, {\em 24}, 481.


\bibitem[Bonatsos and Daskaloyannis, 1999]{Bonatsos:1999xj}
Bonatsos, D.; Daskaloyannis, C.
\newblock Quantum groups and their applications in nuclear physics.
\newblock {\em Prog. Part. Nucl. Phys.} {\bf 1999}, {\em 43}, 537--618.
\newblock {https://doi.org/10.1016/S0146-6410(99)00100-3}.


\bibitem[Barrientos et~al., 2021]{Barrientos:2020kfp}
Barrientos, E.; Mendoza, S.; Padilla, P.
\newblock {Extending Friedmann equations using fractional derivatives using a Last Step Modification technique: The case of a matter dominated accelerated expanding Universe}.
\newblock {\em Symmetry} {\bf 2021}, {\em 13}, 174.

\bibitem[Frederico and Torres, 2008]{10.5555/1466940.1466942}
Frederico, G.S.F.; Torres, D.F.M. 
\newblock Necessary optimality conditions for fractional action-like problems
with intrinsic and observer times.
\newblock {\em WSEAS Trans. Math.} {\bf 2008}, {\em 7}, 6–11.

\bibitem[Agrawal, 2007]{Agrawal-2007}
Agrawal, O.P. 
\newblock Fractional variational calculus in terms of Riesz fractional derivatives.
\newblock {\em J. Phys. A Math. Theor.} {\bf 2007}, {\em 40}, 6287.

\bibitem[Baleanu and Muslih, 2005]{Baleanu-Muslih-2005}
Baleanu, D.; Muslih, S. 
\newblock Lagrangian formulation of classical fields within Riemann-Liouville fractional derivatives.
\newblock {\em Phys. Scripta}  {\bf 2005}, {\em 72}, 119–121.
\newblock {https://doi.org/doi:10.1238/Physica.Regular.072a00119}. 



\bibitem[Baleanu and Trujillo, 2010]{Baleanu-Trujillo-2010}
Baleanu, D.; Trujillo, J. 
\newblock A new method of finding the fractional Euler-Lagrange and Hamilton
equations within caputo fractional derivatives.
\newblock {\em Commun. Nonlinear Sci. Numer. Simul.} {\bf 2010}, {\em 15}, 1111--1115.


\bibitem[El-Nabulsi and Torres, 2008]{El-Nabulsi-Torres-2008}
El-Nabulsi, R.A.; Torres, D.F. 
\newblock Fractional action-like variational problems.
\newblock {\em J. Math. Phys.} {\bf 2008}, {\em 49}, {053521}.


\bibitem[Odzijewicz et~al., 2013a]{Odzijewicz-Malinowska-Torres-2013a}
Odzijewicz, T.; Malinowska, A.; Torres, D. 
\newblock Generalized fractional calculus with applications to the calculus of variations. 
\newblock {\em Comput. Math. Appl.} {\bf 2012}, {\em 64}, 3351--3366.

\bibitem[Odzijewicz et~al., 2013b]{Odzijewicz-Malinowska-Torres-2013b}
Odzijewicz, T.; Malinowska, A.; Torres, D.
\newblock Noether’s theorem for fractional variational problems of variable
order. 
\newblock {\em Open Phys.} {\bf 2013}, {\em 11}, 691--701.

\bibitem[Odzijewicz et~al., 2013c]{Odzijewicz-Malinowska-Torres-2013c}
Odzijewicz, T.; Malinowska, A.; Torres, D. 
\newblock Variable order fractional variational calculus for double integrals.
\newblock In Proceedings of the 2012 IEEE 51st IEEE Conference on Decision and Control (CDC), {Maui, HI, USA, 10--13 December 2012}; IEEE: {Piscataway, NJ, USA}, 2013; pp. 6873--6878.



\bibitem[Hern\'andez-Almada et~al., 2020]{Hernandez-Almada:2020uyr}
Hern\'andez-Almada, A.; Leon, G.; Maga\~na, J.; Garc\'\i{}a-Aspeitia, M.A.; Motta, V. 
\newblock {Generalized Emergent Dark Energy: Observational Hubble data constraints and stability analysis}.
\newblock {\em Mon. Not. R. Astron. Soc.} {\bf 2020}, {\em 497}, 1590--1602.

\bibitem[Hern\'andez-Almada et~al., 2022]{Hernandez-Almada:2021aiw}
Hern\'andez-Almada, A.; Leon, G.; Maga\~na, J.; Garc\'\i{}a-Aspeitia, M.A.; Motta, V.; Saridakis, E.N.; Yesmakhanova, K. 
\newblock {Kaniadakis-holographic dark energy: Observational constraints and
global dynamics}.
\newblock {\em Mon. Not. R. Astron. Soc.} {\bf 2022}, {\em 511}, 4147--4158.

\bibitem[Hern\'andez-Almada et~al., 2021]{Hernandez-Almada:2021rjs}
Hern\'andez-Almada, A.; Leon, G.; Maga\~na, J.; Garc\'\i{}a-Aspeitia, M.A.; Motta, V.; Saridakis, E.N.; Yesmakhanova, K.; Millano, A.D.
\newblock {Observational constraints and dynamical analysis of Kaniadakis
horizon-entropy cosmology}.
\newblock {\em Mon. Not. R. Astron. Soc.} {\bf 2021}, {\em 512}, 5122--5134.
\bibitem[Leon et~al., 2021]{Leon:2021wyx}
Leon, G.; Maga\~na, J.; Hern\'andez-Almada, A.; Garc\'\i{}a-Aspeitia, M.A.; Verdugo, T.; Motta, V.
\newblock {Barrow Entropy Cosmology: An observational approach with a hint of
stability analysis}.
\newblock {\em J. Cosmol. Astropart. Phys.} {\bf 2021}, {\em 12}, 032.




\bibitem[Di~Valentino et~al., 2021]{DiValentino:2021izs}
Di Valentino, E.; Mena, O.; Pan, S.; Visinelli, L.; Yang, W.; Melchiorri, A.; Mota, D.F.; Riess, A.G.; Silk, J. 
\newblock {In the realm of the Hubble tension\textemdash{}A review of
solutions}.
\newblock {\em Class. Quantum Gravity} {\bf 2021}, {\em 38}, 153001.
\bibitem[Efstathiou, 2021]{Efstathiou:2021ocp}
Efstathiou, G. 
\newblock {To H0 or not to H0?}
\newblock {\em Mon. Not. R. Astron. Soc.} {\bf 2021}, {\em 505}, 3866--3872.

\bibitem[Aghanim et~al., 2020]{Planck:2018vyg}
Aghanim, N.; Akrami, Y.; Ashdown, M.; Aumont, J.; Baccigalupi, C.; Ballardini, M.; Banday, A.J.; Barreiro, R.B.; Bartolo, N.; Basak, S.; et~al. 
\newblock {Planck 2018 results. VI. Cosmological parameters}.
\newblock {\em Astron. Astrophys.} {\bf 2020}, {\em 641}, A6;
\newblock Erratum in {\em Astron. Astrophys.} {\bf 2021}, {\em 652}, C4.

\bibitem[Carroll et~al., 2004]{carroll2004spacetime}
Carroll, S.; Addison-Wesley.
\newblock {\em Spacetime and Geometry: An Introduction to General Relativity};
\newblock Addison Wesley: Boston, MA, USA, 2004.

\bibitem[Carroll, 2019]{Carroll}
Carroll, S.M. 
\newblock {\em Spacetime and Geometry};
\newblock Cambridge University Press: Cambridge, UK, 2019.

\bibitem[Wald, 2010]{Wald}
Wald, R.M.
\newblock {\em General Relativity};
\newblock University of Chicago Press: Chicago, {IL, USA}, 2010.

\bibitem[Foreman-Mackey et~al., 2013]{Foreman:2013}
Foreman-Mackey, D.; Hogg, D.W.; Lang, D.; Goodman, J.
\newblock emcee: The {MCMC} hammer.
\newblock {\em Publ. Astron. Soc. Pac.} {\bf 2013}, {\em 125}, 306--312.

\bibitem[Tavakol, 1997]{TWE}
Tavakol, R.
\newblock {\em Introduction to Dynamical Systems, Ch 4. Part One};
\newblock Cambridge University Press: Cambridge, UK, 1997; pp. 84--98.


\bibitem[Wainwright and Ellis, 1997]{wainwrightellis1997}
Wainwright, J.; Ellis, G.F.R. (Eds.) 
\newblock {\em {Dynamical Systems in Cosmology}};
\newblock Cambridge University Press: Cambridge, UK, 1997.

\bibitem[Perko, 2001]{perko}
Perko, L. 
\newblock {\em Differential Equations and Dynamical Systems}; 3rd ed.;
\newblock Springer: Berlin/Heidelberg, Germany, 2001.

\bibitem[Coley, 2003]{Coley:2003mj}
Coley, A.
\newblock {\em {Dynamical Systems and Cosmology}}; 
\newblock Kluwer: Dordrecht, The Netherlands, 2003; Volume 291.


\bibitem[Hirsch et~al., 2004]{Smale}
Hirsch, M.W.; Smale, S.; Devaney, R.
\newblock {\em {Differential Equations, Dynamical Systems, and An Introduction
to Chaos}};
\newblock Academic Press: London, UK; San Diego, CA, USA, 2004.

\bibitem[Wiggins, 2006]{wiggins2006introduction}
Wiggins, S.
\newblock {\em Introduction to Applied Nonlinear Dynamical Systems and Chaos};
\newblock Texts in Applied Mathematics; Springer: Berlin/Heidelberg, Germany, 2006.

\bibitem[Berglund and Gentz, 2006]{Berglund} Berglund, N.; Gentz, B.B. {\em Noise-Induced Phenomena in Slow-Fast Dynamical Systems}; Series: Probability and Applications; Springer: London, UK, 2006.

\bibitem[Fenichel, 1979]{fenichel} Fenichel, N. Geometric singular perturbation theory for ordinary differential equations. {\em J. Differ. Equ.} {\bf 1979}, {\em 31}, 53--98. 

\bibitem[Fusco and Hale, 1988]{Fusco} Fusco, G.; Hale, J.K. {\em J. Dyn. Differ. Equ.} {\bf 1988}, {\em 1}, 75.

\bibitem[Dumortier and Roussarie, 1995]{dumortier} Dumortier, F.; Roussarie, R. {\em Canard Cycles and Center Manifolds}
\newblock{\em Memoirs of the American Mathematical Society} 
{\bf 1996}, {\em 121}, 577. 
\newblock {https://doi.org/10.1090/memo/0577}.

\bibitem[Holmes, 2013]{holmes}
Holmes, M.H. {\em Introduction to Perturbation Methods}; Springer Science+Business Media: New York, NY, USA, 2013; ISBN 978-1-4614-5477-9. {https://doi.org/10.1007/978-1-4614-5477-9}.

\bibitem[Kevorkian, 1981]{Kevorkian1} Kevorkian, J., 
Cole, J.D. {\em Perturbation Methods in Applied Mathematics}; Applied Mathematical Sciences Series; Springer: New York, NY, USA, 1981; Volume 34,
 ISBN 978-1-4757-4213-8. {https://doi.org/10.1007/978-1-4612-3968-0}.

\bibitem[Verhulst, 2000]{Verhulst} Verhulst, F. {\em Methods and Applications of Singular Perturbations: Boundary Layers and Multiple Timescale Dynamics}; Springer: New York, NY, USA, 2000; ISBN 978-0-387-22966-9. {https://doi.org/10.1007/0-387-28313-7}.


\bibitem[Popp, 2000]{POPP2000765}
Popp, K. 
\newblock Non-smooth mechanical systems.
\newblock {\em J. Appl. Math. Mech.} {\bf 2000}, {\em 64}, 765--772.

\bibitem[Abramowitz \& Stegun, 1965]{abramowitz1965handbook}
Abramowitz, M.; Stegun, I.A. {\em Handbook of Mathematical Functions}; Dover Publications: New York, NY, USA, 1965; Volume 361.



\end{thebibliography}
\end{document}